\documentclass[a4paper, 12pt]{article}
\usepackage{style_new_designs}

\usepackage{enumitem}
\usepackage[colorlinks=true, allcolors=blue]{hyperref}
\usepackage{setspace} 
\usepackage{etoolbox}

\AtBeginEnvironment{thebibliography}{\singlespacing}

\date{August 2026}
\title{Finely Stratified Rerandomization Designs}
\author{Max Cytrynbaum\footnote{Yale Department of Economics. Correspondence: max.cytrynbaum@yale.edu}}

\begin{document}
\maketitle

\begin{abstract}
Finely stratified randomization makes unadjusted treatment effect estimation efficient when matches are tight.
However, match quality deteriorates rapidly as covariate dimension increases, attenuating the gains from stratification.
Motivated by this problem, we study designs that finely stratify on a few important covariates, then rerandomize within matched groups to balance the remainder.
We derive the asymptotic distribution for GMM estimators of general causal parameters under such designs, showing that they provide nonparametric control over the stratification covariates and linear control over the rerandomization covariates.
The resulting distribution is generally non-normal, but optimal linear adjustment restores asymptotic normality.
For finite population parameters, we derive upper bounds on the non-identified asymptotic variance, enabling conservative inference that accounts for the efficiency gains from both design stages.
An empirical application to estimating treatment effect heterogeneity among compliers illustrates the gains from adding rerandomization to a stratified design.
\end{abstract}

\noindent{\emph{Keywords}: Matched Pairs, GMM, Finite Population Inference.}

\noindent{\emph{JEL Codes}: C13, C21, C90.}

\onehalfspacing
\newpage

\section{Introduction} \label{section:introduction}

Stratified randomization is commonly used to increase precision in experimental research.
For example, \cite{cytrynbaum2024adjustment} reports a survey of 50 experimental papers in the AER and AEJ from 2018-2023, where 57\% used some form of stratified randomization.
This practice is supported by recent theoretical results showing that finely stratified designs like matched pairs make unadjusted estimators automatically semiparametrically efficient \citep{bai2021inference,bai2024efficiency}.

However, such results rely on tight-matching conditions that require covariate differences within matched groups to vanish asymptotically. 
These conditions rapidly become implausible when the researcher attempts to stratify on more than a few covariates.  
To see why, note that average nearest-neighbor distances among $n$ units with $m$ covariates have characteristic scale $n^{-1/m}$.
Because of this, even large increases in sample size produce only modest improvements in match quality unless covariate dimension $m$ is very small. 

Indeed, \cite{cytrynbaum2026limits} shows that the finite sample excess variance of matched-pairs randomization above the \cite{hahn1998} efficiency bound cannot vanish at a rate faster than $n^{-2/m}$.
At this rate, even with only $m=6$ covariates reducing excess variance by 50\% requires the experiment to be eight times larger.
Because of this, at realistic sample sizes, fine stratification can attain the semiparametric efficiency bound only when matching on very few covariates.

By contrast, rerandomization methods \citep{morgan2012}, which repeatedly draw treatment allocations until selected covariate moments are balanced between the treatment and control groups, can trade off imbalances globally across the covariate space.
They therefore avoid the matching curse of dimensionality that specifically affects finely stratified designs.
Since researchers often collect many covariates in baseline surveys, these observations suggest a hybrid approach in which we first finely stratify to nonparametrically balance a few of the most important covariates, then rerandomize within strata to linearly control imbalances on the remaining covariates.

\begin{figure}[t]
\centering
\includegraphics[width=\textwidth]{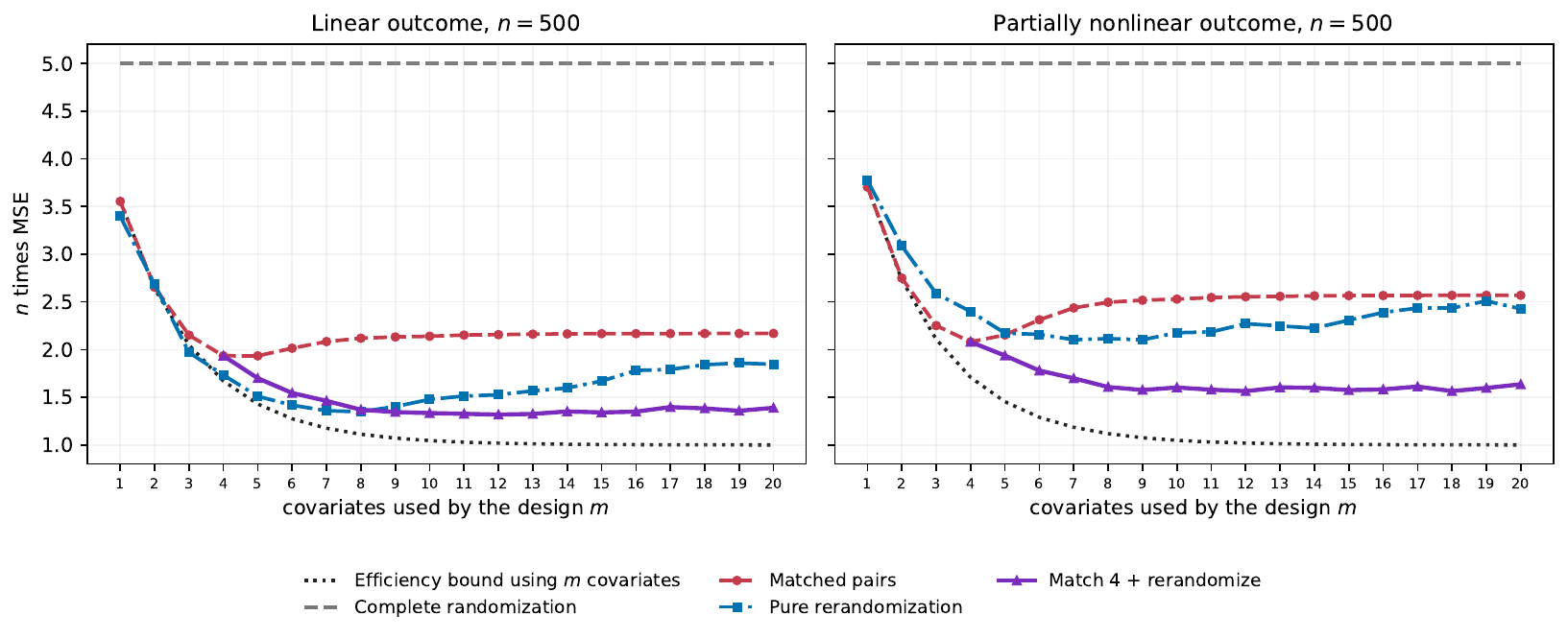}
\caption{Finite-sample efficiency of unadjusted difference in means. The outcome model is linear with geometrically decaying coefficients in the left panel, while nonlinear squares and interaction terms account for 20\% of variance in the right panel. Rerandomization is as in Example \ref{ex:mahalanobis-rerandomization} with acceptance probability $1/2000$. }
\label{fig:hybrid-dimension-introduction}
\end{figure}

In this paper, we develop estimation and inference methods for such finely stratified rerandomization designs.
The benefits of this approach can be seen clearly in Figure \ref{fig:hybrid-dimension-introduction}.
Under a linear outcome model, rerandomization weakly outperforms matched pairs throughout the displayed dimension range.
Under a partially nonlinear model, matched pairs performs best when only a few covariates are used, but its performance deteriorates rapidly as additional covariates reduce match quality.
A hybrid design that finely stratifies on the first four covariates and rerandomizes the remainder achieves the lowest MSE over most of the dimension range under both models.

Versions of finely stratified rerandomization have already been used in prominent experiments.
For example, the OpenResearch Unconditional Income Study matched 3,000 participants into triples using dozens of baseline covariates, then rerandomized until a balance criterion was satisfied \citep{broockman2024income}.
\cite{abaluck2021} similarly rerandomized by selecting the best of 50 matched-pairs assignments for a public health intervention covering more than 340,000 adults.
These applications underscore the need for theoretically grounded guidance on how to combine fine stratification and rerandomization.

We make the following main contributions:
\begin{enumerate}
	\item In Section \ref{section:gmm-asymptotics}, we derive the asymptotic distribution of GMM estimators under finely stratified rerandomization, extending the analysis of GMM under pure fine stratification in \cite{bai2024efficiency}.
	Our analysis accommodates both superpopulation as well as finite population causal parameters. The latter are more appropriate for experiments run in convenience samples, as is the case for most experiments in economics \citep{niehaus2017}.
	
	We show that under finely stratified rerandomization, unadjusted GMM behaves like GMM with partially linear adjustment, providing nonparametric control over the stratified covariates and linear control over the rerandomized covariates, up to a small residual.
	In extensions, we also study optimization of the rerandomization acceptance criteria, as well as nonlinear rerandomization designs, e.g.\ based on comparing covariate density estimates.

	\item In Section \ref{section:covariate-adjustment}, we study optimal ex-post linear adjustment of GMM estimators under finely stratified rerandomization.
	We show that optimal adjustment removes the non-Gaussian residual term that rerandomization introduces into the asymptotic distribution, restoring asymptotic normality and permitting conventional inference methods.
	We also show that combining rerandomization with adjustment provides a form of double robustness to covariate imbalances, which can improve finite-sample performance when the adjustment coefficient is estimated imprecisely.

	\item In Section \ref{section:inference}, we develop inference methods for both finite population and superpopulation GMM parameters that account for the efficiency gains from fine stratification, rerandomization, and optimal adjustment.
	The asymptotic variance for finite population causal parameters is generally not identified \citep{neyman1990}.
	Because of this, we derive identified upper bounds for general causal parameters that yield asymptotically conservative confidence intervals.
	For experiments run in a convenience sample, these intervals can be tighter than intervals for the corresponding superpopulation parameters because they do not need to account for superpopulation sampling uncertainty.
	For superpopulation parameters, the asymptotic variance is identified, and we provide asymptotically exact inference methods.
\end{enumerate}

Finally, we illustrate our results in an empirical application to treatment effect heterogeneity among compliers in \cite{angrist2013}, showing the value of adding a rerandomization step to finely stratified designs.

\subsection{Related Literature}

This paper builds on a recent literature on fine stratification in econometrics as well as the literature on rerandomization in statistics. 
Stratified randomization has a long history in statistics, see \cite{cochran1977} for a survey.
Recent work on fine stratification in econometrics includes \cite{bai2021inference}, \cite{bai2020pairs}, \cite{cytrynbaum2024}, and \cite{bai2024efficiency}. 
A sample of some recent work in the statistics literature on rerandomization includes \cite{morgan2012} and \cite{li2018rerandomization}, \cite{wang2021}, and \cite{wang2022}.
We build on both of these literatures, studying the consequence of rerandomizing treatments within data-adaptive fine strata. 
We show that finely stratified rerandomization does partially linear regression adjustment ``by design,'' providing nonparametric control over a few important variables and linear control over the rest. \medskip

For our main asymptotic theory (Section \ref{section:gmm-asymptotics}), the most closely related previous work is in \cite{wang2021} and \cite{bai2024efficiency}.
\cite{wang2021} study estimation of the sample average treatment effect under stratified rerandomization.
Here, we study rerandomization within data-adaptive fine strata, i.e.\ matched-tuple designs, providing asymptotic theory for generic superpopulation and finite population causal parameters defined by moment equalities.
\cite{wang2021} measure covariate imbalance using the Mahalanobis norm.
We allow much more general forms of rerandomization, including nonlinear criteria based on differences in estimated covariate densities across treatment arms or fitted propensity scores.
We also study how to optimize the rerandomization criterion using researcher beliefs about the relationship between covariates and outcomes.

\cite{bai2024efficiency} study estimation of superpopulation parameters defined by moment equalities under fine stratification without rerandomization.
We extend these results to finely stratified rerandomization, allowing researchers to balance more covariates than can be accommodated by fine stratification alone.  
Relative to that work, we also study GMM estimation of finite population parameters, providing SATE-like versions of the general causal parameters in \cite{bai2024efficiency}. 
In concurrent work, \cite{li2024} study GMM estimation of univariate superpopulation parameters under ``coarse'' stratified rerandomization with fixed, discrete strata.
We study significantly more general forms of stratification and rerandomization criteria than considered in their work, allowing for both finite and superpopulation parameters of arbitrary dimension and fine stratification with continuous covariates. 

For nonlinear rerandomization (Section \ref{section:nonlinear}), we develop a general framework for rerandomization based on nonlinear estimators and derive separate asymptotic results for density-based and propensity-score criteria.
Our propensity-score rerandomization results are related to \cite{zhao2024}, who study balance criteria based on logistic-regression tests, while \cite{li2021} simulate a density-based design without developing asymptotic theory.

For rerandomization criterion optimization (Section \ref{section:acceptance-regions}), \cite{schindl2024} analyze the choice of weighting matrix in quadratic-form rerandomization, while \cite{liu2023} choose a quadratic criterion under a Bayesian model, in both cases without stratification.
Our minimax formulation instead chooses the criterion from a set of researcher beliefs about outcome-covariate relationships, which naturally yields a convex imbalance penalty tailored to those beliefs. 

Our work on optimal adjustment (Section \ref{section:covariate-adjustment}) extends recent work on adjustment for stratified designs, e.g.\ \cite{liu2020}, \cite{cytrynbaum2024adjustment}, \cite{bai2024adjustment}, to stratified rerandomization and GMM parameters. 
Finally our work on inference under data-adaptive fine stratification (Section \ref{section:inference}) builds on previous work by \cite{abadie2008}, \cite{bai2021inference}, \cite{cytrynbaum2024}, and \cite{bai2024efficiency}.  
Other recent work that has considered variance bounds for finite population causal parameters includes \cite{aronow2014}, \cite{fogarty2018b}, \cite{ding2019}, \cite{abadie2020}, and \cite{xu2021}.  

\section{Framework and Designs} \label{section:framework}

Consider data $W_i = (R_i, S_i(1), S_i(0))$ with $(W_i)_{i=1}^n \simiid F$. 
The $S_i(d) \in \mr^{d_S}$ denote potential outcome vectors for a binary treatment $d \in \{0, 1\}$, while $R_i$ denote other pre-treatment variables, such as covariates.
For treatment assignments $\Di \in \{0, 1\}$, the realized outcome $\Si = S_i(\Di) = \Di \Sione + (1-\Di) \Sizero$. 
In what follows, for any array $(a_i)_{i=1}^n$ we denote $\en[a_i] = n\inv \sum_{i=1}^n a_i$, with $\bar a_1 = \en[a_i | \Di=1] = \en[a_i \Di] / \en[\Di]$ and similarly $\bar a_0 = \en[a_i | \Di=0]$. 
Next, we define stratified rerandomization designs.

\begin{defn}[Stratified Rerandomization] \label{defn:rerandomization}
Let treatment proportions $\propfn = l/k$ and suppose that $n$ is divisible by $k$ for notational simplicity.
\begin{enumerate}[label={(\arabic*)}, itemindent=.5pt, itemsep=.4pt] 
    \item (Stratification). Partition the experimental units into $n/k$ disjoint groups (strata) $\group$ with $\{1, \dots, n\} = \cup_{\group} \group$ disjointly and $|\group| = k$.
Let $\psi \in \mr^{d_{\psi}}$ denote a vector of stratification variables, which may be continuous or discrete.
Suppose the groups satisfy the matching condition
\begin{equation} \label{equation:homogeneity}
\frac{1}{n} \sum_{\group} \sum_{i,j \in \group} |\psii - \psij|_2^2 = \op(1).
\end{equation}
Require that the groups only depend on the stratification variables $\psin$ and data-independent randomness $\permn$, so that $\group = \group(\psin, \permn)$ for each $\group$.

\item (Randomization). Independently for each $|\group| = k$, draw treatment variables $(\Di)_{i \in \group}$ by setting $\Di = 1$ for exactly $l$ out of $k$ units, uniformly at random.

\item (Check Balance). For rerandomization covariates $h$, consider an imbalance metric $\imbalance = \rootn(\hbarone  - \hbarzero) + \opone$ under the pure stratified randomization law induced by steps (1) and (2).
For an acceptance region $A \sub \mr^{\dimh}$, check if the balance criterion $\imbalance \in A$ is satisfied. 
If so, accept $\Dn$. 
If not, repeat from the beginning of (2). 
\end{enumerate}
\end{defn}

Intuitively, steps (1) and (2) describe a data-driven ``matched $k$-tuples'' design, while step (3) rerandomizes within $k$-tuples until the balance criterion is satisfied. 
Equation \ref{equation:homogeneity} is a tight-matching condition, requiring that the groups are clustered locally in $\psi$ space. 
Such conditions were introduced by \cite{bai2021inference} for matched-pairs randomization.
\cite{cytrynbaum2024} provides algorithms to match units into groups that satisfy this condition for any fixed $k$.

As mentioned above, \cite{cytrynbaum2026limits} provides finite-sample lower bounds showing that the matching discrepancies underlying Equation \ref{equation:homogeneity} cannot generally vanish faster than $n^{-2/d_\psi}$.
Consequently, tight matching can generally be achieved only in the very low-dimensional regime $d_\psi=o(\log n)$, limiting the plausibility of such conditions to fine stratification on just a few covariates at realistic sample sizes and motivating the rerandomization step to balance additional covariates.

Definition \ref{defn:rerandomization} nests two useful limiting cases. Stratification without rerandomization is obtained by setting $A = \mr^{\dimh}$, so every stratified assignment is accepted. Rerandomization without stratification is obtained by setting $\psi = 1$ and forming the groups at random, which is equivalent to complete randomization \citep{cytrynbaum2024}, before applying the balance criterion $\imbalance \in A$.

Next, we discuss a convenient rerandomization scheme that allows the researcher to select the approximate number of draws until acceptance. 

\begin{ex}[Mahalanobis Rerandomization] \label{ex:mahalanobis-rerandomization}
Here, we show how standard Mahalanobis rerandomization, e.g.\ as in \cite{wang2021} fits into our framework.
Consider matched $k$-tuples randomization, and let $X_i$ denote the raw covariates.
Define the within-tuple demeaned covariates $\check X_i = X_i - k\inv \sum_{j \in \group(i)} X_j$ and variance estimator $\Sigma_n = \var(D)\inv \frac{k}{k-1} \en[\check X_i \check X_i']$.
Rerandomize until
\begin{equation} \label{equation:mahalanobis-rerandomization}
n (\xbarone  - \xbarzero)' \Sigma_n \inv (\xbarone - \xbarzero) \leq \epsilon^2.   
\end{equation}
Equation \ref{equation:mahalanobis-rerandomization} is equivalent to $\imbalance \in A$ for $\imbalance = \Sigma_n\neghalf \rootn (\xbarone - \xbarzero)$ and acceptance region $A = \{x: |x|_2 \leq \epsilon\}$.
Under matched $k$-tuples randomization, $\Sigma_n \convp \Sigma = \var(D)\inv E[\var(X | \psi)]$ \citep{cytrynbaum2024adjustment}, so $\imbalance = \rootn(\hbarone - \hbarzero) + \opone$ for rerandomization variable $h = \Sigma \neghalf X$ consisting of the whitened raw covariates.
Then this design satisfies Definition \ref{defn:rerandomization}. 
\end{ex}

In the example above, one can show that $n (\xbarone  - \xbarzero)' \Sigma_n \inv (\xbarone - \xbarzero) \convwprocess \chi^2_{r}$ for $r = \dim(X)$ under fine stratification.
If $\epsilon(\alpha)^2$ is the $\alpha$ quantile of $\chi^2_r$, then $P(\imbalance \in A) = \alpha + o(1)$.  
Setting $\alpha = 1/m$ gives approximately $m$ expected assignment draws until acceptance for large enough $n$. 

Mahalanobis rerandomization thus uses a convenient choice of acceptance criterion, but the variance normalization and implicit acceptance region in Equation \ref{equation:mahalanobis-rerandomization} are not generally efficient. 
We provide alternative designs that optimize the shape of acceptance region $A$ in Section \ref{section:acceptance-regions} below. \medskip

\emph{Causal Estimands.} Next, we introduce a generic family of causal estimands defined by moment equalities, as in \cite{chenhongtarozzi2008}.
For other examples of this framework, see also \cite{cattaneo2010} and \cite{bai2024efficiency}.
Let $g(D, R, S, \theta) \in \mr^{\dimmom}$ be a score function for generalized method of moments (GMM) estimation.
Recall $W = (R, S(1), S(0))$ and, for $D | W \sim \bern(\propfn)$, define $\mombar(W, \theta) \equiv E[g(D, R, S, \theta) | W] = \propfn \mom(1, R, S(1), \theta) + (1-\propfn) \mom(0, R, S(0), \theta)$.
For now, suppose $\dimmom=\dimtheta$, so the model is exactly identified.

\begin{defn}[Causal Estimands] \label{defn:causal-estimands}
The \emph{superpopulation} estimand $\thetatrue$ is the unique solution to $E[\mombar(W, \theta)] = 0$.
The \emph{finite population} estimand $\thetan$ is the unique solution to $\en[\mombar(W_i, \theta)] = 0$.
\end{defn}

In what follows, we study GMM estimation of both $\thetatrue$ and $\thetan$ under stratified rerandomization designs, showing an asymptotic equivalence between stratified rerandomization and partially linear covariate adjustment. 

This framework also allows us to introduce several useful finite population estimands $\thetan$ that do not appear to have been previously considered, such as in Example \ref{ex:blp-late} below.
The estimand $\thetan$ is a more appropriate target for experiments run in a convenience sample, as is the case for the vast majority of experiments reported in the economics literature (\cite{niehaus2017}).

Inference on $\thetan$, provided in Section \ref{section:inference}, is more powerful than for $\thetatrue$, since we only have to account for estimation uncertainty due to random assignment, without extra variability from sampling into the experiment.

For intuition, first consider the simplest example:

\begin{ex}[ATE and SATE] \label{ex:ate}
Define the Horvitz-Thompson weights $H = \frac{D-\propfn}{\propfn - \propfn^2}$ and let $g(D, Y, \theta) = HY - \theta$, so that $\mombar(W, \theta) = E[HY | W] - \theta = Y(1) - Y(0) - \theta$.
Then $\thetatrue = E[Y(1) - Y(0)] = \ate$, the average treatment effect, and $\thetan = \en[Y_i(1) - Y_i(0)] = \sate$, the sample average treatment effect.   
\end{ex}

For a more interesting example, consider a setting where the researcher wants to estimate a model of treatment effect heterogeneity in an experiment with noncompliance and randomized binary instrument $Z$.
Using the framework above, we can define both finite population and superpopulation local average treatment effect (LATE) \emph{heterogeneity} parameters, applying them in our empirical application to \cite{angrist2013} below.

\begin{ex}[LATE Heterogeneity] \label{ex:blp-late}
Let $D(z)$ be potential treatments for a binary instrument $z \in \{0, 1\}$. 
Let $Y(d)$ be the potential outcomes, with realized outcome $Y = Y(D(Z))$.
Suppose $D(1) \geq D(0)$, and define compliance indicator $C = \one(D(1) > D(0))$, assuming $E[C] > 0$. 
\cite{imbens1994} define the $\late = E[Y(1)-Y(0) | C=1]$.
Let $H = (Z-\propfn)/(\propfn - \propfn^2)$ and consider the score function $g(Z, D, Y, X, \theta) = (HY - HD \cdot f(X, \theta)) \nabla_{\theta} f(X, \theta)$, where $X$ denotes covariates along which treatment effect heterogeneity is of interest.
Then
\[
\mombar(W, \theta) = E[g(Z, D, Y, X, \theta) | W] = C \cdot (Y(1) - Y(0) - f(X, \theta)) \nabla_{\theta} f(X, \theta).
\]
The moment condition $E[\mombar(W, \theta)] = 0$ is the FOC of a treatment effect prediction problem in the complier population $C=1$.
In particular, for $\tau \equiv Y(1) - Y(0)$, the parameter $\thetatrue$ is the best parametric predictor $\thetatrue = \argmin_{\theta} E[(\tau - f(X, \theta))^2 | C=1]$ of treatment effects for compliers. 
Setting $\en[\mombar(W_i, \theta)] = 0$, we obtain
\begin{equation} \label{eqn:late}
\thetan = \argmin_{\theta} \en[(\tau_i - f(X_i, \theta))^2 | C_i=1].
\end{equation}
The finite population LATE was studied in \cite{ren2023noncompliance} under complete randomization, but the more general heterogeneity parameters here appear to be novel. 
We consider both $\thetatrue$ and $\thetan$ when studying treatment effect heterogeneity among compliers in the empirical application to \cite{angrist2013} in Section \ref{section:empirical}.
\end{ex}

\emph{GMM Estimation.}
Let $\momest(\theta) \equiv \en[g(\Di, R_i, \Si, \theta)]$.
In the main text, we focus on the exactly identified case and define the GMM estimator by $\momest(\est) = 0$.

Our theoretical results in the appendix allow for the more general overidentified case.
For a positive-definite weighting matrix $\weightmatpop \in \mr^{\dimmom \times \dimmom}$, the finite population estimand is defined to be $\thetan \in \argmin_{\theta \in \thetaspace}\en[\mombar(W_i, \theta)]'\weightmatpop\en[\mombar(W_i, \theta)]$, and the GMM estimator is $\est \in \argmin_{\theta \in \Theta}\momest(\theta)'\weightmatest\momest(\theta)$, where $\weightmatest \convp \weightmatpop$.
The definition of $\thetan$ reduces to Definition \ref{defn:causal-estimands} whenever the sample moment equality has a unique solution.
In the overidentified case, the formulas below use the GMM linearization matrix $\matgmm = -(G'\weightmatpop G)\inv G'\weightmatpop$.
Under exact identification, this is $\matgmm=-G\inv$.

\section{Asymptotic Theory} \label{section:gmm-asymptotics} 
In this section, we characterize the asymptotic distribution of the GMM estimator $\est$ under the designs in Definition \ref{defn:rerandomization}.
We show that the asymptotic variance of $\est$ is proportional to the residual variance of a partially linear regression model, up to a remainder term due to slackness in the rerandomization criterion. 
In this sense, stratified rerandomization does partially linear regression adjustment ``by design.''
First, we state some technical conditions that are needed for the following results.

\begin{assumption}[Acceptance Region] \label{assumption:linear-rerandomization}
Suppose $A \sub \mrh$ is closed, has non-empty interior, and $\leb(\partial A) = 0$.\footnote{Note that $\partial A$ denotes the boundary of $A$, the limit points of both $A$ and $A^c$.} Let $E[\var(h | \psi)] \succ 0$ and $E[|\psi|_2^2 + |h|_2^2] < \infty$. 
\end{assumption}

Next we state the technical conditions needed for GMM estimation.
Define the matrix $G = E[(\partial / \partial \theta') \mombar(W, \theta)]|_{\theta = \thetatrue} \in \mr^{\dimmom \times \dimtheta}$ and let $\mom_d(W, \theta) = \mom(d, R, S(d), \theta)$ for $d \in \zerone$.
Recall the Frobenius norm $|B|_F^2 = \sum_{ij} B_{ij}^2$ for any matrix $B$.

\begin{assumption}[GMM] \label{assumption-gmm}
The following conditions hold for $d \in \{0, 1\}$: 
\begin{enumerate}[label={(\alph*)}, itemindent=.5pt, itemsep=.4pt] 
\item (Identification). The matrix $G$ is full rank, and $E[\mombar(W, \theta)] = 0$ iff $\theta = \theta_0$.\label{assumption-gmm-identification}
\item We have $E[|\mom_d(W, \thetatrue)|_2^2] < \infty$ and $E[\sup_{\theta \in \thetaspace} |\mom_d(W, \theta)|_2] < \infty$. \label{assumption-gmm-moments}
Also $\theta \to \mom_d(W, \theta)$ is continuous almost surely, and $\thetaspace$ is compact.\footnote{We can formally resolve measurability issues with the sup expressions by either (1) explicitly working with outer probability (e.g. \cite{vandervaart1996}) or (2) requiring that $\{g_d(\cdot, \theta), \theta \in \Theta\}$ is universally separable for $d=0,1$ (\cite{pollard1984}, p.38). 
To focus on the practical design issues, we avoid this formalism, implicitly assuming that all quantities are appropriately measurable.} 
\item There exists a neighborhood $\thetatrue \in U \sub \thetaspace$ such that $G_d (W, \theta) \equiv \partial / \partial \theta' \mom_d(W, \theta)$ exists and is continuous. 
Also $E[\sup_{\theta \in U} |\partial / \partial \theta' \mom_d(W, \theta)|_F] < \infty$. \label{assumption-gmm-deriv}
\end{enumerate} 
\end{assumption}

Compactness could likely be relaxed using concavity assumptions or a VC class condition, but we do not pursue this here.
Recall the GMM linearization matrix $\matgmm = -(G'\weightmatpop G)\inv G'\weightmatpop \in \mr^{\dimtheta \times \dimmom}$.
In the exactly identified case, $\matgmm = -G\inv$.
Let $\vard = \var(D) = p(1-p)$.
Define the sampling and assignment influence functions
\[
\sampif(W, \theta) \equiv \matgmm \mombar(W, \theta),
\qquad
\assignif(W, \theta) \equiv \vard \matgmm (\mom_1(W, \theta) - \mom_0(W, \theta)).
\]
At $\thetatrue$, abbreviate these functions as $\sampif \equiv \sampif(W, \thetatrue)$ and $\assignif \equiv \assignif(W, \thetatrue)$.
The sampling influence function determines variation from sampling the units, while the assignment influence function determines variation from random imbalances between the treatment and control units.

\subsection{Asymptotics for GMM Estimation}
Our first theorem studies GMM estimation of the finite population estimand $\thetan$ in Definition \ref{defn:causal-estimands}.
We extend these results to $\thetatrue$ in Corollary \ref{cor:gmm-superpopulation} below.  
Our main theorem shows that balancing $\psi$ and $h$ ex-ante reduces imbalances in the assignment influence function and thereby improves precision.

\begin{thm}[GMM] \label{thm:gmm-asymptotics}
Let $\Dn$ be as in Definition \ref{defn:rerandomization}.
Require Assumptions \ref{assumption:linear-rerandomization}, \ref{assumption-gmm}.
Then $\rootn(\est - \estn) | \Wn \convwprocess \normal(0, \vassign) + \residualvar$, where the two terms are independent and
\begin{equation} \label{eqn:assignment-variance} 
\vassign = \min_{\gamma \in \mr^{\dimh \times \dimtheta}} \vard\inv E[\var(\assignif - \gamma'h | \psi)].
\end{equation}
Let $\gammaoptmat$ be optimal in Equation \ref{eqn:assignment-variance}.
The term $\residualvar$ is a truncated Gaussian vector 
\begin{equation} \label{eqn:truncated-gaussian} 
\residualvar \sim \gammaoptmat'\zh | \zh \in A,
\qquad
\zh \sim \normal(0, \vard\inv E[\var(h | \psi)]).
\end{equation} 
\end{thm}

Note the variance $\vassign$ is a matrix $\vassign \in \mr^{\dimtheta \times \dimtheta}$, so the minimum should be interpreted in the positive semidefinite sense, $V(\gammaoptmat) = \min_{\gammacoeff} V(\gammacoeff)$ if $V(\gammaoptmat) \preceq V(\gammacoeff)$ for all $\gammacoeff \in \mr^{\dimh \times \dimtheta}$.
Theorem \ref{thm:gmm-asymptotics} shows that $\rootn(\est - \estn)$ is asymptotically distributed as an independent sum of a normal $\normal(0, \vassign)$ and truncated normal vector $\residualvar$.
Both terms depend only on the assignment influence function $\assignif$. 

The normal-plus-truncated-normal form of this limit is familiar from previous work on rerandomization.
\cite{li2018rerandomization} derive this representation for SATE estimation under complete rerandomization with general balance criteria, while \cite{wang2021} derive it under stratified Mahalanobis rerandomization.
Our theorem extends this structure to GMM estimators of general finite population causal parameters under data-adaptive fine stratification.
In particular, by rerandomizing within strata satisfying the tight-matching condition in Equation \ref{equation:homogeneity} we obtain a novel equivalence of this design with ex-post partially linear adjustment.

\emph{Partially Linear Adjustment.}
Let $\ltwoproduct(\psi) = L_2^{\dimtheta}(\psi)$ be the $\dimtheta$-fold Cartesian product of $L_2(\psi)$, the space of square-integrable functions.
Then the variance $\vassign$ in Theorem \ref{thm:gmm-asymptotics} can be written in terms of the residuals of the assignment influence function in a partially linear regression on $\psi$ and $h$:
\begin{equation} \label{eqn:partially-linear-variance}
\vassign = \min_{\substack{\gammacoeff \in \mr^{\dimh \times \dimtheta} \\ t \in \ltwoproduct(\psi)}} \vard\inv \var\left(\assignif - \gammacoeff'h - t(\psi)\right).
\end{equation}
This shows stratified rerandomization does partially linear regression adjustment ``by design,'' providing nonparametric control over $\psi$ and linear control over $h$. 

\medskip 

\emph{Residual Imbalance.} The truncated Gaussian $\residualvar \sim \gammaoptmat'\zh | \zh \in A$ arises from residual covariate imbalances due to slackness in the acceptance criterion, since $A \not = \{0\}$.
If $A$ is symmetric about zero, i.e.\ $x \in A$ iff $-x \in A$, then $E[\residualvar] = 0$, so the GMM estimator $\est$ is first-order unbiased.
In principle, $\residualvar$ can be made negligible relative to $\normal(0, \vassign)$ in large enough samples by choosing very small $A$.
For example, if $A = B(0, \epsilon)$ then $R_{B(0, \epsilon)} \sim (\gammaoptmat'\zh \, | \, |\zh|_2 \leq \epsilon) \convp 0$ as $\epsilon \to 0$. 

However, in finite samples, shrinking $\epsilon$ may make the design computationally infeasible or invalidate the asymptotic approximation.
See \cite{wang2022} for a detailed analysis of complete rerandomization with a sample-size-dependent threshold $\epsilon_n$.
In view of this issue, we develop a minimax criterion to choose a computationally efficient region $A$ in Section \ref{section:acceptance-regions} below.

\medskip

We next extend the asymptotics to the superpopulation estimand $\thetatrue$ solving $E[\mombar(W, \thetatrue)] = 0$.
Targeting $\thetatrue$ adds sampling variance that is invariant to the distribution of treatment assignments $\Dn$.

\begin{cor}[Superpopulation Estimand] \label{cor:gmm-superpopulation}
Suppose $\Dn$ is as in Definition \ref{defn:rerandomization}.
Require Assumption \ref{assumption:linear-rerandomization}, \ref{assumption-gmm}.
We have $\rootn(\est - \thetatrue) \convwprocess \normal(0, \vsamp + \vassign) + \residualvar$, where the normal vector and $\residualvar$ are independent, $\vsamp = \var(\sampif)$.
\end{cor}

The corollary shows that targeting $\thetatrue$ instead of $\thetan$ simply adds the sampling variance $\vsamp$ to the covariance matrix of the normal component.
This variance arises from iid sampling of the influence function $\sampif$.
Stratified rerandomization reduces the assignment variance $\vassign$ but does not affect $\vsamp$.
In this sense, the statistical consequences of different designs and adjustment strategies all happen at the level of the finite population estimand $\thetan$, while targeting the superpopulation $\thetatrue$ just adds extra sampling noise.

\begin{ex}[CATE] \label{ex:heterogeneity-blp}
Consider estimating the best linear predictor of treatment effect heterogeneity in covariates $X$ for an experiment with perfect compliance. 
We can use the score $\mom(D, X, Y, \theta) = (HY - X'\theta)X$.
For $\tau = Y(1)-Y(0)$ we have $\mombar(W, \theta) = (\tau - X'\theta)X$, so the parameters $\thetan, \thetatrue$ are
\[
\thetan = \argmin_{\theta} \en[(\tau_i - X_i'\theta)^2],
\qquad
\thetatrue = \argmin_{\theta} E[(\tau - X'\theta)^2].
\]
The parameter $\thetan$ was studied in \cite{ding2019} under complete randomization.
A calculation shows that $\assignif = E[XX']\inv\ylevel X$ for $\ylevel = (1-\propfn)Y(1) + \propfn Y(0)$, which summarizes each unit's outcome level.
Then for $e = \tau - X'\thetatrue$, the variances are 
\[
\vsamp = \var(\matgmm e X), \qquad \vassign = \min_{\gammacoeff \in \mr^{\dimh \times d_x}} \vard\inv E[\var(\assignif - \gammacoeff'h | \psi)].
\]
\emph{Efficient Design.} The expression for $\vassign$ shows that to precisely estimate heterogeneity parameters $\thetan$ and $\thetatrue$, it is important to balance not only variables that predict outcome levels $\ylevel$ by including these in $\psi$ and $h$, but also their interactions with the desired heterogeneity variable $X$.
We consider such interacted designs in our simulations and empirical application to \cite{angrist2013} below.
\end{ex}

\begin{remark}[Pure Stratification]
The specialization of Corollary \ref{cor:gmm-superpopulation} to fine stratification without rerandomization, obtained by setting $A = \mrh$, is due to \cite{bai2024efficiency}.
They allow more general GMM regularity conditions than those imposed here, accommodating non-smooth scores under a Donsker condition.
\end{remark}

\subsection{Nonlinear Rerandomization} \label{section:nonlinear}

Next, we extend the framework above to allow for stratified rerandomization based on nonlinear covariate imbalance criteria.
Natural examples include rerandomizing until estimated covariate densities in the treated and control arms are similar or until a fitted propensity score has little predictive power for treatment assignment.

Appendix \ref{appendix:gmm-rerandomization} develops a general framework for such nonlinear rerandomization designs. 
We show that a broad class of them are first-order equivalent to linear rerandomization with a certain choice of covariates and acceptance regions.
Here, we focus on two examples.

\medskip

\emph{Density Rerandomization.}
Previous work has proposed assigning treatments to make parametric or kernel density estimates of the covariate distributions similar between treatment arms \citep{endoEtAl2006,maHu2013}.
These papers evaluate such assignment rules through simulation but do not characterize their asymptotic properties.
Here, we provide the first asymptotic theory for rerandomization based on arm-specific parametric density estimates.

Let $f(X, \beta)$ be a possibly misspecified parametric density.
Draw an assignment from the finely stratified design above and, for $d=0,1$, form the density MLE  
\begin{equation} \label{eqn:density-rerandomization}
\wh \beta_d \in \argmax_{\beta} \en[\one(\Di=d)\log f(X_i, \beta)].
\end{equation}
Accept the assignment if $\rootn(\betaestone-\betaestzero) \in B(0,\epsilon)$.
Otherwise, draw another assignment and repeat.
Let $s(X,\beta) \equiv \nabla_{\beta}\log f(X,\beta)$, let $\betatrue$ uniquely solve $E[s(X,\betatrue)]=0$, and define Jacobian $J \equiv E[(\partial/\partial\beta')s(X,\betatrue)]$.

\begin{prop}[Density Rerandomization] \label{prop:density-rerandomization}
Impose Assumption \ref{assumption-gmm}.
Suppose also that the density score $s(X,\beta)$ satisfies Assumption \ref{assumption-gmm}, and the matrices $J$ and $E[\var(s(X,\betatrue) | \psi)]$ are nonsingular.
Then density rerandomization is first-order equivalent to linear rerandomization with covariates $h=J\inv s(X,\betatrue)$ and acceptance region $A=B(0,\epsilon)$.
\end{prop}

For a more general theory of rerandomization using nonlinear criteria constructed from arbitrary GMM estimators, see Appendix \ref{appendix:gmm-rerandomization}.

\medskip

\emph{Propensity Score Rerandomization.}
Under stratified randomization, each unit has the same ex ante treatment propensity $p$.
Covariate imbalance arises when the in-sample treatment proportion departs from $p$ in certain regions of the covariate space.
A fitted propensity score $\wh p(h)$ can be used to detect such local fluctuations, motivating a propensity-based rerandomization approach.
For each candidate assignment drawn from the finely stratified design, estimate $\wh p$ and accept when
\begin{equation} \label{eqn:propensity-rerandomization}
n\en[(\wh p(h_i)-\propfn)^2] \leq \epsilon^2.
\end{equation}
Otherwise, draw another candidate assignment and repeat.
Various choices of estimators are available to construct $\wh p$.
Here, we focus on the MLE for the smooth parametric model $\wh p(h)=\link(\wh\alpha+h'\wh\delta)$, where
\begin{equation} \label{eqn:propensity-estimation}
(\wh\alpha,\wh\delta) \in \argmax_{\alpha \in \mr, \delta \in \mrh} \en[\Di\log \link(\alpha+h_i'\delta)+(1-\Di)\log(1-\link(\alpha+h_i'\delta))].
\end{equation}
\begin{prop}[Propensity Rerandomization] \label{prop:propensity-rerandomization}
Suppose $\Dn$ is first drawn by finely stratified randomization as in steps (1) and (2) of Definition \ref{defn:rerandomization} and then rerandomized according to Equation \ref{eqn:propensity-rerandomization}.
Require Assumptions \ref{assumption-gmm} and \ref{assumption:propensity-rerandomization}.
This design is first-order equivalent to linear rerandomization with covariates $h$ and acceptance region $A=\{x:x'\var(h)\inv x \leq \epsilon^2\vard^{-2}\}$.
\end{prop}

For rerandomization without stratification, \cite{zhao2024} study a related criterion that accepts assignments when the $p$-value from a likelihood-ratio test of the joint significance of covariates in a logistic regression of treatment assignment on the covariates exceeds a prespecified threshold.

\subsection{Optimizing Acceptance Region Shape} \label{section:acceptance-regions}

In this subsection, we study efficient choice of the acceptance region $A \sub \mr^{\dimh}$.
For simplicity, we restrict attention to the case of estimating $\thetan = \sate$.
For the ATE score in Example \ref{ex:ate}, assignment influence $\assignif=\ylevel$.
For $\zhs = \zh | \zh \in A$, Theorem \ref{thm:gmm-asymptotics} gives the asymptotic decomposition
\begin{equation} \label{equation:sate-acceptance-decomposition}
\rootn(\est - \sate) | \Wn \convwprocess \normal(0, \vassign) + \gammaoptmat'\zhs.
\end{equation}
The two terms are independent, with only the second residual term depending on the shape of $A$.
The coefficient $\gammaoptmat$ is the partially linear regression coefficient of the outcome level $\ylevel$ on $(h, \psi)$.
For $E[e | \psi] = E[eh] = 0$, we have
\begin{equation} \label{equation:ylevel-decomposition}
\ylevel = \gammatrue'h + t_0(\psi) + e.
\end{equation}

We would like to choose $A$ to control the residual term $\gammaoptmat'\zhs$ in Equation \ref{equation:sate-acceptance-decomposition}.
To see how to do this, note this term is the limit of $\gammaoptmat'\rootn(\hbarone-\hbarzero)$ under stratified rerandomization.
If $\gammaoptmat$ were known, we could use the oracle criterion $|\gammaoptmat'\rootn(\hbarone-\hbarzero)| \leq \epsilon$, corresponding to acceptance region $A=\{x:|\gammaoptmat'x|\leq\epsilon\}$.
This is a significant dimensionality reduction, but is infeasible because $\gammaoptmat$ is unknown when the experiment is designed.

\emph{Minimax Approach.} In place of the infeasible oracle criterion, we take a minimax approach.
Let $\balancecoeffs \sub \mr^{\dimh}$ be a nonempty, bounded set of plausible values for this partially linear coefficient.
For $\epsilon>0$, consider rerandomizing until the worst case imbalance over $\balancecoeffs$ is small enough:
\begin{equation} \label{eqn:minimax-acceptance-criterion}
\sup_{\gamma \in \balancecoeffs}|\gamma'\rootn(\hbarone-\hbarzero)| \leq \epsilon.
\end{equation}
Equivalently, define the penalty function $\fnbalance(x) \equiv \sup_{\gamma \in \balancecoeffs}|\gamma'x|$.
Equation \ref{eqn:minimax-acceptance-criterion} is equivalent to accepting whenever $\fnbalance(\rootn(\hbarone-\hbarzero)) \leq \epsilon$.
The function $\fnbalance$ is convex as a supremum of linear functions.
It penalizes imbalance directions more heavily when they have a large projection onto some plausible coefficient $\gamma \in \balancecoeffs$, while placing little weight on directions nearly orthogonal to the researcher's beliefs.

\begin{prop}[Acceptance Region] \label{prop:minimax-acceptance-region}
The criterion in Equation \ref{eqn:minimax-acceptance-criterion} is equivalent to $\rootn(\hbarone-\hbarzero) \in A$ for $A=\epsilon\balancecoeffspolar$, where $\balancecoeffspolar \equiv \{x \in \mr^{\dimh}:\sup_{\gamma \in \balancecoeffs}|\gamma'x|\leq1\}$ is the absolute polar set of $\balancecoeffs$.
The set $A$ is symmetric and convex.
If $\balancecoeffs$ is bounded, $A$ is closed, has non-empty interior, and $\leb(\partial A)=0$.
\end{prop}

Proposition \ref{prop:minimax-acceptance-region} shows that the induced region $A$ satisfies the conditions on the acceptance region in Assumption \ref{assumption:linear-rerandomization}.
Since $A$ is symmetric, $E[\zhs]=0$, so the asymptotic distribution in Equation \ref{equation:sate-acceptance-decomposition} is centered at zero under this design.
See \cite{aliprantis2006} for more on polar sets.

\begin{ex}[Rectangle] \label{ex:rectangle}
Suppose $\gamma_{0j} \in [a_j,b_j]$ for each $1 \leq j \leq \dimh$, so $\balancecoeffs=\prod_{j=1}^{\dimh}[a_j,b_j]$.
For $a=(a_j)_j$ and $b=(b_j)_j$, let $\mu=(a+b)/2$ denote the center of the belief set and $\sigma=(b-a)/2$ the coefficient uncertainty vector.
The region is
\begin{equation} \label{eqn:rectangle-acceptance-region}
A=\biggl\{x:|\mu'x|+\sum_{j=1}^{\dimh}\sigma_j|x_j|\leq\epsilon\biggr\}.
\end{equation}
As $\sigma$ shrinks toward zero, $A$ expands in directions approximately orthogonal to $\mu$ and approaches the oracle region $\{x:|\mu'x|\leq\epsilon\}$, with the mean belief $\mu$ in place of $\gammaoptmat$.
By contrast, increasing $\sigma_j$ penalizes imbalances with large $|x_j|$ more heavily, contracting $A$ in those directions, as illustrated in Figure \ref{fig:prior-information}.
See Appendix \ref{section:belief-set-geometry} for a more general characterization covering belief sets formed by translating and linearly transforming $L_p$ balls, including ellipsoidal confidence regions.
\end{ex}

\setlength{\tabcolsep}{1pt}
\begin{figure}[t]
\centering
\begin{tabular}{cc}
\includegraphics[width=0.47\textwidth]{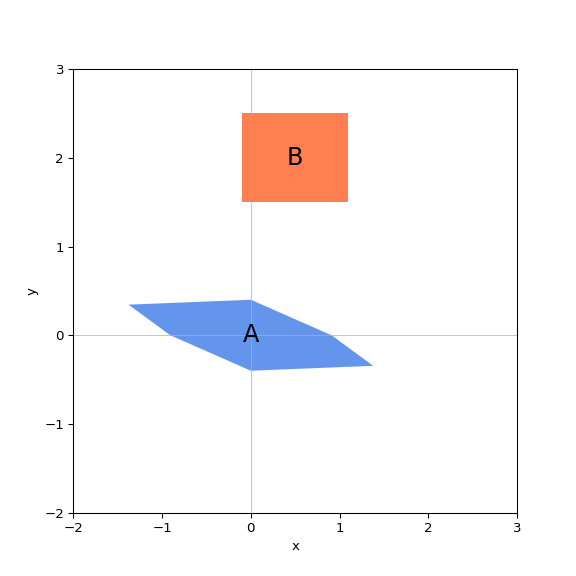} &
\includegraphics[width=0.47\textwidth]{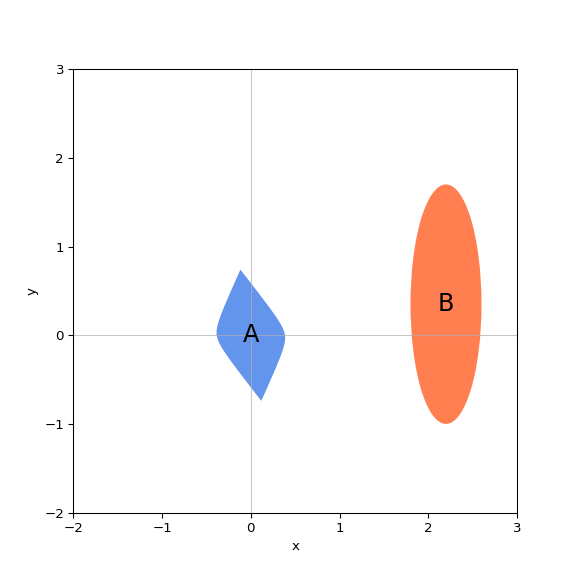} \\
\end{tabular}
\caption{Belief sets and induced acceptance regions $A=\epsilon\balancecoeffspolar$.}
\label{fig:prior-information}
\end{figure}
\setlength{\tabcolsep}{6pt}

Next, we show that such regions minimize the computational cost of rerandomization subject to a uniform bound on the conditional asymptotic bias over the researcher's belief set.

For a belief set coefficient $\gamma \in \balancecoeffs$ and a candidate acceptance region $C$, the asymptotic distribution from Equation \ref{equation:sate-acceptance-decomposition} is $L_{\gamma C}\equiv\normal(0,V(\gamma))+\gamma'Z_{hC}$, where the terms are independent, $Z_{hC}\equiv\zh | \zh \in C$, and $V(\gamma)\equiv\vard\inv E[\var(\ylevel-\gamma'h | \psi)]$.
Given an accepted covariate imbalance $Z_{hC}$, the conditional asymptotic bias is given by $\bias(L_{\gamma C} | Z_{hC})=E[L_{\gamma C} | Z_{hC}]=\gamma'Z_{hC}$.
The expected number of rerandomization draws until acceptance is the inverse acceptance probability, whose asymptotic counterpart is $P(\zh \in C)\inv$.

\begin{thm}[Minimax] \label{thm:minimax}
The acceptance region $A=\epsilon\balancecoeffspolar$ solves
\begin{equation} \label{eqn:minimax-problem}
A = \argmin_{C \sub \mrh}P(\zh \in C)\inv \text{ subject to } \sup_{\gamma \in \balancecoeffs}|\bias(L_{\gamma C} | Z_{hC})|\leq\epsilon.
\end{equation}
Let $\asympdisttrue \equiv \normal(0,\vassign)+\gammatrue'\zhs$ be the true asymptotic distribution.
If the beliefs are well specified in the sense $\gammatrue \in \balancecoeffs$, then $|\bias(\asympdisttrue | \zhs)|\leq\epsilon$ and $\var(\asympdisttrue)\leq\vassign+\epsilon^2$, where $\vassign$ is the partially linear variance in Equation \ref{eqn:partially-linear-variance}.
\end{thm}

Thus, $A$ has the largest acceptance probability among regions that bound the conditional asymptotic bias by $\epsilon$ for every coefficient in the researcher's belief set.
Implicitly, the minimization is over Borel sets $C \in \mc B(\mrh)$ with $P(\zh\in C)>0$.
The solution is unique up to an equivalence class $\{C \in \mc B(\mrh): \leb(C \triangle A) = 0\}$, where $\triangle$ denotes symmetric difference.

Section \ref{section:beliefs-from-pilot} in the appendix shows how to use pilot data to construct the researcher's belief set and obtain $\var(\asympdisttrue | \datapilot)\leq\vassign+\epsilon^2$ with high probability.
Closely related results are also provided by \cite{liu2023}, who derive optimal Mahalanobis-style completely rerandomized designs under a Bayesian criterion.

\section{Restoring Normality} \label{section:covariate-adjustment} 

In this section, we show that optimal ex-post linear adjustment removes the residual term in Theorem \ref{thm:gmm-asymptotics}, restoring asymptotic normality.
We provide a feasible estimator of the optimal adjustment coefficient and explain why combining adjustment with rerandomization has a double robustness property that improves finite-sample performance.
Throughout, suppose $E[|h|_2^2] < \infty$.

For an adjustment coefficient matrix $\gammaest \in \mr^{\dimh \times \dimtheta}$, define the adjusted GMM estimator $\estadj \equiv \est-\gammaest'(\hbarone-\hbarzero)$.
Theorem \ref{thm:gmm-asymptotics} shows that the unadjusted asymptotic distribution contains the residual imbalance $\gammaoptmat'\zhs$.
Since $\rootn(\hbarone-\hbarzero)$ converges to $\zhs$ under stratified rerandomization, adjustment by $\gammaest \convp \gammaoptmat$ cancels this term.
Optimal adjustment uses the coefficient on $h$ in the partially linear projection of $\assignif$ on $(\psi,h)$:
\begin{equation} \label{eqn:adjustment-coefficient}
\gammaoptmat \in \argmin_{\gamma \in \mr^{\dimh \times \dimtheta}}\min_{t \in \ltwoproduct(\psi)}\var(\assignif-\gamma'h-t(\psi)).
\end{equation}
If $E[\var(h | \psi)] \succ 0$, then the unique minimizer of Equation \ref{eqn:adjustment-coefficient} is the regression coefficient matrix $\gammaoptmat = E[\var(h | \psi)]\inv E[\cov(h, \assignif | \psi)]$.
Thus, the optimal coefficient depends on the stratification variables $\psi$, as observed in \cite{cytrynbaum2024adjustment} and \cite{bai2024adjustment} for ATE estimation.

\begin{thm}[Restoring Normality] \label{thm:gmm-adjustment}
Suppose $\Dn$ is as in Definition \ref{defn:rerandomization}.
Require Assumptions \ref{assumption:linear-rerandomization} and \ref{assumption-gmm}.
Suppose $\gammaest \convp \gammaoptmat$.
Then $\rootn(\estadj - \estn) | \Wn \convwprocess \normal(0, \vadj)$ and $\rootn(\estadj - \thetatrue) \convwprocess \normal(0, \vsamp + \vadj)$, where
\[
\vsamp = \var(\sampif), \qquad \vadj = \vassign = \vard\inv E[\var(\assignif - \gammaoptmat'h | \psi)].
\]
\end{thm}

Optimal adjustment therefore removes the only component of the asymptotic distribution that depends on the acceptance region $A$.
For rerandomization without stratification, a version of this equivalence was shown in \cite{li2018rerandomization}.
To construct a feasible estimator $\gammaest\convp\gammaoptmat$, we require a slight strengthening of the moment condition in Assumption \ref{assumption-gmm}.
\begin{assumption} \label{assumption-inference}
There is an open neighborhood $U\sub\thetaspace$ containing $\thetatrue$ such that $E[\sup_{\theta\in U}|(\partial/\partial\theta')\mom_d(W,\theta)|_F^2]<\infty$ for $d\in\{0,1\}$.
\end{assumption}

The coefficient $\gammaoptmat$ may depend on the unknown parameter $\thetatrue$, so adjustment is generally implemented in two steps.
First, estimate $\thetatrue$ by the unadjusted GMM estimator $\est$ and form the scores $\momesti \equiv \mom(D_i,R_i,S_i,\est)$.
For the second step, define the within-group partialled covariates $\hicheck \equiv \hi-|\group(i)|\inv\sum_{j \in \group(i)}h_j$, where $\group(i)$ is the group containing unit $i$, and let $\matgmmest \convp \matgmm$ estimate the GMM linearization matrix.
Then form the feasible estimator
\begin{equation} \label{eqn:feasible-adjustment}
\gammaest = \vard\en[\hicheck \hicheck']\inv\left(\cov_n(\hicheck, \matgmmest \momesti | \Di=1) - \cov_n(\hicheck, \matgmmest \momesti | \Di=0)\right).
\end{equation}
Under Assumptions \ref{assumption:linear-rerandomization}, \ref{assumption-gmm}, and \ref{assumption-inference}, if $E[\var(h | \psi)]\succ0$, Theorem \ref{thm:adjustment-coefficient} in Appendix \ref{proofs:covariate-adjustment} shows that $\gammaest=\gammaoptmat+\opone$.
Corollary \ref{cor:one-step-adjustment} in the appendix also formalizes a simpler case in which the optimal adjustment coefficient does not depend on $\theta$, so a one-step adjustment is available without a preliminary estimate of $\thetatrue$.
In particular, this occurs when the assignment influence function has the form $\assignif(W,\theta)=\assignif^{(1)}(\psi,\theta)+\assignif^{(2)}(W)$.

\begin{ex}[Adjusting CATE Estimate] \label{ex:cate-adjustment}
Recall from Example \ref{ex:heterogeneity-blp} the treatment effect heterogeneity BLP parameter $\thetan=\argmin_{\theta}\en[(Y_i(1)-Y_i(0)-X_i'\theta)^2]$.
For the score $\mom(D,X,Y,\theta)=(HY-X'\theta)X$, the assignment influence function is $\assignif(W,\theta)=E[XX']\inv\ylevel X$, which does not depend on $\theta$.
Hence, Corollary \ref{cor:one-step-adjustment} gives the feasible one-step adjustment $\gammaest=\gammaoptmat+\opone$ for
\begin{equation*}
\begin{aligned}
\gammaest=\en[\hicheck\hicheck']\inv\Bigl(&(1-p)\cov_n(\hicheck,Y_iX_i | \Di=1) \\
&+p\cov_n(\hicheck,Y_iX_i | \Di=0)\Bigr)\en[X_iX_i']\inv.
\end{aligned}
\end{equation*}
\end{ex}

\emph{Double Robustness.} Theorem \ref{thm:gmm-adjustment} establishes a first-order equivalence between rerandomization and optimal ex-post adjustment, but combining the two methods can provide additional finite-sample gains.
To see the mechanism, write the partially linear projection as $\assignif=\gammaoptmat'h+t(\psi)+e$, where $E[e | \psi]=0$ and $E[eh']=0$.
For $d\in\{0,1\}$, let $\bar e_d=\en[e_i | \Di=d]$.
The expansion underlying Theorem \ref{thm:gmm-adjustment} gives
\begin{equation} \label{eqn:double-robustness-expansion}
\rootn(\estadj-\thetan)=(\gammaoptmat-\gammaest)'\rootn(\hbarone-\hbarzero)+\rootn(\ebarone-\ebarzero)+\opone.
\end{equation}
The first term is the product of adjustment-coefficient estimation error and accepted covariate imbalance.
It can be small when either the coefficient is estimated accurately or rerandomization produces a small imbalance, and smaller still when both occur.
This provides a form of double robustness to covariate imbalances and explains why combining rerandomization with adjustment can outperform either method alone in finite samples.

The mechanism is particularly relevant when $\gammaoptmat$ is difficult to estimate, such as in small samples or when $\dim(h)$ is large, since Equation \ref{eqn:feasible-adjustment} estimates it using within-group variation in $h$.
Figure \ref{fig:simulation-efficiency} illustrates the robustness provided by rerandomization when the adjustment coefficient is difficult to estimate.

\section{Variance Bounds and Inference Methods} \label{section:inference} \label{section:inference-bounds} \label{section:inference-finite} \label{section:inference-population}

This section develops inference for the optimally adjusted GMM estimator $\estadj$.
For the finite population causal parameter $\thetan$, the asymptotic variance contains an unidentified cross-arm covariance, so we construct an identified upper bound and asymptotically conservative confidence intervals.
For the superpopulation parameter $\thetatrue$, the total variance is identified, allowing asymptotically exact inference.

To build intuition, first recall the classical problem of inference on $\thetan=\sate$ under complete randomization.
Let $\hk_d=\var(Y(d))$ and $\tau=Y(1)-Y(0)$.
Since the assignment variance depends on the covariance between $Y(1)$ and $Y(0)$, which are never jointly observed, it is not identified.
By Cauchy-Schwarz inequality, 
\begin{equation} \label{equation:neyman-bound}
\vassign = \frac{\hk_1}{p} + \frac{\hk_0}{1-p} - \var(\tau) \leq \frac{\hk_1}{p} + \frac{\hk_0}{1-p} - (\hksd_1 - \hksd_0)^2 \leq \frac{\hk_1}{p} + \frac{\hk_0}{1-p}.
\end{equation}
Both upper bounds are due to \cite{neyman1990}.
We next extend the sharper bound to a scalar contrast $c'\thetan$ of the general GMM parameter, where $c\in\mr^{\dimtheta}$, accounting for stratification, rerandomization, and optimal adjustment.

For $d\in\{0,1\}$, define arm-specific assignment influences $\assignifarm{d}=\vard\matgmm\mom_d(W,\thetatrue)$, so that $\assignif=\assignifarm{1}-\assignifarm{0}$.
Recall from Section \ref{section:covariate-adjustment} that the optimal adjustment coefficient is $\gammaoptmat=E[\var(h | \psi)]\inv E[\cov(h,\assignif | \psi)]$.
Define its arm-specific versions by $\gammaoptarm{d}=E[\var(h | \psi)]\inv E[\cov(h,\assignifarm{d} | \psi)]$.
The asymptotic variance depends on $h$-adjusted arm-specific influence functions $\assignifarmh{d}=\assignifarm{d}-\gammaoptarm{d}'h$.
By Theorem \ref{thm:gmm-adjustment}
\begin{equation} \label{eqn:contrast-adjusted-variance}
\vadj(c)\equiv c'\vadj c=\vard\inv E[\var(c'(\assignifarmh{1}-\assignifarmh{0}) | \psi)].
\end{equation}
As in the $\sate$ example above, the covariance between $c'\assignifarmh{1}$ and $c'\assignifarmh{0}$ is not identified because the two influence functions can never be jointly observed.
Define residual variance $\sigma_d^2(c)=E[\var(c'\assignifarmh{d} | \psi)]$ and $\sigma_d(c)=\sigma_d^2(c)\half$.
Cauchy-Schwarz bounds the unidentified covariance using these two identified variance components.

\begin{thm}[Variance Bounds] \label{thm:variance-bounds}
Under the conditions of Theorem \ref{thm:gmm-adjustment}, we have
\begin{equation}
\vadj(c)\leq\vadjbound(c)\equiv\vard\inv(\sigma_1(c)+\sigma_0(c))^2.
\end{equation}
\end{thm}

We now construct implementable estimators of the finite population variance bound and the superpopulation variance.
Estimating the within-arm conditional variances $\sigma_d^2(c)$ requires at least two treated and two control units in each stratum.
Let $\groupset_n$ denote the original strata in Definition \ref{defn:rerandomization}.
When treatment propensity $\propfn = l/k$ with $2\leq l\leq k-2$, the original strata can be used directly.

\emph{Collapsed Strata.} When $l=1$ or $l=k-1$, we instead match original strata together to form larger strata using the collapsed-strata method of \cite{hansen1953}. This is also known as pairs-of-pairs inference in later work including \cite{abadie2008}, \cite{bai2021inference}, \cite{cytrynbaum2024}, and \cite{bai2024efficiency}.
To do so, define the centroid of each original stratum $s\in\groupset_n$ by $\bar\psi_s=k\inv\sum_{i\in s}\psii$.
Let $\groupmatching:\groupset_n\to\groupset_n$ be a pairing with $\groupmatching(s)\ne s$ and $\groupmatching^2(s)=s$, satisfying
\begin{equation} \label{eqn:collapsed-strata-matching}
\frac{1}{n}\sum_{s\in\groupset_n}|\bar\psi_s-\bar\psi_{\groupmatching(s)}|_2^2=\op(1).
\end{equation}
We form this pairing in practice by applying the non-bipartite matching algorithm of \cite{derigs1988} to the centroids.
See \cite{cytrynbaum2024} for the guarantee in Equation \ref{eqn:collapsed-strata-matching}.
Let $\widetilde{\groupset}_n$ denote the resulting strata, with $\widetilde{\groupset}_n=\groupset_n$ when $2\leq l\leq k-2$ and $\widetilde{\groupset}_n=\{s\cup\groupmatching(s):s\in\groupset_n\}$ when $l=1$ or $l=k-1$.

To estimate $\sigma_d^2(c)$, we first estimate the observed $h$-adjusted assignment influence function.
Recall from Section \ref{section:covariate-adjustment} that $\momesti=\mom(D_i,R_i,S_i,\est)$ and let $\matgmmest\convp\matgmm$.
Using the within-group partialled covariates $\hicheck$ defined in that section, set $\gammaestarm{d}=\vard\en[\hicheck\hicheck']\inv\cov_n(\hicheck,\matgmmest\momesti | \Di=d)$ for $d\in\{0,1\}$.
The observed influence estimate is then $\assignifarmhesti=\vard\matgmmest\momesti-\gammaestarm{\Di}'h_i$.
For each $s\in\widetilde{\groupset}_n$, write $a_s=\sum_{i\in s}\Di$.
Define
\begin{align*}
\varestone&=\frac{1}{n}\sum_{s\in\widetilde{\groupset}_n}\frac{1}{\propfn(a_s-1)}\sum_{\substack{i\ne j\\i,j\in s}}\Di\Dj\assignifarmhesti(\assignifarmhestj)', \\
\varestzero&=\frac{1}{n}\sum_{s\in\widetilde{\groupset}_n}\frac{1}{(1-\propfn)(|s|-a_s-1)}\sum_{\substack{i\ne j\\i,j\in s}}(1-\Di)(1-\Dj)\assignifarmhesti(\assignifarmhestj)'.
\end{align*}
For a fixed contrast $c$, define the estimated within-arm variance components
\begin{equation*}
\widehat\sigma_1^2(c)=\en\left[\frac{\Di}{\propfn}(c'\assignifarmhesti)^2\right]-c'\varestone c, \quad \widehat\sigma_0^2(c)=\en\left[\frac{1-\Di}{1-\propfn}(c'\assignifarmhesti)^2\right]-c'\varestzero c.
\end{equation*}
The finite population variance estimator is $\widehat{\bar V}_a(c)\equiv\vard\inv(\widehat\sigma_1(c)+\widehat\sigma_0(c))^2$.
For superpopulation inference, we also estimate the cross-arm product of conditional means.
Using the original strata $\groupset_n$, define
\begin{equation} \label{eqn:cross-arm-estimator}
\varestcross=\frac{1}{n}\sum_{s\in\groupset_n}\frac{k}{l(k-l)}\sum_{i,j\in s}\Di(1-\Dj)\assignifarmhesti(\assignifarmhestj)'.
\end{equation}
Let $H_i=(\Di-\propfn)/(\propfn-\propfn^2)$ and $\scoreadjesti=\matgmmest\momesti-H_i\gammaest'h_i$ be the estimated adjusted GMM score.
The identified superpopulation variance estimator is
\begin{equation} \label{eqn:varest-superpopulation}
\widehat V\equiv\var_n(\scoreadjesti)-\vard\inv(\varestone+\varestzero-\varestcross-\varestcross').
\end{equation}

For a significance level $\alpha\in(0,1)$, define the finite population confidence interval $\cifin=[c'\estadj\pm z_{1-\alpha/2}\widehat{\bar V}_a(c)\half/\rootn]$ and the superpopulation confidence interval $\cipop=[c'\estadj\pm z_{1-\alpha/2}(c'\widehat Vc)\half/\rootn]$.

\begin{thm}[Inference] \label{thm:conservative-inference} \label{thm:exact-inference}
Fix $c\in\mr^{\dimtheta}$, suppose $\Dn$ is as in Definition \ref{defn:rerandomization}, and impose Assumptions \ref{assumption:linear-rerandomization}, \ref{assumption-gmm}, and \ref{assumption-inference}.
Then the following statements hold.
\begin{enumerate}[label=(\alph*), itemindent=.5pt, itemsep=.4pt]
\item $\widehat{\bar V}_a(c)\convp\vadjbound(c)\geq\vadj(c)$ and $P(c'\thetan\in\cifin)\geq1-\alpha-o(1)$.
\item $\widehat V\convp\vsamp+\vadj$ and $P(c'\thetatrue\in\cipop)=1-\alpha+o(1)$.
\end{enumerate}
\end{thm}

Thus, inference on the finite population parameter is conservative, while inference on the superpopulation parameter is asymptotically exact.

\section{Simulations} \label{section:simulations}

In this section, we use simulations to study the finite-sample properties of the designs and estimators analyzed above.
We find that ex-post adjustment performs well when there are few covariates and the outcome model is approximately linear.
Stratified rerandomization becomes more valuable when there are many covariates or important covariates enter the outcome model nonlinearly, providing meaningful gains over fine stratification and greater robustness than adjustment alone.

We estimate the sample average treatment effect $\thetan=\sate$ under data generated as $Y(d)=m_d(r)+e_d$, where $r\sim\normal(0,I_m)$ and $m=\dim(r)$.
The residuals $(e_1,e_0)\sim\normal(0,\widetilde\Sigma)$ satisfy $\var(e_d)=4$ and $\corr(e_1,e_0)=0.8$, independently of $r$.
For Models 1--3, the outcome function is $m_d(r)=c_d+r'\linearcoeff_d+r'\quadcoeff_dr$.
We consider four specifications:

\begin{enumerate}[itemsep=.4pt]
\item \emph{Diffuse linear.} Set $\linearcoeff_1=\one_m/\sqrt m$, $\beta_0=0$, $\quadcoeff_d=0$, and $c_d=0$ for $d\in\{0,1\}$.
\item \emph{Dominant variable.} As in 1, but $\beta_{1,1}=4$, $\beta_{0,1}=0$, $\beta_{d,2:m}=\one_{m-1}/\sqrt{m-1}$.
\item \emph{Quadratic.} As in 2, but $\quadcoeff_1=\diag(\alpha)$, $\alpha_1=2$, $\alpha_{2:m}=1/(2\sqrt{m-1})$.
\item \emph{Smooth nonlinear.} As in 2, but set $m_d(r)=2\arctan(r'\linearcoeff_d)$.
\end{enumerate}

All covariates are equally important in Model 1.
In Models 2 to 4, $r_1$ represents a baseline outcome that is more predictive than $r_{2:m}$, as is common when baseline and endline outcomes measure the same object.

We set $p=1/2$ and compare three designs.
Design \textbf{C} is complete randomization.
Design \textbf{S} is full fine stratification, using $\psi=r$ in Model 1 and weighting the first covariate more heavily in Models 2-4 by setting $\psi_1=\sqrt 2r_1$ and $\psi_{2:m}=r_{2:m}$.\footnote{We match using the algorithm in \cite{bai2021inference} for $p=1/2$.}
Design \textbf{SR} is stratified rerandomization, with $\psi=r_1$ and $h=r_{2:m}$.
For \textbf{SR}, we use the Mahalanobis criterion in Example \ref{ex:mahalanobis-rerandomization} with acceptance probability $1/500$.
For each design, $\est$ is the difference in mean outcomes between the treated and control arms, while $\estadj$ is its adjusted counterpart from Section \ref{section:covariate-adjustment}.

\begin{figure}[t]
\centering
\includegraphics[width=\textwidth]{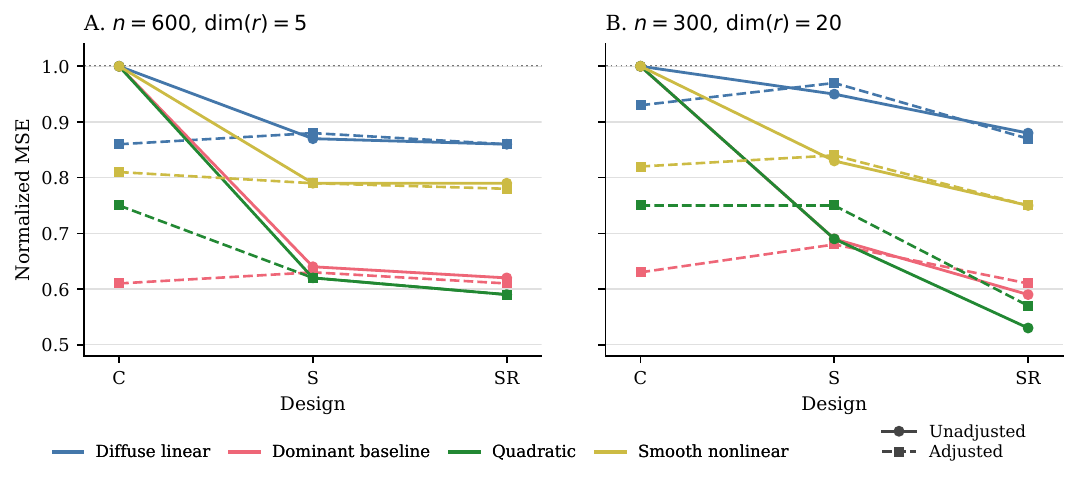}
\caption{Normalized MSE under complete randomization (\textbf{C}), full fine stratification (\textbf{S}), and stratified rerandomization (\textbf{SR}). Colors distinguish the four outcome models. Solid lines with circles show the unadjusted estimator, while dashed lines with squares show the adjusted estimator. MSE is normalized within each model, sample size, and dimension by the MSE of the unadjusted estimator under \textbf{C}.}
\label{fig:simulation-efficiency}
\end{figure}

Figure \ref{fig:simulation-efficiency} compares MSE under two combinations of sample size and dimension.
Within each model, sample size, and dimension, MSE is normalized by the MSE of the unadjusted estimator under \textbf{C}.
When $n=600$ and $\dim(r)=5$, adjustment under \textbf{C} substantially reduces MSE in every model and is competitive with rerandomization in three of the four models.
When $n=300$ and $\dim(r)=20$, unadjusted \textbf{SR} has lower MSE than adjusted \textbf{C} in every model.
Thus, rerandomization provides robustness when the adjustment coefficient is difficult to estimate.
Across the complete results in Appendix \ref{section:additional-simulation-results}, \textbf{SR} weakly minimizes MSE among the three designs for both estimators in every specification.

\begin{figure}[t]
\centering
\includegraphics[width=\textwidth]{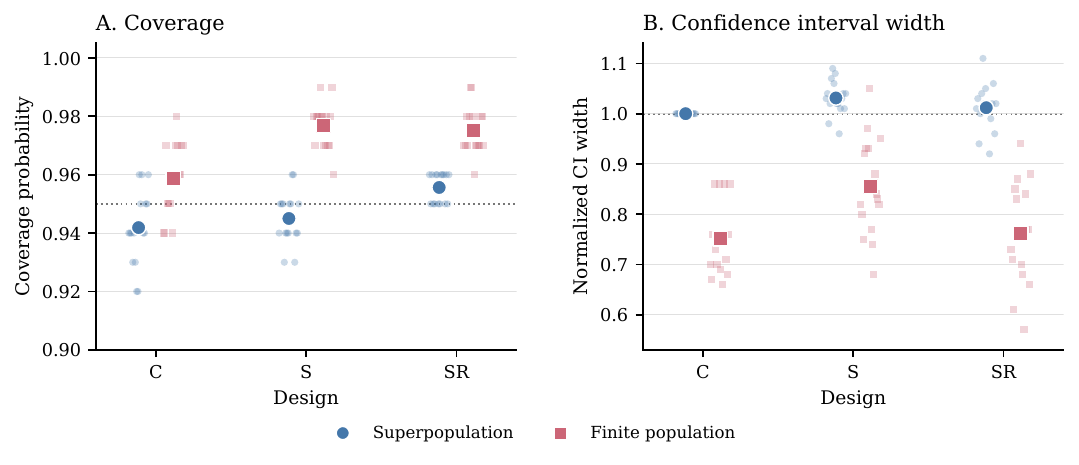}
\caption{Coverage and normalized confidence interval width for the adjusted estimator under \textbf{C}, \textbf{S}, and \textbf{SR}. Faint markers report the 16 combinations of outcome model, sample size, and dimension for each design. Large markers report their averages. Circles refer to superpopulation inference for $\thetatrue$, while squares refer to finite population inference for $\thetan$. The dotted lines mark $0.95$ coverage in Panel A and the normalized width of the superpopulation interval under \textbf{C} in Panel B.}
\label{fig:simulation-inference}
\end{figure}

Figure \ref{fig:simulation-inference} summarizes inference across all combinations of the four models, $n\in\{300,600\}$, and $\dim(r)\in\{5,20\}$.
All intervals use the adjusted estimator $\estadj$.
The superpopulation intervals cover $\thetatrue$ using $\widehat V$, while the finite population intervals cover $\thetan$ using the conservative variance bound $\widehat{\bar V}_a(c)$.
Interval widths are normalized within each simulation specification by the width of the superpopulation interval under \textbf{C}.
Coverage under \textbf{S} and \textbf{SR} is close to or above the nominal level.
Finite population coverage is generally conservative, and these intervals are shorter than their superpopulation counterparts in every specification shown.

Table \ref{table:efficiency} in Appendix \ref{section:additional-simulation-results} reports the numerical results for every simulation specification, and Table \ref{table:exotics} provides additional comparisons of different rerandomization criteria.

\section{Empirical Application} \label{section:empirical}

In this section, we apply our methods to data from the ``Opportunity Knocks'' experiment in \cite{angrist2013}. 
The authors randomized eligibility to receive payment for academic performance to first and second year students at a large Canadian university.
They estimated the effect of the program on future student GPA, graduation, and other outcomes.
They measured baseline covariates including high school GPA, sex, age, native language, and parent's education.  
Randomization was coarsely stratified on year in college, sex, and quartiles of high school GPA within year-sex cells, with approximately $p = 3/10$ of $n=1203$ students assigned to receive incentives. 

Some students assigned treatment $Z = 1$ did not engage with the program either by checking their earnings or making contact with the program advisor. 
The authors view this as noncompliance with the instrument $Z$ and estimate both intention-to-treat (ITT) effects and effects on compliers ($\late$). 
Let $D \in \{0, 1\}$ denote the endogenous decision to engage with the program, with $D(z)$ the potential treatments, $Y(d)$ the potential outcomes, and $T(z) = Y(D(z))$ the ITT potential outcomes with realized outcome $T = Y(D(Z)) = Y$. 
\cite{angrist2013} also estimate ITT-style treatment effect heterogeneity along several dimensions, such as gender and student-reported financial stress. 

In what follows, we use this data to study the efficiency and inference properties of complete randomization, the original coarse stratification, and fine stratification on different variable sets, along with rerandomized versions of each design.
Following the approach in \cite{li2018rerandomization} and \cite{bai2020pairs}, we impute missing ITT potential outcomes and potential treatments using cross-validated LASSO and random forests applied to $11$ baseline covariates and their pairwise interactions.
This allows us to simulate RMSE, coverage, and confidence interval width under the counterfactual designs.
Appendix \ref{section:imputation-details} describes the imputation procedure.

Given imputed data $(X_i, \wh T_i(z), \wh D_i(z))$ for $i=1,\dots,1203$, each replication samples $n$ units with replacement and draws instrument assignments $(\tilde Z_i)_{i=1}^n$ under one of the designs below.
We then observe treatments $\tilde D_i = \wh D_i(\tilde Z_i)$ and outcomes $\tilde Y_i = \wh T_i(\tilde Z_i)$, and form estimators $\est$ and $\estadj$ and confidence intervals $\wh C_{fin}$ and $\wh C_{pop}$ for the parameters $\late$ and $\clate$ described below. 

\begin{table}[htbp]
\begin{adjustbox}{scale=0.95, keepaspectratio, center}
  \centering
    \begin{tabular}{cccccccc}
          &       & \multicolumn{2}{c}{RMSE} & \multicolumn{2}{c}{Cover} & \multicolumn{2}{c}{CI Width} \\
    \midrule
    $\thetan$ (LATE) & Design & $\est$ & $\estadj$ & Pop. & Fin. & Pop. & Fin. \\
    \midrule
            & C    & 1.18 & 1.00 & 0.94 & 0.98 & 1.00 & 0.96 \\
            & CR   & 1.02 & 0.97 & 0.95 & 0.98 & 1.00 & 0.96 \\
            & S    & 0.98 & 0.97 & 0.95 & 0.98 & 1.00 & 0.95 \\
    LATE    & SR   & 1.00 & 0.99 & 0.94 & 0.98 & 1.00 & 0.95 \\
            & F    & 1.03 & 0.97 & 0.95 & 0.98 & 0.98 & 0.93 \\
            & FR   & 0.95 & 0.94 & 0.93 & 0.98 & 0.98 & 0.93 \\
            & F+   & 0.97 & 0.97 & 0.94 & 0.98 & 0.97 & 0.91 \\
            & FR+  & 0.97 & 0.97 & 0.95 & 0.98 & 0.97 & 0.91 \\
\cmidrule{2-8}
            & C    & 3.37 & 1.00 & 0.94 & 0.98 & 1.00 & 1.02 \\
            & CR   & 2.02 & 0.95 & 0.95 & 0.98 & 0.98 & 0.98 \\
            & S    & 3.41 & 0.98 & 0.95 & 0.98 & 0.98 & 1.00 \\
    CLATE   & SR   & 1.86 & 0.89 & 0.95 & 0.99 & 0.97 & 0.97 \\
    (Stress) & F    & 3.15 & 0.99 & 0.93 & 0.98 & 0.97 & 0.98 \\
            & FR   & 1.88 & 0.86 & 0.95 & 0.99 & 0.96 & 0.95 \\
            & F+   & 3.06 & 0.97 & 0.92 & 0.97 & 0.95 & 0.94 \\
            & FR+  & 1.68 & 0.84 & 0.96 & 0.99 & 0.94 & 0.92 \\
\cmidrule{2-8}
            & C    & 3.13 & 1.00 & 0.94 & 0.97 & 1.00 & 1.04 \\
            & CR   & 1.88 & 0.85 & 0.96 & 0.99 & 0.97 & 0.99 \\
            & S    & 1.49 & 1.02 & 0.93 & 0.97 & 0.97 & 1.02 \\
    CLATE   & SR   & 1.07 & 0.87 & 0.95 & 0.99 & 0.95 & 0.99 \\
    (GPA)   & F    & 0.91 & 1.02 & 0.96 & 0.98 & 1.12 & 1.00 \\
            & FR   & 0.83 & 0.84 & 0.96 & 0.98 & 1.07 & 0.96 \\
            & F+   & 0.87 & 1.00 & 0.95 & 0.97 & 1.09 & 0.98 \\
            & FR+  & 0.84 & 0.86 & 0.96 & 0.99 & 1.05 & 0.96 \\
    \bottomrule
    \end{tabular}
\end{adjustbox}
\caption{LATE Parameters}
\label{table:angrist-late}
\end{table}

We use the same design-specific covariate vector $h$ for rerandomization and adjustment.
We begin with the $11$ baseline covariates above and the interactions of HS GPA, sex, year, and mother's and father's education with both financial stress $F \in \{0, 1\}$ and HS GPA $G \in \mr$, for a total of $21$ covariates.
The interactions are motivated by our desire to estimate treatment effect heterogeneity along the dimensions $F$ and $G$, as discussed in Example \ref{ex:heterogeneity-blp}.
We compare four base designs and their rerandomized versions.
\textbf{C}/\textbf{CR} use complete randomization without and with rerandomization.
\textbf{S}/\textbf{SR} use the original coarse strata without and with rerandomization.
\textbf{F}/\textbf{FR} use fine stratification on HS GPA, while \textbf{F+}/\textbf{FR+} use fine stratification on HS GPA, sex, and year.
Each rerandomized design uses the stratified Mahalanobis criterion in Example \ref{ex:mahalanobis-rerandomization} with nominal acceptance probability $1/500$.
For each design, $h$ excludes directions perfectly determined by the stratification variables.

Table \ref{table:angrist-late} presents efficiency and inference results for LATE-style treatment effects on compliers.
In particular, if $C_i = \one(D_i(1) - D_i(0) > 0)$ is a compliance indicator then $\late = \en[Y_i(1)-Y_i(0) | C_i=1]$ and CLATE (Example \ref{ex:blp-late}) is the coefficient on $x_i$ in the infeasible regression   
\[
\thetan = \argmin_{\theta} \en[(Y_i(1) - Y_i(0) - \theta'(1, x_i))^2 | C_i=1].
\]
We consider heterogeneity variables $x_i = F_i \in \{0, 1\}$, an indicator for student financial stress, and $x_i = G_i \in \mr$, the student's HS GPA.  
For $x_i = F_i$, the CLATE has a simple interpretation as the difference in treatment effects for compliers with and without financial stress:
\[
\clate = \en[Y_i(1)-Y_i(0)|F_i=1, C_i=1] - \en[Y_i(1)-Y_i(0)|F_i=0, C_i=1].
\]

Cover Pop.\ and CI Width Pop.\ refer to inference on the superpopulation estimands $\thetatrue$ corresponding to $\thetan$, i.e.\ $\thetatrue = \argmin_{\theta} E[(Y(1) - Y(0) - \theta'(1, x))^2 | C=1]$ for $\thetan = $ CLATE and $\thetatrue = E[Y(1)-Y(0)|C=1]$ for LATE.
Both RMSE columns are normalized by the RMSE of $\estadj$ under design \textbf{C}, while confidence interval widths are normalized by the width of $\wh C_{pop}$ under \textbf{C}. 

Adjustment produces large gains for the heterogeneity parameters under complete and coarse stratification, while rerandomization produces substantial additional gains across the designs.
These gains remain sizable even after adjustment.
For the scalar LATE, adjusted RMSE ranges from $0.94$ to $1.00$ across designs.
For financial-stress heterogeneity, rerandomizing the original coarse strata reduces normalized adjusted RMSE from $0.98$ to $0.89$.
For GPA heterogeneity, rerandomizing the fine-GPA design reduces normalized adjusted RMSE from $1.02$ to $0.84$.
Mean coverage is $0.95$ for superpopulation intervals and $0.98$ for finite population intervals.
Finite population intervals are slightly shorter on average than their superpopulation counterparts, with the largest reductions for GPA heterogeneity.

\section{Recommendations for Practice}

We recommend experimenters finely stratify on a few variables expected to be most predictive of outcomes, while rerandomizing to balance the remaining baseline covariates.
This can be done using the stratified Mahalanobis design in Example \ref{ex:mahalanobis-rerandomization} or the optimized designs in Section \ref{section:acceptance-regions}, if the researcher has a strong prior. 
Doing so balances linear functions of less important covariates without degrading match quality on the most important covariates, such as baseline outcomes.

We recommend pairing stratified rerandomization with the feasible adjustment procedure in Section \ref{section:covariate-adjustment}, which provides a form of double robustness to covariate imbalance.
This combination was especially effective in our simulations when the number of covariates was large relative to sample size and yielded additional gains in our empirical application.

Most experiments in economics are conducted in convenience samples rather than random samples from a much larger superpopulation \citep{niehaus2017}.
When performing inference on finite population causal parameters within such samples, researchers can use our general variance bounds to construct conservative confidence intervals that are shorter than the corresponding superpopulation intervals.

\clearpage 

\appendix
\renewcommand{\thesection}{\Alph{section}}

\section{Proof of Main Asymptotic Results} \label{section:main-proofs}

Below, we carefully distinguish between $P_n$, the law of the data $(\Wn, \Dn)$ under ``pure'' stratified randomization, and $Q_n$, the law under rerandomized stratification. 
We suppress the $n$ subscript in what follows. 

\begin{defn}[Pure Stratification] \label{defn:pure-stratification}
For $(W_i)_{i=1}^n \simiid F$, let $P$ denote the law of $(\Wn, \Dn)$ under the design in steps (1) and (2) of Definition \ref{defn:rerandomization}.
\end{defn}

We slightly generalize the rerandomization design introduced in Definition \ref{defn:rerandomization}, which will be useful for our results on nonlinear rerandomization in Section \ref{section:nonlinear}. 

\begin{defn}[Rerandomization] \label{defn:rerandomization-general}
Consider the following: 
\begin{enumerate}[label={(\alph*)}, itemindent=.5pt, itemsep=.4pt] 
\item (Acceptance Regions). Let $\rerandparamest = \rerandparampop + \op(1)$ for $\rerandparampop \in \mr^{d_{\tau}}$ under $P$.
Define sample acceptance region $\rerandsetest = \{x: \setfn(x, \rerandparamest) \leq 0\}$ and population region $\rerandsetpop = \{x: \setfn(x, \rerandparampop) \leq 0\}$ for $\setfn(x, y)$ a jointly continuous function. 
Accept $\Dn$ if $\imbalance \in \rerandsetest$ for $\imbalance = \rootn(\hbarone - \hbarzero) + \op(1)$ under $P$.

\item (Rerandomization Distribution). Let $\filtrationcand = \sigma(\Wn, \permn)$, where $\permn \indep \Wn$ is possibly used to break ties in matching (Equation \ref{equation:homogeneity}). 
For any event $B$ and $P$ as in Definition \ref{defn:pure-stratification}, define the rerandomization distribution $Q(B | \filtrationcand) = P(B | \filtrationcand, \imbalance \in \rerandsetest)$ and $Q(B) = E[Q(B | \filtrationcand)]$. 
\item (Assumptions). Assume $P(\setfn(\zh, \rerandparampop) = 0) = 0$ for $\zh \sim \normal(0, \vard\inv E[\var(h | \psi)])$.
Require $P(\zh \in \rerandsetpop) > 0$. 
Suppose $E[|h|_2^2] < \infty$.
\end{enumerate}
\end{defn}

The proof of Theorem \ref{thm:gmm-asymptotics} shows that when $A=\mrh$, Definition \ref{defn:rerandomization} specializes the above by taking $\setfn(x,y)=-1$.
Otherwise, take $\setfn(x,y)=\setfn(x)=d(x,A)-d(x,A^c)$ for $d(x,S)=\inf_{z\in S}|x-z|_2$.
Next, the following essential lemma shows that the high level properties (e.g.\ convergence in probability) of $P$ are inherited by the rerandomized version $Q$. 

\begin{lem}[Dominance] \label{lemma:dominance}
Let $(B_n)_{n \geq 1}$ and $(R_n)_{n \geq 1}$ events and random variables.
Suppose that the rerandomization measure $Q$ is as in Definition \ref{defn:rerandomization-general}. 
\begin{enumerate}[label={(\alph*)}, itemindent=.5pt, itemsep=.4pt] 
\item If $B_n \in \filtrationcand$ then $P(B_n) = Q(B_n)$.
    If $R_n$ is $\filtrationcand$-measurable then $R_n = \op(1) / \Op(1)$ under $P$ $\iff$ $R_n = \op(1) / \Op(1)$ under $Q$.
\item $Q(B_n) = o(1)$ if $P(B_n) = o(1)$. 
If $R_n = \op(1) / \Op(1)$ under $P$ then $R_n = \op(1) / \Op(1)$ under $Q$.
\end{enumerate}
\end{lem}

\begin{proof}[Proof of Lemma \ref{lemma:dominance}]
(a) follows since $Q = P$ on $\filtrationcand$ by definition.
Let $c = P(\zh \in \rerandsetpop) > 0$ by assumption.  
Define $S_n = \{P(\imbalance \in \rerandsetest | \filtrationcand) \geq c / 2 \}$. 
Then by Lemma \ref{lemma:characteristic-convergence}, applied with $\diff=0$ and $t=0$, $P(\imbalance \in \rerandsetest | \filtrationcand) \convp P(\zh \in \rerandsetpop) = c$, so $P(S_n) \to 1$.
We have the upper bound 
\begin{align*}
&\one(S_n) Q(B_n | \filtrationcand) =  \one(S_n) P(B_n | \imbalance \in \rerandsetest, \filtrationcand)= \one(S_n) \frac{P(B_n, \imbalance \in \rerandsetest | \filtrationcand)}{P(\imbalance \in \rerandsetest | \filtrationcand)} \\
\leq \; &(c/2)\inv \one(S_n) P(B_n, \imbalance \in \rerandsetest | \filtrationcand) \leq (c/2)\inv P(B_n | \filtrationcand).
\end{align*}
The first equality by definition of $Q$.
The first inequality by the definition of $S_n$.
The final inequality by additivity of measures.
Then for $r_n \equiv (1-\one(S_n)) Q(B_n | \filtrationcand)$, we have $Q(B_n | \filtrationcand) = \one(S_n) Q(B_n | \filtrationcand) + r_n$.
Note that $|r_n| \leq 1$ and $r_n \convp 0$, so $E_Q[r_n] = o(1)$ by modes of convergence.
Then expand $Q(B_n)$ as
\begin{align*}
&\Eq[Q(B_n | \filtrationcand)] = \Eq[\one(S_n) Q(B_n | \filtrationcand)] + \Eq[r_n] \leq (c/2)\inv \Eq[P(B_n | \filtrationcand)] + o(1) \\
&= (c/2)\inv E_P[P(B_n | \filtrationcand)] + o(1) = (c/2)\inv P(B_n) + o(1).    
\end{align*}
The second equality follows from part (a), and the final equality by tower law.
The $\op(1)$ results follow by setting $B_n = \{R_n > \epsilon\}$.
The $\Op(1)$ results follow by the $\op(1)$ statement and Lemma \ref{lemma:big-op}.
\end{proof}

\textbf{Proof Strategy.} Equipped with this lemma, we will take the following approach to prove Theorem \ref{thm:gmm-asymptotics}: (1) show linearization of the GMM estimator $\est$ about $\thetan$ and $\thetatrue$ under $P$, (2) invoke Lemma \ref{lemma:dominance} to show these properties still hold under $Q$, then (3) prove distributional convergence of the simpler linearized quantities directly under $Q$. 

\subsection{Rerandomization Asymptotics} \label{proofs:rerandomization-asymptotics}
In this subsection, we develop the necessary tools for step (3) in our proof strategy. 
First, we state a conditional CLT for pure fine stratification, conditional on the data $\Wn$ and tie-breaking randomness in the matching procedure $\permn$.

\begin{thm}[CLT] \label{thm:clt}
Suppose $E[|\diff(W)|_2^2] < \infty$. 
Define $\filtrationcand = \sigma(\Wn, \permn)$. 
Suppose $\Dn$ is assigned according to Definition \ref{defn:pure-stratification}.
Then $X_n \equiv \rootn \en[\Hi \diff(\Wi)]$ has $X_n | \filtrationcand \convwprocess \normal(0, V)$. 
In particular, for each $t \in \mr^{\dimdiff}$ we have $E[e^{it' X_n} | \filtrationcand] = \varphi_V(t) + \op(1)$ with $\varphi_V(t) = e^{-t'V t / 2}$ and $V = \vard \inv E[\var(\diff | \psi)]$.
The same statements hold conditional on $\Wn$ alone.
\end{thm}

\begin{proof}
First consider the case $\dimdiff = 1$.
Define $\ui = \diffi - E[\diffi | \psii]$.
By Lemma \lemmabalance{} in \cite{cytrynbaum2024}, since $E[\diffi^2] < \infty$ we have $\rootn \en[(\Di - \propfn)E[\diffi | \psii]] = \op(1)$.
Then it suffices to study $\rootn \en[(\Di - \propfn)\ui]$.
To do so, we will use a martingale difference sequence (MDS) CLT.
Fix an ordering $(s_j)_{j=1}^{n/k}$ of $\groupsetn$ and let $D_{s_j} = (D_i)_{i \in s_j}$.
Define $\filtrationh_{0,n} = \filtrationcand$ and $\filtrationh_{j,n} = \sigma(\filtrationcand, (D_{s_q})_{q=1}^j)$ for $j \geq 1$.
Let $\zjn = \negrootn \sum_{i \in s_j}(\Di - \propfn)\ui$ and $S_{j,n} = \sum_{q=1}^j Z_{q,n}$.
Conditional on $\filtrationcand$, assignments are independent across strata and $E[\Di | \filtrationcand] = \propfn$.
Since $s_j$ and $(u_i)_{i \in s_j}$ are $\filtrationcand$-measurable, $E[\zjn | \filtrationh_{j-1,n}] = 0$.
Thus, $(S_{j,n},\filtrationh_{j,n})_{j \geq 1}$ is an MDS.
Next, we compute its variance process.
By independence across strata conditional on $\filtrationcand$,
\begin{align*}
\hk_n &\equiv \sum_{j=1}^{n/k} E[\zjn^2 | \filtrationh_{j-1,n}] = n\inv \sum_{j=1}^{n/k}\Big(\sum_{r \not = t \in s_j}\ur\ut\cov(D_r,D_t | \filtrationcand) + \sum_{i \in s_j}\ui^2\var(\Di | \filtrationcand)\Big).
\end{align*}
By Lemma \lemmadesignproperties{} of \cite{cytrynbaum2024}, if $r \not = t$ belong to the same stratum, then $\cov(D_r,D_t | \filtrationcand) = -l(k-l)/k^2(k-1) \equiv c$, while $\var(\Di | \filtrationcand) = \vard$.
Then we may expand $\hk_n$ as
\[
\hk_n = c n\inv \sum_{j=1}^{n/k}\sum_{r \not = t \in s_j}\ur\ut + \vard\en[u_i^2] \equiv c n\inv \sum_{j=1}^{n/k}v_j + \vard\en[u_i^2] \equiv T_{n1} + T_{n2}.
\]
First consider $T_{n1}$.
Our plan is to apply the WLLN in Lemma \lemmalln{} of \cite{cytrynbaum2024} to show $T_{n1} = \op(1)$.
Define $\filtrationcand^{\psi} = \sigma(\psin, \permn)$ so that $\groupsetn \in \filtrationcandpsi$.
For $r \not = t$ we have $E[u_r u_t | \psin, \permn] = E[u_r E[u_t | \psin, u_r, \permn] | \psin, \permn] = E[u_r E[u_t | \psi_t] | \psin, \permn] = 0$.
The second equality follows by applying $(A, B) \indep C \implies A \indep C | B$ with $A = u_t$, $B = \psi_t$ and $C = (\psi_{-t}, u_r, \permn)$.
Then $E[v_j | \filtrationcandpsi] = 0$ for $j \in [n/k]$.
Next, observe that for any positive constants $(a_k)_{k=1}^m$ we have $\sum_k a_k \one(\sum_k a_k > c) \leq m \sum_k a_k \one(a_k > c / m)$ and $ab \one(ab > c) \leq a^2 \one(a^2 > c) + b^2 \one(b^2 > c)$.
Then for $c_n \to \infty$ with $c_n = o(\rootn)$ we have
\begin{align*}
|v_j|\one(|v_j| > c_n) &\leq \sum_{r \not = t \in s_j}|\ur\ut|\one\Big(\sum_{r \not = t \in s_j}|\ur\ut| > c_n\Big) \\
&\leq k^2\sum_{r \not = t \in s_j}|\ur\ut|\one(|\ur\ut| > c_n/k^2) \leq 2k^3\sum_{r \in s_j}\ur^2\one(\ur^2 > c_n/k^2).
\end{align*}
Then we have
\[
n\inv E\Big[\sum_{j=1}^{n/k}E[|v_j|\one(|v_j| > c_n) | \filtrationcandpsi]\Big] \leq 2k^3\en\Big[E[\ui^2\one(\ui^2 > c_n/k^2) | \psin,\permn]\Big] \equiv A_n.
\]
Then $E[A_n] = 2k^3 E[\en[E[\ui^2\one(\ui^2 > c_n/k^2) | \psii]]] = 2k^3 E[\ui^2\one(\ui^2 > c_n/k^2)] \to 0$ as $n \to \infty$.
The first equality is by the conditional independence argument above, the second equality is tower law, and the limit by dominated convergence since $E[\ui^2] \leq E[\diffi^2] < \infty$ by the contraction property of conditional expectation.
Then $A_n = \op(1)$ by Markov inequality.
The conclusion $c n\inv \sum_{j=1}^{n/k} v_j = \op(1)$ now follows by Lemma \lemmalln{} of \cite{cytrynbaum2024}.
For $T_{n2}$, we have $\en[\ui^2] \convp E[\ui^2] = E[\var(\diff | \psi)]$ by vanilla WLLN.
Thus, $\hk_n \convp \vard E[\var(\diff | \psi)]$.

Independence across strata gives $E[\zjn^2\one(|\zjn| > \epsilon) | \filtrationh_{j-1,n}] = E[\zjn^2\one(|\zjn| > \epsilon) | \filtrationh_{0,n}]$ for every $j$.
Thus, it suffices to show that the sum of latter terms is $\op(1)$.
\begin{align*}
&\zjn^2\one(|\zjn| > \epsilon) \leq n\inv\sum_{r,t \in s_j}|\ur\ut|\one\Big(n\inv\sum_{r,t \in s_j}|\ur\ut| > \epsilon^2\Big) \\
&\leq k^2n\inv\sum_{r,t \in s_j}|\ur\ut|\one(|\ur\ut| > n\epsilon^2/k^2) \leq k^3n\inv\sum_{r \in s_j}\ur^2\one(\ur^2 > n\epsilon^2/k^2).
\end{align*}
Then using the inequality above we compute
\begin{align*}
&E\Big[\sum_{j=1}^{n/k}E[\zjn^2\one(|\zjn| > \epsilon) | \filtrationh_{0,n}]\Big] \leq k^3E\Big[n\inv\sum_{j=1}^{n/k}\sum_{r \in s_j}E[\ur^2\one(\ur^2 > n\epsilon^2/k^2) | \filtrationcandpsi]\Big] \\
&= k^3E\Big[\en[E[\ui^2\one(\ui^2 > n\epsilon^2/k^2) | \psii]]\Big] = k^3E[\ui^2\one(\ui^2 > n\epsilon^2/k^2)] = o(1).
\end{align*}
The first equality by the conditional independence argument above.
The second equality by dominated convergence.
Then $\sum_{j=1}^{n/k} E[\zjn^2 \one(|\zjn| > \epsilon) | \filtrationh_{0, n}] = \op(1)$ by Markov.
This proves the Lindeberg condition.

Let $Y_n = \rootn\en[(\Di-\propfn)\diffi]$.
Since $\filtrationh_{0,n} = \filtrationcand$, Theorem \thmmdsclt{} in \cite{cytrynbaum2024} gives $E[e^{itY_n} | \filtrationcand] = e^{-t^2V_Y/2} + \op(1)$, where $V_Y = \vard E[\var(\diff | \psi)]$.
The statistic in the theorem is $X_n = \vard\inv Y_n$, so in the scalar case $E[e^{itX_n} | \filtrationcand] = e^{-t^2V/2} + \op(1)$ with $V = \vard\inv E[\var(\diff | \psi)]$.

Finally, consider vector-valued $\diff$.
For any $t \in \mr^{\dimdiff}$, apply the scalar result to $\diffbar(W_i) = t'\diff(W_i)$.
Since $t'X_n = \rootn\en[\Hi\diffbar(W_i)]$, this gives $E[e^{it'X_n} | \filtrationcand] = e^{-t'Vt/2} + \op(1)$, as claimed.
Finally, the conditional characteristic functions are bounded, so this convergence also holds in $L_1$.
Since $\sigma(\Wn) \sub \filtrationcand$, the tower property and Jensen's inequality give the same characteristic-function convergence conditional on $\Wn$ alone.
\end{proof}

We use this CLT to establish the following key lemma, which will shortly allow us to compute the asymptotic distribution of $\rootn \en[\Hi \diff(\Wi)]$ directly under the rerandomization law $Q$.

\begin{lem} \label{lemma:characteristic-convergence}
Let Definition \ref{defn:rerandomization-general} hold.
Suppose $E[|\diff|_2^2] < \infty$.
Let $\imbdiff = \en[\hti \diffi]$ and $\diffh = (\diff',h')'$.
Fix $t \in \mr^{\dimdiff}$.
Let $(\zdiff, \zh) \sim \normal(0, \covjoint)$ for $\covjoint = \vard \inv E[\var(\diffh | \psi)]$. 
Then under $P$ in Definition \ref{defn:pure-stratification}, we have $E[e^{it'\rootn\imbdiff}\one(\imbalance \in \rerandsetest) | \filtrationcand] = E[e^{it'\zdiff}\one(\zh \in \rerandsetpop)] + \op(1)$.
\end{lem}

\begin{proof}
Define $B_n = (\rootn\imbdiff,\imbalance,\rerandparamest)$.
Fix $r = (r_1',r_2',r_3')' \in \mr^{\dimdiff + \dimh + \dimrerandparam}$ and let $\bar r = (r_1',r_2')'$.
The conditional characteristic function satisfies
\begin{align*}
\varphi_{B_n}(r) &= E[e^{ir_1'\rootn\imbdiff + ir_2'\imbalance + ir_3'\rerandparamest} | \filtrationcand] \\
&= e^{ir_3'\rerandparampop}E[e^{ir_1'\rootn\imbdiff + ir_2'\rootn\en[\Hi h_i]} | \filtrationcand] + \op(1) \\
&= e^{ir_3'\rerandparampop}e^{-\bar r'\covjoint\bar r/2} + \op(1) \equiv \varphi_B(r) + \op(1).
\end{align*}
The second equality follows because $\rerandparamest = \rerandparampop + \op(1)$ and $\imbalance = \rootn\en[\Hi h_i] + \op(1)$.
The resulting exponential remainders are bounded by two and converge to zero in probability, so their conditional expectations are $\op(1)$ by Lemma \ref{lemma:modes-of-convergence}.
The final equality applies Theorem \ref{thm:clt} to $\rootn\en[\Hi\diffh_i]$.
The function $\varphi_B$ is the characteristic function of $B = (\zdiff,\zh,\rerandparampop)$.
Then we have shown that $B_n | \filtrationcand \convwprocess B$ in the sense of Proposition \ref{prop:levy}.
Define $G(z_1,z_2,x) = e^{it'z_1}\one(\setfn(z_2,x) \leq 0)$, so
\[
G(B_n) = e^{it'\rootn\imbdiff}\one(\setfn(\imbalance,\rerandparamest) \leq 0) = e^{it'\rootn\imbdiff}\one(\imbalance \in \rerandsetest).
\]
Let $E_G = \{w:G\text{ is discontinuous at }w\}$.
Joint continuity of $\setfn$ implies that $G$ is continuous at every $(z_1,z_2,x)$ such that $\setfn(z_2,x) \not = 0$.
Therefore, $P(B \in E_G) \leq P(\setfn(\zh,\rerandparampop) = 0) = 0$ by Definition \ref{defn:rerandomization-general}.
Proposition \ref{prop:levy} now gives $E[G(B_n) | \filtrationcand] = E[G(B)] + \op(1)$, which is the required claim.
\end{proof}

Finally, we come to the core asymptotic result for step (3) above.  

\begin{thm}[Asymptotic Distribution] \label{thm:rerand-asymptotics}
Let Definition \ref{defn:rerandomization-general} hold.
Suppose $E[|\diff|_2^2] < \infty$ and let $(\zdiff,\zh) \sim \normal(0,\covjoint)$ for $\covjoint = \vard\inv E[\var((\diff,h) | \psi)]$.
Let $\gammaoptmat \in \mr^{\dimh \times \dimdiff}$ solve $E[\var(h | \psi)]\gammaoptmat = E[\cov(h,\diff | \psi)]$, and define $\vardiff = \vard\inv E[\var(\diff-\gammaoptmat'h | \psi)]$.
Let $R \sim \gammaoptmat'\zh | \zh \in \rerandsetpop$.
Then under $Q$ in Definition \ref{defn:rerandomization-general}: 
\begin{enumerate}[label={(\alph*)}, itemindent=.5pt, itemsep=.4pt] 
\item We have $\rootn\en[\hti\diff(W_i)] | \filtrationcand \convwprocess [\zdiff | \zh \in \rerandsetpop] \sim \normal(0,\vardiff)+R$, where the two summands are independent and $\vardiff = \min_{\gammacoeff \in \mr^{\dimh \times \dimdiff}}\vard\inv E[\var(\diff(W)-\gammacoeff'h | \psi)]$.
The same convergence holds conditional on $\Wn$ alone.
\item Suppose $\phi(W) \in \mr^{\dimdiff}$ and $E[|\phi(W)|_2^2] < \infty$.
Let $X_n = \en[\phi(\Wi)] + \en[\hti\diff(W_i)]$ and $\vphi = \var(\phi(W))$.
Then $\rootn(X_n-E[\phi(W)]) \convwprocess \normal(0,\vphi)+\normal(0,\vardiff)+R$, where the three summands are independent.
\end{enumerate}
\end{thm}

\begin{proof}
Consider (a). 
Let $\imbdiff = \en[\hti \diff(W_i)]$.
Let $t \in \mr^{\dimdiff}$. 
By definition of $Q$,
\begin{align*}
\Eq[e^{it'\rootn\imbdiff} | \filtrationcand] &= E[e^{it'\rootn\imbdiff} | \imbalance \in \rerandsetest,\filtrationcand] \\
&= \frac{E[e^{it'\rootn\imbdiff}\one(\imbalance \in \rerandsetest) | \filtrationcand]}{P(\imbalance \in \rerandsetest | \filtrationcand)} \equiv \frac{\an}{\bn}.
\end{align*}
Define $\ainf = E[e^{it'\zdiff}\one(\zh \in \rerandsetpop)]$ and $\binf = P(\zh \in \rerandsetpop)$.
By Lemma \ref{lemma:characteristic-convergence}, $\an \convp \ainf$ and $\bn \convp \binf$, with $\binf > 0$ by assumption in Definition \ref{defn:rerandomization-general}. 
Both $\an$ and $\bn$ are $\filtrationcand$-measurable, so the same convergences hold under $Q$ by Lemma \ref{lemma:dominance}(a).
Therefore, $\an/\bn = \ainf/\binf + \op(1)$, and hence $\Eq[e^{it'\rootn\imbdiff} | \filtrationcand] = E[e^{it'\zdiff} | \zh \in \rerandsetpop] + \op(1)$.
Proposition \ref{prop:levy} proves the first convergence statement.

It remains to characterize the conditional Gaussian law.
Define $\varphi_{\rerandsetpop}(t) = E[e^{it'\zdiff} | \zh \in \rerandsetpop]$.
The existence and optimality of $\gammaoptmat$ follow from Lemma \ref{lemma:orthogonalization}.
Letting $\zindep = \zdiff - \gammaoptmat'\zh$, by Lemma \ref{lemma:orthogonalization} $\zindep \indep \zh$ and $\zindep$ is Gaussian. 
Then $\zindep \indep (\zh,\one(\zh \in \rerandsetpop))$.
Therefore, $\zindep \indep \zh | \zh \in \rerandsetpop$.
Then for any $t \in \mr^{\dimdiff}$,
\begin{align*}
\varphi_{\rerandsetpop}(t) &= E[e^{it'\zindep}e^{it'\gammaoptmat'\zh} | \zh \in \rerandsetpop] \\
&= E[e^{it'\zindep}]E[e^{it'\gammaoptmat'\zh} | \zh \in \rerandsetpop].
\end{align*}
By Proposition \ref{prop:levy}, $\zdiff | \zh \in \rerandsetpop \equaldist \zindep + [\gammaoptmat'\zh | \zh \in \rerandsetpop]$, where the two terms on the right are independent.
Moreover, $E[\zindep] = 0$ and $\var(\zindep) = \vard\inv E[\var(\diff-\gammaoptmat'h | \psi)] = \vardiff$.
The conditional characteristic functions are bounded, so their convergence also holds in $L_1(Q)$.
The tower property and Jensen's inequality therefore give the same characteristic-function convergence conditional on $\Wn$ alone.
This finishes (a). 

Next we prove (b). 
Write $\rootn(X_n-E[\phi(W)]) = A_n+B_n$, where $A_n = \rootn(\en[\phi(W_i)]-E[\phi(W)])$ and $B_n = \rootn\imbdiff$.
The random variable $A_n$ is $\filtrationcand$-measurable.
For $t \in \mr^{\dimdiff}$, part (a) and bounded convergence give $\Eq[e^{it'B_n} | \filtrationcand] = \varphi_{\rerandsetpop}(t) + \op(1)$ in $L_1(Q)$.
Therefore,
\[
\Eq[e^{it'(A_n+B_n)}] = \varphi_{\rerandsetpop}(t)\Eq[e^{it'A_n}] + o(1).
\]
The marginal law of $(\Wn,\permn)$ is identical under $P$ and $Q$, so the ordinary CLT gives $\Eq[e^{it'A_n}] = e^{-t'\vphi t/2} + o(1)$.
Combining this result with the expression for $\varphi_{\rerandsetpop}(t)$ above yields
\[
\Eq[e^{it'\rootn(X_n-E[\phi(W)])}] = e^{-t'(\vphi+\vardiff)t/2}E[e^{it'\gammaoptmat'\zh} | \zh \in \rerandsetpop] + o(1).
\]
This finishes the proof of (b).
\end{proof}

\subsection{Proof of Main Results}

Next, we prove Theorem \ref{thm:gmm-asymptotics} and Corollary \ref{cor:gmm-superpopulation}.
See Section \ref{proofs:gmm-linearization} below for the proof of the following lemma, which uses a novel ULLN for GMM estimation under fine stratification. 

\begin{lem}[Linearization] \label{lemma:linearization}
Suppose Definition \ref{defn:rerandomization-general} and Assumption \ref{assumption-gmm} hold.
Let $\matgmm = -(G'\weightmatpop G)\inv G' \weightmatpop$.
Then $\rootn(\est - \estn) = \rootn\en[\Hi\assignif(W_i)] + \op(1)$.
Moreover,
\[
\rootn(\est - \thetatrue) = \rootn\en[\sampif(W_i) + \Hi\assignif(W_i)] + \op(1).
\]
\end{lem}

\begin{proof}[Proof of Theorem \ref{thm:gmm-asymptotics}]
We claim that the conditions of Definition \ref{defn:rerandomization-general} hold.
This will allow us to apply our general rerandomization asymptotics in Theorem \ref{thm:rerand-asymptotics} and linearization in Lemma \ref{lemma:linearization}. 
To check part (a), if $A=\mrh$, take $\setfn(x,y)=-1$, in which case the claim is immediate.
Otherwise, define $\setfn(x,y)=\setfn(x)=d(x,A)-d(x,A^c)$, where $d(x,S)=\inf_{s\in S}|x-s|_2$.
It is well known that $x \to d(x,S)$ is continuous for any nonempty set $S$.
Since $\setfn$ does not depend on $y$, it is jointly continuous in $(x,y)$.
The sample and population regions $\rerandsetest = \rerandsetpop = \{x: \setfn(x) \leq 0\}$. 
If $\setfn(x) \leq 0$ then $d(x, A) = 0$, so $x \in A \cup \partial A \sub A$ by closedness.
If $\setfn(x) > 0$ then $x \not \in A$.
This shows $\rerandsetest = A$, so $\{\imbalance \in \rerandsetest\} = \{\imbalance \in A\}$.
Then our criterion is of the form in Definition \ref{defn:rerandomization-general}.
For part (b), $P(\setfn(\zh) = 0) = P(\zh \in \partial A) = 0$ since $\leb(\partial A) = 0$ and by absolute continuity of $\zh$ relative to Lebesgue measure. 
We also have $P(\zh \in \rerandsetpop) = P(\zh \in A) > 0$ because $E[\var(h | \psi)] \succ 0$ implies that $\zh$ has an everywhere-positive density and $A$ has non-empty interior.
This proves the claim.
Then by Lemma \ref{lemma:linearization}, $\rootn(\est - \estn) = \rootn\en[\Hi\assignif(W_i)] + \op(1)$.
The result now follows immediately by Slutsky and Theorem \ref{thm:rerand-asymptotics}(a), letting $\diff \to \assignif$.
Likewise, Corollary \ref{cor:gmm-superpopulation} follows from Theorem \ref{thm:rerand-asymptotics}(b), letting $\phi \to \sampif$.
\end{proof}

\section{General GMM Rerandomization} \label{appendix:gmm-rerandomization}

\setlength{\tabcolsep}{1pt}
\begin{figure}[t]
\centering
\begin{tabular}{cc}
\includegraphics[width=0.50\textwidth]{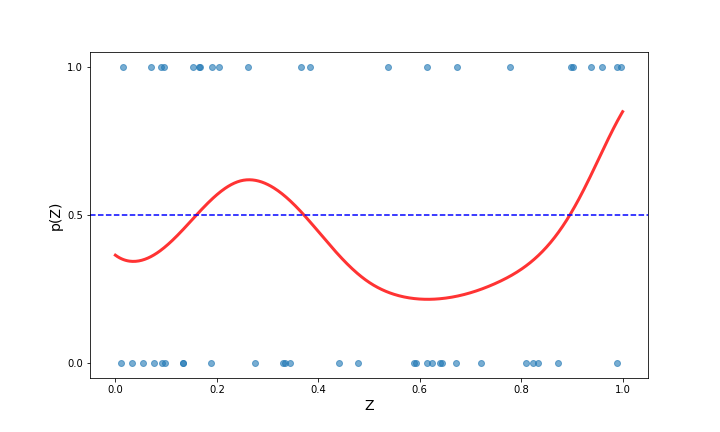} &
\includegraphics[width=0.50\textwidth]{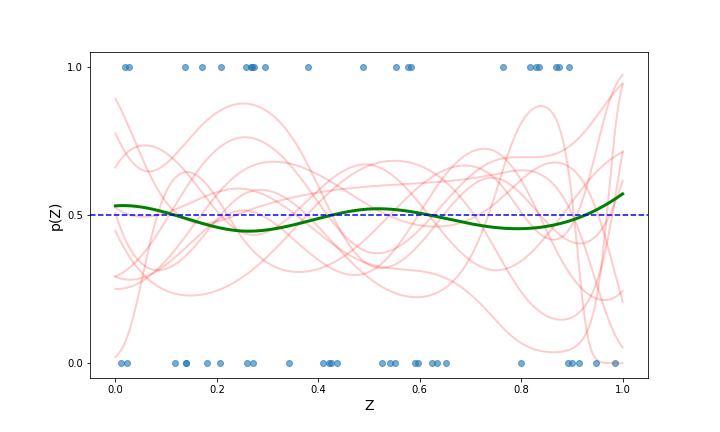} \\
\end{tabular}
\caption{Propensity rerandomization from Equation \ref{eqn:propensity-rerandomization} with $\propfn=1/2$ for $Z \sim \unif[0,1]$ and $X=B(Z)$ a B-spline basis.
LHS: $\Dn$ and estimated propensity with $\wh p(Z) \ll 1/2$ for $Z \in [0.4,0.9]$, showing imbalance.
RHS: Accepted allocation $\Dn$ with $\imbalancesquare \leq \epsilon^2$.}
\label{fig:propensity-rerandomization}
\end{figure}
\setlength{\tabcolsep}{6pt}

We provide a general version of the nonlinear rerandomization results in Section \ref{section:nonlinear}.
Let $\rerandmom(X,\beta) \in \mr^{d_{\rerandmom}}$ be a GMM score, separate from the causal score $g$ defining the outcome estimand.
Suppose $\rerandmom$ satisfies the analog of Assumption \ref{assumption-gmm} for parameter $\beta$ and let the exactly identified within-arm estimators solve
\begin{equation} \label{eqn:rerand-gmm}
\en[\Di\rerandmom(X_i,\betaestone)]=0, \qquad \en[(1-\Di)\rerandmom(X_i,\betaestzero)]=0.
\end{equation}

\begin{defn}[GMM Rerandomization] \label{defn:gmm-rerandomization}
Let $A \sub \mr^{\dimbeta}$ be a closed symmetric acceptance region with non-empty interior and $\leb(\partial A)=0$.
Form groups as in Definition \ref{defn:rerandomization}, draw $\Dn$ by stratified randomization, and estimate Equation \ref{eqn:rerand-gmm}.
Accept the assignment if $\rootn(\betaestone-\betaestzero) \in A$ and otherwise repeat the randomization and estimation steps.
\end{defn}

When $\rerandmom(X_i,\beta)=X_i-\beta$, this criterion reduces to linear rerandomization based on the difference of covariate means.
More generally, let $\betatrue$ be the unique solution to $E[\rerandmom(X,\betatrue)]=0$, define $\rerandmom_i^* \equiv \rerandmom(X_i,\betatrue)$, and let $\rerandjacob \equiv E[(\partial/\partial\beta')\rerandmom(X,\betatrue)]$.
The following theorem shows that the general criterion is first-order equivalent to linear rerandomization on its score.

\begin{thm}[GMM Rerandomization] \label{thm:nonlinear-rerandomization}
Suppose $\Dn$ is as in Definition \ref{defn:gmm-rerandomization} and the causal score $g$ satisfies Assumption \ref{assumption-gmm}.
Require $E[\var(\rerandmom_i^*|\psi)]\succ0$.
Then $\rootn(\est-\estn)|\Wn \convwprocess \normal(0,\vassign)+R$, where the two terms are independent and
\begin{equation} \label{eqn:gmm-rerand-variance}
\vassign=\min_{\gammacoeff \in \mr^{d_{\rerandmom} \times \dimtheta}}\vard\inv E[\var(\assignif-\gammacoeff'\rerandmom_i^*|\psi)].
\end{equation}
The residual satisfies $R \sim \gammaoptmat'\zm|\zm \in \rerandjacob A$ for $\zm \sim \normal(0,\vard\inv E[\var(\rerandmom_i^*|\psi)])$, where $\gammaoptmat$ is optimal in Equation \ref{eqn:gmm-rerand-variance}.
\end{thm}

Thus GMM rerandomization implicitly balances $-\rerandjacob\inv\rerandmom(X,\betatrue)$, the influence function for the difference between the within-arm estimators.

\begin{assumption}[Propensity Rerandomization] \label{assumption:propensity-rerandomization}
To state the technical conditions, write $\beta=(\alpha,\delta)$, $\betaest=(\wh\alpha,\wh\delta)$, and $X=(1,h)$.
Let $\imbalancesquare \equiv n\en[(\propfn-\link(X_i'\betaest))^2]$.
The following conditions hold:
\begin{enumerate}[label={(\alph*)}, itemindent=.5pt, itemsep=.4pt]
\item Let $\link$ be twice differentiable, with $|L'|_{\infty}, |L''|_{\infty}<\infty$.
For each $\propfn \in (0,1)$, there is a unique $c$ with $L(c)=\propfn$ and $|L'(c)|>0$.
\item The score $m(\Di,X_i,\beta)=\Di\frac{\link'(X_i'\beta)X_i}{\link(X_i'\beta)}-(1-\Di)\frac{\link'(X_i'\beta)X_i}{1-\link(X_i'\beta)}$ satisfies Assumption \ref{assumption-gmm} for parameter $\beta$.
The maximizer in Equation \ref{eqn:propensity-estimation} exists.
\item We have $E[|h|_2^2] < \infty$, and $E[\var(h | \psi)]$ and $\var(h)$ are full rank.
\end{enumerate}
\end{assumption}

{\small
\bibliography{design_references.bib}}

\clearpage

\begin{center}
{\LARGE Online Appendix}
\end{center}

\section{Additional Results} \label{section:appendix}

\pagenumbering{arabic}\renewcommand{\thepage}{\arabic{page}}

\subsection{Belief-Set Geometry} \label{section:belief-set-geometry}

For $p \in [1,\infty)$, let $|x|_p=(\sum_j|x_j|^p)^{1/p}$, and let $|x|_\infty=\max_j|x_j|$.
Define $\ballp(0,1)=\{x:|x|_p\leq1\}$.
The following result characterizes the minimax acceptance region for translated $L_p$ belief sets.

\begin{lem}[Belief Specification] \label{lemma:polar-sets}
For $p \in [1,\infty]$, let $1/p+1/q=1$.
Suppose $\balancecoeffs=\bar\gamma+U\ballp(0,1)$ for $\bar\gamma\in\mrh$ and invertible $U$.
Then the acceptance region is $A=\{x:|x'\bar\gamma|+|U'x|_q\leq\epsilon\}$.
\end{lem}

\subsection{Beliefs From Pilot Data} \label{section:beliefs-from-pilot}
Here, we discuss an alternative to the a priori beliefs in Section \ref{section:acceptance-regions} that uses pilot data to specify the set $B$ in a data-driven way. 
Suppose we have access to a dataset $\datapilot \indep (\Wn, \Dn)$ of size $m$. 
Suppose $\sqrt{\npilot} (\wh \gamma_{pilot} - \gammatrue) \approx \normal(0, \Sigmapilot)$ for some pilot estimator $\wh \gamma_{pilot}$, discussed below.
Consider forming the Wald region $\coeffpilot = \{\gamma: m (\wh \gamma_{pilot} - \gamma)' \Sigmapilot \inv (\wh \gamma_{pilot} - \gamma) \leq c_{\alpha}\}$ using critical value $P(\chi^2_{\dimh} \leq c_{\alpha}) = 1-\alpha$ for $\alpha \in (0, 1)$.
Equivalently, one can write this Wald region as
\begin{equation}
\coeffpilot = \wh \gamma + c_{\alpha}^{1/2} m^{-1/2} \cdot \Sigmapilot^{1/2} B_2(0, 1).
\end{equation}
Viewing this $1-\alpha$ confidence region as a belief set, Lemma \ref{lemma:polar-sets} implies that the corresponding minimax acceptance region is 
\begin{equation} \label{eqn:pilot-wald-region}
\wh A_{pilot} = \epsilon \coeffpilot^{\circ} = \{x: |x'\wh \gamma_{pilot}| + \frac{c_{\alpha}^{1/2} |\wh \Sigma^{1/2} x|_2}{m \half} \leq \epsilon \}.
\end{equation}

Note that the acceptance region $\wh A_{pilot}$ expands as the pilot size $m$ is larger.
This reflects smaller uncertainty about the true parameter $\gammatrue$, and thus less adversarial worst case imbalance $\sup_{\gamma \in \coeffpilot} |\gamma' \rootn(\hbarone - \hbarzero)|$. 
Conversely, $\wh A_{pilot}$ shrinks as the confidence parameter $\alpha$ and variance estimate $\Sigmapilot$ increase, reflecting greater uncertainty and a more conservative approach to covariate imbalances. 
Our next result shows that rerandomization with acceptance region $\wh A_{pilot}$ controls the variance of the residual imbalance $\residualvar = \gammatrue'\zh | \zh \in \wh A_{pilot}$ with high probability, marginally over the realizations of the pilot data.    
The result is an immediate consequence of Theorem \ref{thm:gmm-asymptotics} and Theorem \ref{thm:minimax}.  

\begin{cor}[Pilot Data] \label{cor:pilot-acceptance}
Suppose $P(\gammatrue \in \coeffpilot) \geq 1-\alpha$, for $\datapilot \indep (\Wn, \Dn)$. 
Let $\Dn$ as in Definition \ref{defn:rerandomization} with $A = \wh A_{pilot} = \epsilon \coeffpilot^{\circ}$. 
If Assumptions \ref{assumption:linear-rerandomization}, \ref{assumption-gmm} hold, then $\rootn(\est - \thetan)| \datapilot \convwprocess \normal(0, \vard \inv E[\var(\ylevel - \gammaoptmat'h | \psi)]) + \residualvar$, where $\var(\residualvar | \datapilot) \leq \epsilon^2$ with probability $\geq 1-\alpha$.
\end{cor}

Formally, the pilot estimate of $\gammatrue$ and Wald region could be constructed as in \cite{robinson88}. 
A simpler practical approach suggested by the theory is to let $\gammapilot, \Sigmapilot$ be point and variance estimators from the regression $Y_{T} \sim 1 + h + \psi$, for the ``tyranny of the minority'' (\cite{lin2013}) outcomes $Y_T = (1-\propfn)DY / \propfn +  p(1-D)Y / (1-p)$, noting that $E[Y_T | W] = (1-p)Y(1) + pY(0) = \ylevel$. \medskip 

\subsection{Additional Simulation Results} \label{section:additional-simulation-results} \label{section:simulations-exotic}

Table \ref{table:efficiency} reports the complete numerical results underlying Figures \ref{fig:simulation-efficiency} and \ref{fig:simulation-inference}.
\begin{table}[htbp]
\begin{adjustbox}{scale=0.82, keepaspectratio, center}
  \centering
    \begin{tabular}{ccccccccc|cccccc}
          &       &       &       &       & \multicolumn{2}{c}{$n=300$} &       & \multicolumn{1}{r}{} &       &       & \multicolumn{2}{c}{$n=600$} &       &  \\
          &       &       & \multicolumn{2}{c}{MSE} & \multicolumn{2}{c}{Cover} & \multicolumn{2}{c}{CI Width} & \multicolumn{2}{c}{MSE} & \multicolumn{2}{c}{Cover} & \multicolumn{2}{c}{CI Width} \\
    \midrule
    $\dim(r)$   & Mod.    & Design & $\est$ & $\estadj$   & Pop.  & Fin.  & Pop.  & Fin.  & $\est$ & $\estadj$   & Pop.  & Fin.  & Pop.  & Fin. \\
    \midrule
          &       & C     & 1.00  & 0.89  & 0.94  & 0.94  & 1.00  & 0.70  & 1.00  & 0.86  & 0.96  & 0.96  & 1.00  & 0.69 \\
          & 1     & S     & 0.85  & 0.87  & 0.94  & 0.98  & 1.03  & 0.82  & 0.87  & 0.88  & 0.95  & 0.97  & 1.02  & 0.77 \\
          &       & SR    & 0.81  & 0.81  & 0.96  & 0.97  & 1.01  & 0.73  & 0.86  & 0.86  & 0.95  & 0.96  & 1.01  & 0.70 \\
\cmidrule{3-15}          &       & C     & 1.00  & 0.62  & 0.94  & 0.94  & 1.00  & 0.67  & 1.00  & 0.61  & 0.95  & 0.97  & 1.00  & 0.66 \\
          & 2     & S     & 0.62  & 0.62  & 0.95  & 0.97  & 1.04  & 0.80  & 0.64  & 0.63  & 0.95  & 0.97  & 1.02  & 0.74 \\
    5     &       & SR    & 0.55  & 0.55  & 0.95  & 0.97  & 1.03  & 0.71  & 0.62  & 0.61  & 0.96  & 0.97  & 1.01  & 0.68 \\
\cmidrule{3-15}          &       & C     & 1.00  & 0.73  & 0.94  & 0.97  & 1.00  & 0.76  & 1.00  & 0.75  & 0.96  & 0.98  & 1.00  & 0.76 \\
          & 3     & S     & 0.60  & 0.64  & 0.95  & 0.98  & 0.98  & 0.75  & 0.62  & 0.62  & 0.96  & 0.98  & 0.96  & 0.68 \\
          &       & SR    & 0.53  & 0.53  & 0.96  & 0.98  & 0.94  & 0.61  & 0.59  & 0.59  & 0.96  & 0.97  & 0.92  & 0.57 \\
\cmidrule{3-15}          &       & C     & 1.00  & 0.80  & 0.93  & 0.95  & 1.00  & 0.86  & 1.00  & 0.81  & 0.94  & 0.97  & 1.00  & 0.86 \\
          & 4     & S     & 0.73  & 0.74  & 0.95  & 0.98  & 1.02  & 0.92  & 0.79  & 0.79  & 0.96  & 0.97  & 1.01  & 0.88 \\
          &       & SR    & 0.70  & 0.71  & 0.95  & 0.97  & 1.00  & 0.85  & 0.79  & 0.78  & 0.96  & 0.97  & 0.99  & 0.84 \\
\cmidrule{2-15}          &       & C     & 1.00  & 0.93  & 0.94  & 0.95  & 1.00  & 0.73  & 1.00  & 0.85  & 0.94  & 0.96  & 1.00  & 0.71 \\
          & 1     & S     & 0.95  & 0.97  & 0.93  & 0.98  & 1.07  & 0.93  & 0.93  & 0.95  & 0.93  & 0.97  & 1.03  & 0.84 \\
          &       & SR    & 0.88  & 0.87  & 0.95  & 0.98  & 1.04  & 0.83  & 0.85  & 0.83  & 0.95  & 0.97  & 1.02  & 0.77 \\
\cmidrule{3-15}          &       & C     & 1.00  & 0.63  & 0.93  & 0.95  & 1.00  & 0.70  & 1.00  & 0.65  & 0.95  & 0.96  & 1.00  & 0.68 \\
          & 2     & S     & 0.69  & 0.68  & 0.94  & 0.99  & 1.09  & 0.97  & 0.74  & 0.71  & 0.94  & 0.98  & 1.04  & 0.83 \\
    20    &       & SR    & 0.59  & 0.61  & 0.96  & 0.99  & 1.11  & 0.87  & 0.65  & 0.64  & 0.96  & 0.98  & 1.06  & 0.77 \\
\cmidrule{3-15}          &       & C     & 1.00  & 0.75  & 0.92  & 0.96  & 1.00  & 0.76  & 1.00  & 0.78  & 0.95  & 0.97  & 1.00  & 0.76 \\
          & 3     & S     & 0.69  & 0.75  & 0.94  & 0.98  & 1.06  & 0.93  & 0.76  & 0.76  & 0.94  & 0.99  & 1.01  & 0.82 \\
          &       & SR    & 0.53  & 0.57  & 0.96  & 0.99  & 1.02  & 0.76  & 0.59  & 0.60  & 0.95  & 0.98  & 0.96  & 0.66 \\
\cmidrule{3-15}          &       & C     & 1.00  & 0.82  & 0.92  & 0.94  & 1.00  & 0.86  & 1.00  & 0.84  & 0.96  & 0.97  & 1.00  & 0.86 \\
          & 4     & S     & 0.83  & 0.84  & 0.94  & 0.98  & 1.08  & 1.05  & 0.94  & 0.91  & 0.95  & 0.96  & 1.04  & 0.95 \\
          &       & SR    & 0.75  & 0.75  & 0.95  & 0.98  & 1.05  & 0.94  & 0.83  & 0.82  & 0.96  & 0.97  & 1.02  & 0.88 \\
    \bottomrule
    \end{tabular}%
\end{adjustbox}
\caption{Efficiency and inference across simulation specifications.}
\label{table:efficiency}%
\end{table}%

\emph{Comparing Rerandomization Types.}
\begin{table}[htbp]
\begin{adjustbox}{scale=0.73, keepaspectratio, center}
  \centering
    \begin{tabular}{cccrrrrrr}
          &       &       & \multicolumn{2}{c}{MSE} & \multicolumn{2}{c}{Cover} & \multicolumn{2}{c}{CI Width} \\
    \midrule
    $\thetan$ & Mod.    & \textbf{SR} Type & \multicolumn{1}{c}{$\est$} & \multicolumn{1}{c}{$\estadj$} & \multicolumn{1}{c}{Pop.} & \multicolumn{1}{c}{Fin.} & \multicolumn{1}{c}{Pop.} & \multicolumn{1}{c}{Fin.} \\
    \midrule
          &       & MH    & 1.00  & 1.03  & 0.96  & 0.99  & 1.00  & 0.82 \\
          &       & Prop  & 1.04  & 1.05  & 0.95  & 0.98  & 1.00  & 0.81 \\
          & 2     & Best1 & 0.99  & 1.02  & 0.96  & 0.98  & 1.00  & 0.81 \\
          &       & Best2 & 0.99  & 1.07  & 0.95  & 0.98  & 1.00  & 0.82 \\
          &       & Opt1  & 1.00  & 1.08  & 0.95  & 0.98  & 1.00  & 0.82 \\
    SATE  &       & Opt2  & 1.01  & 1.02  & 0.95  & 0.98  & 1.00  & 0.82 \\
\cmidrule{3-9}          &       & MH    & 1.00  & 1.06  & 0.96  & 0.98  & 1.00  & 0.77 \\
          &       & Prop  & 1.02  & 1.06  & 0.95  & 0.98  & 1.00  & 0.77 \\
          & 3     & Best1 & 0.99  & 1.04  & 0.96  & 0.99  & 1.00  & 0.77 \\
          &       & Best2 & 1.01  & 1.11  & 0.95  & 0.99  & 1.00  & 0.77 \\
          &       & Opt1  & 1.00  & 1.08  & 0.95  & 0.99  & 1.00  & 0.77 \\
          &       & Opt2  & 0.99  & 1.03  & 0.96  & 0.99  & 1.00  & 0.77 \\
    \midrule
          &       & MH    & 1.00  & 1.03  & 0.97  & 0.98  & 1.00  & 1.00 \\
          &       & Prop  & 0.99  & 1.01  & 0.97  & 0.99  & 1.00  & 1.01 \\
          & 2     & Best1 & 1.00  & 1.03  & 0.97  & 0.98  & 1.00  & 1.01 \\
          &       & Best2 & 1.04  & 1.06  & 0.97  & 0.97  & 1.00  & 1.01 \\
          &       & Opt1  & 1.00  & 1.03  & 0.98  & 0.98  & 1.00  & 1.01 \\
    CATE  &       & Opt2  & 0.97  & 1.00  & 0.98  & 0.98  & 1.00  & 1.01 \\
\cmidrule{3-9}          &       & MH    & 1.00  & 1.09  & 0.97  & 0.99  & 1.00  & 0.81 \\
          &       & Prop  & 0.96  & 1.03  & 0.97  & 0.99  & 0.99  & 0.81 \\
          & 3     & Best1 & 1.00  & 1.08  & 0.96  & 0.99  & 1.00  & 0.81 \\
          &       & Best2 & 1.02  & 1.09  & 0.96  & 0.99  & 1.01  & 0.82 \\
          &       & Opt1  & 1.00  & 1.08  & 0.96  & 0.99  & 1.00  & 0.81 \\
          &       & Opt2  & 0.99  & 1.09  & 0.97  & 0.99  & 1.01  & 0.82 \\
    \bottomrule
    \end{tabular}%
    \end{adjustbox}
  \caption{Stratified Rerandomization Types}
\label{table:exotics}%
\end{table}%

In Table \ref{table:exotics} we compare different types of stratified rerandomization acceptance criteria. 
\textbf{MH} is Mahalanobis rerandomization, as in Table \ref{table:efficiency}. 
\textbf{Prop} is the propensity-based rerandomization in Equation \ref{eqn:propensity-rerandomization}, using Logit $L(x) = (1 + e^{-x}) \inv$ and $X = (1, w)$.
Designs \textbf{Opt1} and \textbf{Opt2} refer to the optimal acceptance regions in Section \ref{section:acceptance-regions}.
The belief sets are both well-specified, with either high uncertainty $\balancecoeffs_1 = \{x: |x-\gammaoptmat|_2 \leq 1\}$ or low uncertainty $\balancecoeffs_2 = \{x: |x-\gammaoptmat|_2 \leq 1/10\}$, respectively.
In all designs, we set the balance threshold $\epsilon(\alpha)$ so $P(\zh \in A) = 1/500$. 
Finally, in \textbf{Best1} and \textbf{Best2} we rerandomize by implementing the best allocation out of either $k = 500$ or $k=2500$ stratified draws, according to the minimal Mahalanobis imbalance metric.
Note that such ``best-of-k'' stratified rerandomization designs are not formally covered by our theory.\footnote{Recent work by \cite{wang2024} provided the first formal results for ``best-of-k'' designs in the case without stratification.}
In addition to $\thetan = \sate$, we also provide efficiency and inference results for the  treatment effect heterogeneity parameter from Example \ref{ex:heterogeneity-blp}. 
In particular, let $\alpha_n = \argmin_{\alpha} \en[(Y_i(1) - Y_i(0) - \alpha'(1, r_{1i}))^2]$. 
We define $\thetan$ to be the coefficient on $r_{1}$, denoting $\thetan = \cate$ in the table.
Cover Pop.\ and CI Width Pop. refer to inference on the corresponding superpopulation parameter $\thetatrue$. 
Next, we summarize a few findings from Table \ref{table:exotics}. 
Proposition \ref{prop:propensity-rerandomization} showed that \textbf{Prop} was first-order equivalent to \textbf{MH}, and this is supported by finite-sample evidence in the table.
We find that best of $k$ style rerandomization and Mahalnobis rerandomization with acceptance probability $\alpha \approx 1/k$ are indistinguishable in practice. 
In particular, our inference methods also work well for this design.
We don't find major finite sample efficiency improvements from using the optimal acceptance regions in Section \ref{section:acceptance-regions} in this experiment.

\subsection{Empirical Application Details} \label{section:imputation-details}

The full set of covariates from the baseline survey in \cite{angrist2013} used in our imputation procedure is HS GPA, sex, year in college, mother's and father's education, whether survey question 1 was answered correctly, expected degree, age, native language, attempted credits, and financial stress.
The vector $X$ consists of these basic covariates and all of their pairwise interactions. 
For the ITT potential outcomes, we set $\wh T_i(z)=T_i$ if $Z_i=z$ and impute $\wh T_i(z)=\wh m_z^T(X_i)+\hksdest_z^T(X_i)\epsilon_{iz}$ otherwise, where $\epsilon_{iz}\sim\normal(0,1)$.
We estimate $\wh m_z^T(X)$ using LASSO, regressing $TZ/p$ on $X$ for $z=1$ and $T(1-Z)/(1-p)$ on $X$ for $z=0$, with the regularization parameter chosen by cross-validation.
We estimate the corresponding conditional residual variances using random forests applied to $(T_i-\wh m_1^T(X_i))^2Z_i/p$ and $(T_i-\wh m_0^T(X_i))^2(1-Z_i)/(1-p)$.
The function $\hksdest_z^T(X)$ is the square root of the nonnegative fitted conditional variance.

Because ineligible students could not engage with the program, we impose one-sided noncompliance and set $\wh D_i(0)=0$.
We set $\wh D_i(1)=D_i$ for students with $Z_i=1$.
For students with $Z_i=0$, we impute $\wh D_i(1)=\one(\wh m_1^D(X_i)+\hksdest_1^D(X_i)u_i\geq1/2)$, where $u_i\sim\normal(0,1)$ and the conditional mean and variance functions are estimated by cross-validated random forests.
For units with $\wh D_i(1)=0$, we impose exclusion by setting $\wh T_i(0)=\wh T_i(1)=T_i$.

\section{Additional Proofs} \label{section:proofs}

\subsection{GMM Linearization} \label{proofs:gmm-linearization}

This section collects proofs needed for the key linearization result in Lemma \ref{lemma:linearization}.
Define the score-level assignment component $\diff(W, \theta) \equiv \vard\{\mom_1(W, \theta)-\mom_0(W, \theta)\}$.
The Horvitz--Thompson weight $H=(D-p)/\vard$ gives the decomposition
\begin{equation} \label{eqn:orthog-decomposition}
g(D,R,S,\theta)=\mombar(W,\theta)+H\diff(W,\theta).
\end{equation}
First, define the following curves and objective functions
\begin{align*}
\momtrue(\theta) &= E[\mombar(W_i, \theta)], \quad \momn(\theta) = \en[\mombar(\Wi, \theta)], \quad \momest(\theta) = \en[\mombar(\Wi, \theta)] + \en[\hti \diff(\Wi, \theta)]. \\
\objtrue(\theta) &= \momtrue(\theta)'\weightmatpop\momtrue(\theta), \quad \objn(\theta) = \momn(\theta)'\weightmatpop\momn(\theta), \quad  \objest(\theta) = \momest(\theta)'\weightmatest\momest(\theta)
\end{align*}
Define $\wh G(\theta) = (\partial/\partial \theta')\momest(\theta)$ and $G_n(\theta) = (\partial/\partial \theta') \momn(\theta)$ and $G_0(\theta) = (\partial/\partial \theta')\momtrue(\theta)$. 
Define $G = G_0(\thetatrue)$. 
For each $d \in \{0, 1\}$, define $\mom_d(W, \theta) = g(d, R, S(d), \theta)$.

\begin{lem}[ULLN] \label{lemma:ulln}
Working under $P$ in Definition \ref{defn:pure-stratification}: 
\begin{enumerate}[label={(\alph*)}, itemindent=.5pt, itemsep=.4pt] 
\item If Assumption \ref{assumption-gmm}\ref{assumption-gmm-moments} holds, $\|\momest - \momtrue\|_{\infty, \thetaspace} = \op(1)$, $\|\momn - \momtrue\|_{\infty, \thetaspace} = \op(1)$, and $\momtrue(\theta)$ is continuous.  
If also $\weightmatest \convp \weightmatpop$ then $|\objn - \objtrue|_{\infty, \thetaspace} = \op(1)$ and $|\objest - \objtrue|_{\infty, \thetaspace} = \op(1)$. 
\item If Assumption \ref{assumption-gmm}\ref{assumption-gmm-deriv} holds, then there is an open ball $U \sub \thetaspace$ with $\thetatrue \in U$ and $\|\wh G - G_0\|_{\infty, U} = \op(1)$ and $\|G_n - G_0\|_{\infty, U} = \op(1)$.
Also, $G_0(\theta)$ is continuous on $U$ for $G_0(\theta) = \partial / \partial \theta' E[\mombar(W, \theta)]$.
\end{enumerate}
\end{lem}
\begin{proof}
Consider (a). 
Set $\compact=\thetaspace$.
First we show $\|\momest - \momtrue\|_{\infty, \thetaspace} = \op(1)$, modifying the approach used in the iid setting in \cite{tauchen1985}. 
It suffices to prove the statement componentwise.
Then without loss assume $\dimmom = 1$ and fix $\epsilon > 0$.
Note also that $\mombar, \diff$ are linear combinations of $\mom_d$ for $d \in \zerone$, so $\mombar$ and $\diff$ inherit the properties in Assumption \ref{assumption-gmm}.
We have $(\momest - \momn)(\theta) = \en[\hti \diff(\Wi, \theta)]$. 
For each $\theta \in \compact$ define $\utm = B(\theta, m\inv)\cap\compact$ and $\barvm(\Di, \Wi) = \sup_{\bar \theta \in \utm} \hti \diff(\Wi, \bar\theta)$. 
Then $\barvm(\Di, \Wi)$ may be expanded 
\begin{align*}
&\sup_{\bar \theta \in \utm} \hti \diff(\Wi, \bar \theta) = \frac{\Di}{\propfn} \sup_{\bar \theta \in \utm} \diff(\Wi, \bar \theta) + \frac{1-\Di}{1-\propfn} \sup_{\bar \theta \in \utm} - \diff(\Wi, \bar \theta) \\
= \; &\sup_{\bar \theta \in \utm} \diff(\Wi, \bar \theta) + \sup_{\bar \theta \in \utm} -\diff(\Wi, \bar \theta) \\
+ \; &\hti ((1-\propfn) \sup_{\bar \theta \in \utm} \diff(\Wi, \bar \theta) + \propfn \inf_{\bar \theta \in \utm} \diff(\Wi, \bar \theta)) \equiv f_{\theta m}(W_i) + \hti r_{\theta m}(W_i). 
\end{align*}
In particular, $E[\barvm(\Di,\Wi)] = E[f_{\theta m}(W_i)]$. 
Note both expectations exist by the envelope condition in Assumption \ref{assumption-gmm}.
By continuity at $\theta$ relative to $\compact$, $f_{\theta m}(W_i) \to \diff(\Wi, \theta) - \diff(\Wi, \theta) = 0$ as $m \to \infty$. 
Also $|f_{\theta m}(\Wi)| \lesssim \sup_{\bar \theta \in \utm} | \diff(\Wi, \bar \theta) | \leq \sup_{\theta \in \thetaspace} | \diff(\Wi, \theta) |$. 
Then by our envelope assumption $\sup_m |f_{\theta m}(W_i)| \in L_1(P)$, so $\lim_m E[\barvm(\Di, \Wi)] = \lim_m E[f_{\theta m}(W_i)] = 0$ by dominated convergence.  
For each $\theta$, let $m(\theta)$ be such that $E[f_{\theta m(\theta)}(W_i)] \leq \epsilon$. 
Then $\{U_{\theta m(\theta)}: \theta \in \compact\}$ is a relatively open cover of $\compact$, so by compactness it admits a finite subcover $\{U_{\theta_l, m(\theta_l)}\}_{l=1}^{L(\epsilon)} \equiv \{U_l\}_{l=1}^{L(\epsilon)}$. 
Next, for each $(\theta, m)$ we claim $\en[\barvm(D_i, W_i)] = E[f_{\theta m}(\Wi)] + \op(1)$. 
We have $\en[f_{\theta m}(W_i)] = E[f_{\theta m}(W_i)] + \op(1)$ by WLLN since $E[f_{\theta m}(W_i)] < \infty$ as just shown. 
Similarly, we have
\[
|r_{\theta m}(W_i)| = |(1-\propfn) \sup_{\bar \theta \in \utm} \diff(\Wi, \bar \theta) + \propfn \inf_{\bar \theta \in \utm} \diff(\Wi, \bar \theta)| \leq \sup_{\bar \theta \in \utm} |\diff(\Wi, \bar \theta)| \in L_1(P).
\]
Then $\en[\hti r_{\theta m}(W_i)] = \op(1)$ by Lemma \lemmastochasticbalance \, in \cite{cytrynbaum2024}. 
This proves the claim.
Define $f_l(W)$ and $r_l(W)$ to be the functions above evaluated at $(\theta_l, m(\theta_l))$.
For each $l$, let $T_{nl}\equiv\en[\bar v_{\theta_l m(\theta_l)}(\Di,\Wi)]-E[f_{\theta_l m(\theta_l)}(W_i)]=\op(1)$ by the preceding claim.
Putting this all together, we have 
\begin{align*}
\sup_{\theta \in \compact} \en[\hti \diff(\Wi, \theta)] &\leq \max_{l=1}^{L(\epsilon)} \sup_{\theta \in U_l} \en[\hti \diff(\Wi, \theta)] \leq \max_{l=1}^{L(\epsilon)} \en[\bar v_{\theta_l m(\theta_l)}(\Di, \Wi)] \\
&= \max_{l=1}^{L(\epsilon)} (E[f_{\theta_l m(\theta_l)}(\Wi)] + T_{nl}) \leq \epsilon + \max_{l=1}^{L(\epsilon)} T_{nl} = \epsilon + \op(1).   
\end{align*}
By symmetry, also $\sup_{\theta \in \compact} -\en[\hti \diff(\Wi, \theta)] \leq \epsilon + \op(1)$.
Then $\sup_{\theta \in \compact} |\en[\hti \diff(\Wi, \theta)]| \leq 2\epsilon + \op(1)$.  
Since $\epsilon > 0$ was arbitrary, this finishes the proof of (1). \medskip

Next we show $\|\momn - \momtrue\|_{\infty, \thetaspace} = \op(1)$.
We have $(\momn - \momtrue)(\theta) = \en[\mombar(\Wi, \theta)] - E[\mombar(W, \theta)]$.
Under our assumptions, $|\en[\mombar(\Wi, \theta)] - E[\mombar(W, \theta)]|_{\infty, \thetaspace} = \op(1)$ and $\momtrue(\theta) = E[\mombar(W, \theta)]$ is continuous by Lemma 2.4 of \cite{newey1994}.
This proves the second claim.
The statement about objective functions now follows by algebra, since $|\objest(\theta) - \objn(\theta)| \lesssim |\momest - \momn|_{\infty, \thetaspace}||\weightmatest|_2 |\momest|_{\infty, \thetaspace} +|\momn|_{\infty, \thetaspace}|\weightmatest - \weightmatpop|_2 |\momest|_{\infty, \thetaspace} + |\momn|_{\infty, \thetaspace} |\weightmatpop|_2 |\momest - \momn|_{\infty, \thetaspace} $.
We have $|\momn|_{\infty, \thetaspace}, |\momest|_{\infty, \thetaspace} = \op(1) + |\momtrue|_{\infty, \thetaspace} = \Op(1)$ since $|\momtrue|_{\infty, \thetaspace} \leq E[\sup_{\theta \in \thetaspace} |\mombar(W, \theta)|_2] < \infty$.
Also $|\weightmatest|_2 = \Op(1)$ and $|\weightmatest - \weightmatpop|_2 = \op(1)$ by continuous mapping.
Taking $\sup_{\theta \in \thetaspace}$ on both sides gives the result.
The proof that $|\objn - \objtrue|_{\infty, \compact} = \op(1)$ is identical.
By triangle inequality, this proves the claim.
The proof of (2) is similar.
\end{proof}

\begin{lem}[Consistency] \label{lemma:consistency}
Under the distribution $P$ in Definition \ref{defn:pure-stratification}, if Assumption \ref{assumption-gmm} holds then $\est - \thetatrue = \op(1)$ and $\estn - \thetatrue = \op(1)$. 
\end{lem}

\begin{proof}
By definition, $\est \in \argmin_{\theta \in \Theta} \objest(\theta)$ and $\estn \in \argmin_{\theta \in \Theta} \objn(\theta)$.
Since $\momtrue(\thetatrue) = 0$ uniquely and $\weightmatpop \succ 0$, the objective $\objtrue(\theta)$ is uniquely minimized at $\thetatrue$.
Then by uniform convergence of $\objest, \objn$ to $\objtrue$, extremum consistency (e.g.\ Theorem 2.1 in \cite{newey1994}) implies that $\thetan \convp \thetatrue$ and $\wh \theta \convp \thetatrue$.
\end{proof}

\begin{proof}[Proof of Lemma \ref{lemma:linearization}]
By Lemma \ref{lemma:dominance}, it suffices to show the result under $P$ in Definition \ref{defn:pure-stratification}.
By Lemma \ref{lemma:consistency} and since $\thetatrue$ is in the interior of $\thetaspace$, both $\est$ and $\estn$ are interior with probability approaching one.
The first-order conditions are therefore $\wh G(\est)' \weightmatest \momest(\est) = 0$ and $G_n(\estn)' \weightmatpop \momn(\estn) = 0$.
By differentiability in Assumption \ref{assumption-gmm} and applying Taylor's Theorem componentwise, for each $k \in [\dimmom]$ and some $\tilde \theta_k \in [\thetatrue, \est]$ we have $\momest(\est) = \momest(\thetatrue) + \frac{\partial \wh \mom_{k}}{\partial \theta'}(\tilde \theta_k)_{k=1}^{\dimmom}(\est - \thetatrue)$. 
Arguing exactly as in \cite{newey1994}, we find $\rootn(\est - \thetatrue) = -(G'\weightmatpop G)\inv G' \weightmatpop \rootn \momest(\thetatrue) + \op(1) = \matgmm \rootn \momest(\thetatrue) + \op(1)$.
This proves the second statement of Lemma \ref{lemma:linearization}.
For the finite population target, applying Taylor's Theorem componentwise gives $\momn(\estn)=\momn(\thetatrue)+\widetilde G_n(\estn-\thetatrue)$, where the rows of $\widetilde G_n$ are evaluated between $\estn$ and $\thetatrue$.
The first-order condition can therefore be written as
\[
0=G_n(\estn)'\weightmatpop\Big(\momn(\thetatrue)+\widetilde G_n(\estn-\thetatrue)\Big).
\]
By Lemmas \ref{lemma:ulln} and \ref{lemma:consistency}, $G_n(\estn)=G+\opone$.
The intermediate values defining the rows of $\widetilde G_n$ also converge to $\thetatrue$, so uniform convergence in Lemma \ref{lemma:ulln} and continuity of $G_0$ give $\widetilde G_n=G+\opone$.
Since $G'\weightmatpop G$ is nonsingular and $\rootn\momn(\thetatrue)=\Opone$, solving the previous display gives $\rootn(\estn - \thetatrue) = \matgmm \rootn \momn(\thetatrue) + \op(1)$.
Then we have $\rootn(\est - \estn) = \rootn(\est - \thetatrue + \thetatrue - \estn) = \matgmm \rootn (\momest(\thetatrue) - \momn(\thetatrue)) + \op(1) = \matgmm \rootn \en[\hti \diff(\Wi, \thetatrue)] + \opone$. 
This finishes the proof.
\end{proof}

\subsection{Nonlinear Rerandomization} \label{proofs:nonlinear-rerandomization}

\begin{proof}[Proof of Theorem \ref{thm:nonlinear-rerandomization}]
We first prove a slightly more general result, allowing for over-identified GMM estimation with positive definite weighting matrix $\rerandweightmat_n \convp \rerandweightmat$. 
For $|x|_{2, A}^2 = x'Ax$, define $\wh \beta_d \in \argmin_{\beta \in \mr^{\dimbeta}} |\en[\one(\Di = d) \rerandmom(X_i, \beta)]|_{2, \rerandweightmat_n}^2$.
Define $\mom^1(D, X, \beta) = D \rerandmom(X, \beta)$ and $\mom^0(D, X, \beta) = (1-D) \rerandmom(X, \beta)$.
Under the expansion in Equation \ref{eqn:orthog-decomposition}, we have $\mombar^1(X, \beta) = \propfn g^1(1, X, \beta) = \propfn m(X, \beta)$ and $\diff^1(X, \beta) = \vard g^1(1, X, \beta) = \vard \rerandmom(X, \beta)$.
Similarly, $\mombar^0(X, \beta) = (1-\propfn) g^0(0, X, \beta) = (1-\propfn) m(X, \beta)$ and $\diff^0(X, \beta) = -\vard g^0(0, X, \beta) = -\vard \rerandmom(X, \beta)$.
Note that $E[\mom^1(D, X, \beta)] = \propfn E[\rerandmom(X, \beta)]$ and $E[\mom^0(D, X, \beta)] = (1-\propfn) E[\rerandmom(X, \beta)]$, so the GMM parameters $\beta_1 = \beta_0 = \betatrue$, where $\betatrue$ uniquely solves $E[\rerandmom(X, \betatrue)] = 0$.
Let $\rerandjacob = E[(\partial / \partial \beta') m(X, \betatrue)]$, full rank by assumption.
Then $G^1 = E[(\partial / \partial \beta')\mom^1(D, X, \betatrue)] = \propfn E[(\partial / \partial \beta')m(X, \betatrue)] = \propfn \rerandjacob$ and $\matgmm^1 = -((G^1)' \rerandweightmat G^1) \inv (G^1)' \rerandweightmat = -\propfn \inv (\rerandjacob' \rerandweightmat \rerandjacob)\inv \rerandjacob' \rerandweightmat \equiv \propfn \inv \rerandmatgmm$. 
By symmetry, we have $\matgmm^0 = (1-\propfn) \inv \rerandmatgmm$.
Observe that 
\begin{align*}
(\matgmm^1 \mombar^1 - \matgmm^0 \mombar^0)(X, \beta) &= \propfn \inv \rerandmatgmm \propfn \rerandmom(X, \beta) - (1-\propfn) \inv \rerandmatgmm (1-\propfn) \rerandmom(X, \beta) = 0, \\
(\matgmm^1 \diff^1 - \matgmm^0 \diff^0)(X, \beta) &= \propfn \inv \rerandmatgmm \vard \rerandmom(X, \beta) - (1-\propfn) \inv \rerandmatgmm \vard (-\rerandmom(X, \beta)) \\
&= (1-\propfn) \rerandmatgmm \rerandmom(X, \beta) + \propfn \rerandmatgmm \rerandmom(X, \beta) = \rerandmatgmm \rerandmom(X, \beta). 
\end{align*}
Then applying Lemma \ref{lemma:linearization} to GMM estimation using $g^1$ and $g^0$, under the measure $P$ in Definition \ref{defn:pure-stratification} we have
\begin{align*}
&\rootn(\betaestone - \betaestzero) = \rootn(\betaestone - \betatrue - (\betaestzero - \betatrue)) = \rootn \matgmm^1 \en[\mombar^1(X_i, \betatrue) + \hti \diff^1(X_i, \betatrue)]  \\
&- \rootn \matgmm^0 \en[\mombar^0(X_i, \betatrue) + \hti \diff^0(X_i, \betatrue)] + \op(1) = \rootn \rerandmatgmm \en[\hti \rerandmom(X, \betatrue)] + \op(1).
\end{align*}
Thus Definition \ref{defn:gmm-rerandomization} is first-order equivalent to Definition \ref{defn:rerandomization} with covariates $h^*=\rerandmatgmm\rerandmom(X,\betatrue)$ and acceptance region $A$.
In the exactly identified case, $h^*=-\rerandjacob\inv\rerandmom(X,\betatrue)$.
Let $\zm\sim\normal(0,\vard\inv E[\var(\rerandmom_i^*|\psi)])$.
Since $\rerandjacob$ and $E[\var(\rerandmom_i^*|\psi)]$ are nonsingular, $E[\var(h^*|\psi)]\succ0$.
Moreover, $\rerandjacob A$ has non-empty interior and $\partial(\rerandjacob A)=\rerandjacob\partial A$, which has Lebesgue measure zero.
Since $\zm$ has a nonsingular Gaussian distribution, $P(\zm\in\rerandjacob A)>0$ and $P(\zm\in\partial(\rerandjacob A))=0$.
Let $\widetilde\gamma$ be the optimal coefficient on $h^*$ from Theorem \ref{thm:gmm-asymptotics} and define $\gamma=-(\rerandjacob')\inv\widetilde\gamma$.
Then $\widetilde\gamma'h^*=\gamma'\rerandmom(X,\betatrue)$, so the normal component has covariance
\begin{equation}
\vassign=\min_{\gamma \in \mr^{\dimbeta \times \dimtheta}}\vard\inv E[\var(\assignif-\gamma'\rerandmom(X,\betatrue)|\psi)].
\end{equation}
For the residual component, let $\zh$ denote the Gaussian limit of the imbalance in $h^*$ and take $\zm=-\rerandjacob\zh$, which has the distribution above.
Then $\widetilde\gamma'\zh=\gamma'\zm$.
Moreover, $\zh \in A$ if and only if $\zm \in -\rerandjacob A$.
Since $A$ is symmetric, $-\rerandjacob A=\rerandjacob A$, which gives the residual distribution in the theorem.
\end{proof}

\begin{proof}[Proof of Proposition \ref{prop:density-rerandomization}]
Apply Theorem \ref{thm:nonlinear-rerandomization} with $\rerandmom(X,\beta)=s(X,\beta)$ and $A=B(0,\epsilon)$.
The within-arm likelihood first-order conditions are Equation \ref{eqn:rerand-gmm}, so the associated linear covariates are $-J\inv s(X,\betatrue)$.
Since $B(0,\epsilon)$ is symmetric, replacing these covariates by $J\inv s(X,\betatrue)$ does not change the acceptance event.
\end{proof}

\begin{proof}[Proof of Proposition \ref{prop:propensity-rerandomization}]
By assumption, $\wh \beta$ is a GMM estimator for $m(\Di, X_i, \beta) = \Di \frac{\link'(X_i'\beta) X_i}{\link(X_i'\beta)} - (1-\Di) \frac{\link'(X_i'\beta) X_i}{1-\link(X_i'\beta)}$.
Let $c$ such that $\link(c) = \propfn$.
Then $\betatrue = (c, 0)$ has $E[m(D, X, \betatrue)] = E[\hti \link'(c) X_i] = 0$.
Relative to the decomposition in Equation \ref{eqn:orthog-decomposition}, we have 
Define $\bar m(X, \beta) = \propfn \frac{\link'(X_i'\beta) X_i}{\link(X_i'\beta)} - (1-\propfn) \frac{\link'(X_i'\beta) X_i}{1-\link(X_i'\beta)}$ and $\diff(X, \beta) = \vard (\frac{\link'(X_i'\beta) X_i}{\link(X_i'\beta)} + \frac{\link'(X_i'\beta) X_i}{1-\link(X_i'\beta)})$.
Since $\link(X_i'\betatrue) = \link(c) = \propfn$, we have $\bar m(X, \betatrue) = 0$ and $\diff(X, \betatrue) = L'(c) X_i$.
The condition $\var(h) \succ 0$ implies $E[XX'] \succ 0$ for $X = (1, h)$.
A calculation shows that $\rerandjacob = E[\frac{\partial}{\partial \beta'} \bar m(X, \betatrue)] = - \vard \inv \link'(c)^2 E[X_i X_i']$, so $\rerandmatgmm = -\rerandjacob \inv = \frac{\vard}{\link'(c)^2} E[X_i X_i'] \inv$.
By Lemma \ref{lemma:linearization}, we have shown 
\begin{align*}
\rootn(\betaest - \betatrue) &= \rootn \rerandmatgmm \en[\bar m(X_i, \betatrue) + \hti \diff(X_i, \betatrue)] + \op(1) \\
&= \vard \frac{\rootn}{\link'(c)} E[X_i X_i'] \inv \en[\hti X_i] + \op(1).
\end{align*}
Consider rerandomizing until $\imbalancesquare = n \en[(\propfn - \link(X_i'\betaest))^2] \leq \epsilon^2$. 
Then for $\betatrue$ s.t.\ $L(x'\betatrue) = \propfn$, the above quantity is $n \en[(\link(X_i'\betaest) - \link(X_i'\betatrue))^2]$.
By Taylor's Theorem, $\link(X_i'\betaest) - \link(X_i'\betatrue) = \link'(\xii)(X_i'\betaest - X_i'\betatrue) = \link'(\xii)X_i'(\betaest - \betatrue)$ for some $\xii \in [X_i'\betatrue, X_i'\betaest]$. 
Then we have 
\[
\imbalancesquare = n (\betaest - \betatrue)'\en[X_i X_i' \link'(\xii)^2](\betaest - \betatrue).
\]
Claim that $\en[X_i X_i' \link'(\xii)^2] = \en[X_i X_i' \link'(X_i'\betatrue)^2] + \op(1)$. 
If so, then $\en[X_i X_i' \link'(\xii)^2] = \link'(c)^2 \en[X_i X_i'] +\opone = \link'(c)^2 E[X_i X_i'] +\opone$. 
To see this, note that $|\link'(X_i'\betatrue)^2 - \link'(\xii)^2| = |\link'(X_i'\betatrue) - \link'(\xii)||\link'(X_i'\betatrue) + \link'(\xii)| \leq 2 |\link'|_{\infty} |\link''|_{\infty} |X_i'\betatrue -\xii|_2 \lesssim |X_i'\betatrue -X_i'\betaest|_2 \leq |X_i|_2|\betatrue -\betaest|_2$. 
Then we have 
\begin{align*}
&|\en[X_i X_i' \link'(\xii)^2] - \en[X_i X_i' \link'(X_i'\betatrue)^2]|_2 \leq \en[|X_i|_2^2 |\link'(X_i'\betatrue)^2 - \link'(\xii)^2|] \\
&\lesssim \en[|X_i|_2^3] |\betatrue -\betaest|_2 = \op(1).
\end{align*}
The last equality if $\en[|X_i|_2^3] = \op(n^{1/2})$.
Note that $\en[|X_i|_2^3] \leq \en[|X_i|_2^2] \max_{i=1}^n |X_i|_2 = \Op(1) \op(n^{1/2})$ since $E[|X_i|_2^2] < \infty$ by assumption, using Lemma \lemmamaximal of \cite{cytrynbaum2024}.
Using the claim, $\rootn(\betaest - \betatrue) = \Opone$, and the linear expansion of $\rootn(\betaest - \betatrue)$, $\imbalancesquare = \link'(c)^2 n (\betaest - \betatrue)'E[X_i X_i'](\betaest - \betatrue) + \opone$, which is
\begin{align*}
&= \vard^2 \link'(c)^2 (\link'(c) \inv E[X_i X_i'] \inv \rootn \en[\hti X_i])' E[X_i X_i'] \\
& \times (\link'(c) \inv E[X_i X_i'] \inv \rootn \en[\hti X_i]) + \opone \\
&= \vard^2 \rootn \en[\hti X_i]' E[X_i X_i'] \inv \rootn \en[\hti X_i] + \opone. 
\end{align*}
Note $\en[\hti] = \Op(n \inv)$ by stratification.
Since $X = (1, h)$, $\rootn \en[\hti X_i]' = (0, \rootn \en[\hti \hi]') + \Op(\negrootn)$.
Also, by block inversion $(E[X_i X_i'] \inv)_{hh} = \var(\hi) \inv$.
For some $\xi_n = \opone$ 
\begin{align*}
\imbalancesquare &= \vard^2 (0, \rootn \en[\hti \hi]')E[X_i X_i'] \inv (0, \rootn \en[\hti \hi]')' + \opone \\     
&=\vard^2 \rootn \en[\hti \hi]'(E[X_i X_i'] \inv)_{hh}\rootn \en[\hti \hi] + \opone \\
&=\vard^2 \rootn \en[\hti \hi]'\var(\hi) \inv\rootn \en[\hti \hi] + \xi_n. 
\end{align*}

Define the function $\setfn(x, y) = \vard^2 x'\var(h)\inv x + y - \epsilon^2$.
Then $\imbalancesquare \leq \epsilon^2 \iff \setfn(\imbalance, \xi_n) \leq 0$ for $\imbalance = \rootn \en[\hti \hi]$ and $\xi_n \convp 0$.
Clearly, $\setfn$ is jointly continuous. 
Also note $E[|h|_2^2] < \infty$ by assumption.
Finally, for $\zh \sim \normal(0, \vard\inv E[\var(h | \psi)])$, we have $P(\setfn(\zh, 0) = 0) = P(\zh'\var(h)\inv \zh = \epsilon^2\vard^{-2}) = 0$ since $E[\var(h | \psi)]$ is full rank.
Then this rerandomization satisfies all the conditions in Definition \ref{defn:rerandomization-general}. 
By Lemma \ref{lemma:linearization}, the GMM estimator satisfies $\rootn(\est - \estn) = \rootn\en[\hti\assignif(W_i)] + \opone$ under this rerandomization.
By Theorem \ref{thm:rerand-asymptotics}, $\rootn\en[\hti\assignif(W_i)] | \filtrationcand \convwprocess \normal(0, \vassign) + R$ with residual variable
\[
R \sim \gammaoptmat' \zh | \zh \in T \sim \gammaoptmat' \zh | \vard^2 \zh' \var(h)\inv \zh \leq \epsilon^2.
\]
The acceptance region is $T = \{x: \setfn(x, 0) \leq 0\} = \{x: x'\var(h)\inv x \leq \epsilon^2\vard^{-2}\}$, and
\[
\vassign = \min_{\gammacoeff \in \mr^{\dimh \times \dimtheta}} \vard\inv E[\var(\assignif - \gammacoeff'h | \psi)].
\]
This gives the claimed elliptical acceptance region and proves the first-order equivalence.
\end{proof}

\subsection{Covariate Adjustment} \label{proofs:covariate-adjustment}

For the proofs in this subsection, let $w$ be a general adjustment vector containing $h$.
The main-text statements specialize these results to $w=h$.
For $\hti=(\Di-\propfn)/(\propfn-\propfn^2)$, define the generalized adjusted estimator $\estadj=\est-\en[\hti\alphaest'\wi]$.
In this subsection, let $\alphaopt\in\argmin_{\alpha\in\mr^{\dimw\times\dimtheta}}E[\var(\assignif-\alpha'w | \psi)]$.
Define $\wicheck = \wi - |\group(i)|\inv\sum_{j\in\group(i)} w_j$.

For completeness, we first record the adjustment benchmark under stratification without rerandomization.

\begin{prop}[Linear Adjustment] \label{prop:pure-stratification-adjustment}
Suppose $\Dn$ is as in Definition \ref{defn:rerandomization} with $A=\mr^{\dimh}$, and suppose $\alphaest\convp\alpha\in\mr^{\dimh\times\dimtheta}$ and $E[|h|_2^2]<\infty$.
Require Assumption \ref{assumption-gmm}.
Then $\rootn(\estadj-\estn) | \Wn \convwprocess \normal(0,\vassign(\alpha))$ and $\rootn(\estadj-\thetatrue) \convwprocess \normal(0,\vsamp+\vassign(\alpha))$.
Here, $\vassign(\alpha)=\vard\inv E[\var(\assignif-\alpha'h | \psi)]$.
\end{prop}

A version of Proposition \ref{prop:pure-stratification-adjustment} was given in \cite{cytrynbaum2024adjustment} for the special case $\thetatrue=\ate$.

\begin{proof}[Proof of Proposition \ref{prop:pure-stratification-adjustment}]
Set $w=h$.
Since $\alphaest\convp\alpha$ and $\en[\hti\hi]=\Op(\negrootn)$ by Theorem \ref{thm:clt}, we have $\estadj-\estn=(\est-\estn)-\en[\hti\alpha'\hi]+\op(\negrootn)=\en[\Hi(\matgmm\diff(W_i,\thetatrue)-\alpha'\hi)]+\op(\negrootn)$, where the final equality follows from Lemma \ref{lemma:linearization}.
The first statement now follows from Slutsky and Theorem \ref{thm:clt}.  
The second statement follows by the same argument used in the proof of Corollary \ref{cor:gmm-superpopulation}.
\end{proof}

\begin{proof}[Proof of Theorem \ref{thm:gmm-adjustment}]
Set $w=h$, $\alphaest=\gammaest$, and $\alphaopt=\gammaoptmat$.
By the same argument in the proof of Proposition \ref{prop:pure-stratification-adjustment}, we have $\estadj-\estn = \en[\Hi (\matgmm \diff(W_i, \thetatrue) - \gammaoptmat'h_i)] + \op(\negrootn)$.
Then by Theorem \ref{thm:rerand-asymptotics}, $\rootn(\estadj - \thetan) | \filtrationcand \convwprocess \normal(0, V) + R$, independent with 
\[
V = \vard \inv E[\var(\matgmm \diff(W) - \gammaoptmat'h - \beta_0'h | \psi)] = \min_{\beta \in \mr^{\dimh \times \dimtheta}} \vard \inv E[\var(\matgmm \diff(W) - \gammaoptmat'h - \beta'h | \psi)].
\]
The residual term $R \sim \beta_0'\zh | \zh \in A$.
Then it suffices to show that $\beta_0 = 0$.
By the definition of $\gammaoptmat$, $E[\cov(h, \matgmm\diff-\gammaoptmat'h | \psi)]=0$.
Lemma \ref{lemma:orthogonalization} therefore gives $\beta_0=0$.
Thus $V=\vard\inv E[\var(\assignif-\gammaoptmat'h | \psi)]=\vassign=\vadj$, proving the statement for $\thetan$.
The result for $\thetatrue$ follows as in Corollary \ref{cor:gmm-superpopulation}.
\end{proof}

For the case of general $w$, define $b_d=E[\var(w | \psi)]\inv E[\cov(w,\vard\matgmm\mom_d(W,\thetatrue) | \psi)]$ and $\adjcoeffest_d=\vard\en[\wicheck\wicheck']\inv\cov_n(\wicheck,\matgmmest\momesti | \Di=d)$.
By definition, $\alphaopt=\adjcoeffone-\adjcoeffzero$ and $\alphaest=\adjcoeffoneest-\adjcoeffzeroest$.
Thus, consistency of the adjustment coefficient follows from consistency of the two within-arm coefficients.

\begin{thm}[Adjustment Coefficients] \label{thm:adjustment-general}
Suppose $\Dn$ is as in Definition \ref{defn:rerandomization}.
Require Assumptions \ref{assumption:linear-rerandomization}, \ref{assumption-gmm}, and \ref{assumption-inference}.
Assume $E[|w|_2^2]<\infty$ and $E[\var(w | \psi)] \succ 0$.
Then $\wh b_d=b_d+\opone$ for $d\in\{0,1\}$.
\end{thm}

\begin{proof}[Proof of Theorem \ref{thm:adjustment-general}]
By Lemma \ref{lemma:dominance}, it suffices to show the result under $P$ in Definition \ref{defn:pure-stratification}.
First consider estimating $\adjcoeffone$.
By Lemma \ref{lemma:conditional-variance-check}, $\en[\wicheck \wicheck'] = k\inv (k-1) E[\var(w | \psi)] + \op(1)$.
Then if $E[\var(w | \psi)] \succ 0$, $\en[\wicheck \wicheck']\inv \convp k(k-1)\inv E[\var(w | \psi)] \inv$ by continuous mapping. 
$\matgmmest \convp \matgmm$ by assumption.
Then it suffices to show
\begin{align*}
\cov_n(\wicheck, \momesti | \Di=1) &= \en[(\Di/p) \wicheck \momesti'] - \en[\wicheck | \Di=1] \en[\momesti | \Di=1] \\
&= \frac{k-1}{k} E[\cov(w, \momone(\thetatrue) | \psi)] + \opone.
\end{align*}
That $\en[\wicheck | \Di=1] = \opone$ can be shown similar to Lemma \ref{lemma:conditional-variance-check} below.
Then consider the first term.
First, claim that $\en[(\Di / \propfn) \wicheck \momesti'] = \en[(\Di / \propfn) \wicheck \momi'] + \op(1)$. 
By Taylor's theorem, $|\momi(\est) - \momi(\thetatrue)|_2 \leq |\frac{\partial \momi}{\partial \theta'}(\tilde \theta_i)|_2 |\est - \thetatrue|_2$, where $\tilde \theta_i \in [\thetatrue, \est]$ may change by row.
Then using $|xy'|_2 \leq |x|_2 |y|_2$, we have $|\en[(\Di/\propfn) \wicheck (\momi(\est) - \momi(\thetatrue))']|_2 \leq \en[|\wicheck|_2 |\momi(\est) - \momi(\thetatrue)|_2] \leq |\est - \thetatrue|_2 \en[|\wicheck|_2 |\frac{\partial \momi}{\partial \theta'}(\tilde \theta_i)|_2 ] \leq |\est - \thetatrue|_2 (\en[|\wicheck|_2^2] + \en[|\frac{\partial \momi}{\partial \theta'}(\tilde \theta_i)|_2^2])$ by Young's inequality. 
Choose an open ball $U_0$ containing $\thetatrue$ with $U_0\sub U$.
Since $U_0$ is convex and $\est\convp\thetatrue$, every $\tilde\theta_i$ lies in $U$ with probability approaching one.
Assumption \ref{assumption-inference} and the WLLN therefore give $\en[|\frac{\partial \momi}{\partial \theta'}(\tilde \theta_i)|_2^2] = \Op(1)$.
Similarly, $\en[|\wicheck|_2^2] \leq \en[|\wi|_2^2] = \Op(1)$ by the bound in Lemma \ref{lemma:conditional-variance-check}.
Since $|\est - \thetatrue|_2 = \op(1)$ by Theorem \ref{thm:gmm-asymptotics}, this proves the claim. 
Next, we calculate 
\begin{align*}
\en[(\Di / \propfn) \wicheck \momi'] &= \en[(\Di / \propfn) \wicheck \momonei'] = \propfn \inv \en[(\Di - \propfn) \wicheck \momonei'] + \en[\wicheck \momonei'] \\
&= \en[\wicheck \momonei'] + \opone = k \inv (k-1) E[\cov(w, \momonei | \psi)] + \opone.
\end{align*}
The first equality since $\momonei = \mom(1, R_i, S_i(1), \thetatrue)$.
The third and fourth equalities by Lemma \ref{lemma:conditional-variance-check}, since $E[|w|_2^2 + |\mom_1|_2^2] < \infty$
Then we have shown $\adjcoeffoneest \convp \adjcoeffone$, and $\adjcoeffzeroest \convp \adjcoeffzero$ by symmetry. 
\end{proof}

\begin{thm}[Feasible Adjustment] \label{thm:adjustment-coefficient}
Suppose $\Dn$ is as in Definition \ref{defn:rerandomization}.
Require Assumptions \ref{assumption:linear-rerandomization}, \ref{assumption-gmm}, and \ref{assumption-inference}.
Assume that $E[|h|_2^2]<\infty$ and $E[\var(h | \psi)]\succ0$.
Then the estimator in Equation \ref{eqn:feasible-adjustment} satisfies $\gammaest=\gammaoptmat+\opone$.
\end{thm}

\begin{proof}
Set $w=h$.
Theorem \ref{thm:adjustment-general} gives $\adjcoeffest_d=\adjcoeff_d+\opone$ for $d=0,1$.
For $w=h$, we have $\alphaopt=\gammaoptmat$.
The result follows from $\gammaoptmat=\adjcoeffone-\adjcoeffzero$ and $\gammaest=\adjcoeffoneest-\adjcoeffzeroest$.
\end{proof}

\begin{cor}[One-step Adjustment] \label{cor:one-step-adjustment}
Suppose $\Dn$ is as in Definition \ref{defn:rerandomization}.
Require Assumptions \ref{assumption:linear-rerandomization} and \ref{assumption-gmm}.
Assume that $E[|h|_2^2]<\infty$ and $E[\var(h | \psi)]\succ0$.
Suppose $\assignif(W,\theta)=\assignif^{(1)}(\psi,\theta)+\assignif^{(2)}(W)$.
For any $\theta\in\Theta$ with $E[|\mom_d(W,\theta)|_2^2]<\infty$ for $d\in\{0,1\}$, replacing $\momesti$ in Equation \ref{eqn:feasible-adjustment} by $\momi=\mom(\Di,R_i,S_i,\theta)$ gives $\gammaest=\gammaoptmat+\opone$.
\end{cor}

\begin{proof}
For fixed $\theta$ satisfying the stated moment condition, the proof of Theorem \ref{thm:adjustment-general}, omitting the plug-in step, gives $\gammaest\convp E[\var(h | \psi)]\inv E[\cov(h,\assignif(W,\theta) | \psi)]$.
The conditional covariance of $h$ with $\assignif^{(1)}(\psi,\theta)$ is zero, so the decomposition implies that this limit is the same at every $\theta$, and at $\thetatrue$ it equals $\gammaoptmat$.
\end{proof}

\begin{lem} \label{lemma:conditional-variance-check}
Let $E[w_i^2 + v_i^2] < \infty$ with $\wi, \vi \in \sigma(\Wi)$.
Then under $P$ in Definition \ref{defn:pure-stratification}, $\en[(\Di - \propfn) \wicheck \vicheck] = \op(1)$ and $\en[(\Di - \propfn) \wicheck \vi] = \op(1)$.
Also $\en[\wicheck \vicheck] = \frac{k-1}{k} E[\cov(w, v | \psi)] + \op(1)$.
\end{lem}

We omit the proof, since this is a slight restatement of Lemma A.8 in \cite{cytrynbaum2024adjustment}.

\subsection{Acceptance Region Optimization} \label{proofs:acceptance-region-optimization}
\begin{proof}[Proof of Proposition \ref{prop:minimax-acceptance-region}]
Define the function $f(a) = \sup_{b \in \balancecoeffs} |b'a|$.
As the supremum of convex functions, $f$ is convex (e.g.\ \cite{rockafellar1996}).
Thus $\balancecoeffspolar=\{a:f(a)\leq1\}$ is convex.
Since $f(a)=f(-a)$, the set $\balancecoeffspolar$ is symmetric.
For $a_n=\rootn\en[\hti\hi]$, positive homogeneity gives $f(a_n)\leq\epsilon$ if and only if $a_n\in\epsilon\balancecoeffspolar=A$.

Suppose $\balancecoeffs$ is bounded.
Let $M=\sup_{b\in\balancecoeffs}|b|_2<\infty$.
By Cauchy--Schwarz, $f(a)\leq M|a|_2$, so $f$ is finite and continuous by Corollary 10.1.1 of \cite{rockafellar1996}.
It follows that $\balancecoeffspolar$ is closed.
If $M=0$, then $\balancecoeffspolar=\mrh$.
Otherwise, $B(0,M\inv)\sub\balancecoeffspolar$, so $\balancecoeffspolar$ has non-empty interior.
The boundary of a convex set with non-empty interior has Lebesgue measure zero.
These properties are preserved by scaling by $\epsilon>0$, proving the claims for $A$.

\end{proof}

\begin{proof}[Proof of Theorem \ref{thm:minimax}]
For any $C$, the definition of $L_{\gamma C}$ and independence give $\bias(L_{\gamma C} | Z_{hC})=E[L_{\gamma C} | Z_{hC}]=\gamma'Z_{hC}$.
If $C=A$, then $Z_{hC}\in\epsilon\balancecoeffspolar$ almost surely.
By the definition of the absolute polar, the conditional-bias identity above implies $\sup_{\gamma\in\balancecoeffs}|\bias(L_{\gamma C} | Z_{hC})|=\sup_{\gamma\in\balancecoeffs}|\gamma'Z_{hC}|\leq\epsilon$.
Thus $A$ is feasible.

Now let $C$ be any feasible region.
For every $x\in C\setminus A$, the definition of $A$ gives $\sup_{\gamma\in\balancecoeffs}|\gamma'x|>\epsilon$.
Feasibility and the conditional-bias identity therefore imply $P(\zh\in C\setminus A)=0$.
It follows that $P(\zh\in C)\leq P(\zh\in A)$, proving optimality.
If equality holds, then $P(\zh\in A\setminus C)=0$ as well.
Since $\zh$ has an everywhere positive Gaussian density, equality is possible only if $\leb(C\triangle A)=0$.

Taking $C=A$ and $\gammatrue\in\balancecoeffs$, the conditional-bias identity gives $|\bias(\asympdisttrue | \zhs)|\leq\epsilon$.
The region $A$ is symmetric, so $E[\gammatrue'\zhs]=0$.
Independence and $|\gammatrue'\zhs|\leq\epsilon$ then give $\var(\asympdisttrue)=\vassign+\var(\gammatrue'\zhs)\leq\vassign+\epsilon^2$.
\end{proof}

\begin{proof}[Proof of Lemma \ref{lemma:polar-sets}]
Write $x=\bar\gamma$ and $\Sigma=U$.
For $\balancecoeffs = x + \Sigma \ballp(0,1)$ we compute
\begin{align*}
\sup_{b \in \balancecoeffs} |a'b| &= \sup_{u \in \Sigma \ballp(0,1)} |a'x + a'u| \leq |a'x| + \sup_{u \in \Sigma \ballp(0,1)} |a' \Sigma \Sigma \inv u| \\
&= |a'x| + \sup_{v \in \ballp(0,1)} |(\Sigma' a)' v| = |a'x| + |\Sigma' a|_q.
\end{align*}
Before proceeding, we claim that for any $z \in \mr^{\dimh}$, we have $\max_{v \in \ballp(0,1)} v'z = \max_{v \in \ballp(0,1)} |v'z|$.
Clearly $\max_{v \in \ballp(0,1)} v'z \leq \max_{v \in \ballp(0,1)} |v'z|$.
Since $\ballp(0,1)$ is compact and $v \to v'z$ continuous, $v^* \in \argmax_{v \in \ballp(0,1)} |v'z|$ exists.
Then $\max_{v \in \ballp(0,1)} |v'z| = |z'v^*| = z'v^* \sgn(z'v^*) = z'w$ for $w = v^* \sgn(z'v^*) \in \ballp(0,1)$ since $v^* \in \ballp(0,1)$.
Then $\max_{v \in \ballp(0,1)} |v'z| = z'w \leq \max_{w \in \ballp(0,1)} z'w$.
This proves the claim.
Choose $s(a)\in\{-1,1\}$ such that $s(a)a'x=|a'x|$.
Define $b(a) = x + s(a)\Sigma v(a)$ with $v(a) \in \argmax_{v \in \ballp(0,1)} v' \Sigma' a$, which exists by compactness and continuity.
Note $b(a) \in \balancecoeffs$ by construction.
We may calculate $|a'b(a)| = |a'x + s(a)a'\Sigma v(a)|$.
By the claim, $a'\Sigma v(a) \geq 0$. 
Then by matching signs, $|a'x + s(a)a'\Sigma v(a)| = |a'x| + |a'\Sigma v(a)|$.
By the claim again, this is $|a'x| + a'\Sigma v(a) = |a'x| + \max_{v \in \ballp(0,1)} |a'\Sigma v| = |a'x| + |\Sigma' a|_q$.
Combining with the upper bound above, we have shown that $\sup_{b \in \balancecoeffs} |a'b| = |a'x| + |\Sigma' a|_q$. 
\end{proof}

\subsection{Inference} \label{proofs:inference}

Throughout this subsection, use the arm-specific adjusted influences $\assignifarmh{d}$ and their observed estimate $\assignifarmhesti$ defined in Section \ref{section:inference}.
For each $i$, define the oracle observed influence $\assignifarmhoraclei=\vard\matgmm\momi(\thetatrue)-\gammaoptarm{\Di}'h_i$.
Suppose $E[\var(h | \psi)]\succ0$ throughout this subsection.

\begin{proof}[Proof of Theorem \ref{thm:variance-bounds}]
By two applications of Cauchy--Schwarz,
\begin{align*}
|E[\cov(c'\assignifarmh{1}, c'\assignifarmh{0} | \psi)]| &\leq E[\var(c'\assignifarmh{1} | \psi)\half \var(c'\assignifarmh{0} | \psi) \half ] \\
&\leq E[\var(c'\assignifarmh{1} | \psi)]\half E[\var(c'\assignifarmh{0} | \psi)]\half.
\end{align*}
Substituting this bound into Equation \ref{eqn:contrast-adjusted-variance} gives $\vadj(c)\leq\vard\inv(\sigma_1(c)+\sigma_0(c))^2$.
\end{proof}

\begin{proof}[Proof of Theorem \ref{thm:conservative-inference}]
By Lemma \ref{lemma:dominance}, it suffices to show the result under $P$ in Definition \ref{defn:pure-stratification}.
For part (a), Lemmas \ref{lemma:inference-variance-convergence} and \ref{lemma:inference-matching} give
\begin{align*}
\widehat\sigma_1^2(c)&=c'\left(\en\left[\frac{\Di}{\propfn}\assignifarmhesti(\assignifarmhesti)'\right]-\varestone\right)c=E[\var(c'\assignifarmh{1} | \psi)]+\opone=\sigma_1^2(c)+\opone, \\
\widehat\sigma_0^2(c)&=c'\left(\en\left[\frac{1-\Di}{1-\propfn}\assignifarmhesti(\assignifarmhesti)'\right]-\varestzero\right)c=E[\var(c'\assignifarmh{0} | \psi)]+\opone=\sigma_0^2(c)+\opone.
\end{align*}
Thus, $\widehat{\bar V}_a(c)=\vadjbound(c)+\opone$ by continuous mapping.
Theorem \ref{thm:gmm-adjustment}, Theorem \ref{thm:variance-bounds}, and Slutsky's theorem give the coverage statement in part (a).

For part (b), define $\scoreadji(\thetatrue)=\matgmm\momi(\thetatrue)-H_i\gammaoptmat'\hi=\sampif+H_i(\assignif-\gammaoptmat'h)$.
Therefore,
\begin{align*}
\var(\scoreadji) &= \vsamp + \vard\inv E[(\assignif-\gammaoptmat'h)(\assignif-\gammaoptmat'h)'] \\
&= \vsamp + \vadj + \vard\inv E[E[\assignif-\gammaoptmat'h | \psi] E[\assignif-\gammaoptmat'h | \psi]'].
\end{align*}
This shows that $\vsamp + \vadj = \var(\scoreadji) - \vard\inv E[E[\assignif-\gammaoptmat'h | \psi] E[\assignif-\gammaoptmat'h | \psi]']$.
By definition, $\gammaoptmat=\gammaoptarm{1}-\gammaoptarm{0}$ and $\assignif=\assignifarm{1}-\assignifarm{0}$, so $\assignif-\gammaoptmat'h=\assignifarmh{1}-\assignifarmh{0}$.
Lemmas \ref{lemma:inference-variance-convergence} and \ref{lemma:inference-matching} now give
\begin{align*}
\vsamp + \vadj &= \var(\scoreadji) - \vard\inv E[E[\assignifarmh{1} - \assignifarmh{0} | \psi] E[\assignifarmh{1} - \assignifarmh{0} | \psi]'] \\
&= \var_n(\scoreadjesti) - \vard\inv(\varestone+\varestzero-\varestcross-\varestcross')+\opone=\widehat V+\opone.
\end{align*}
The coverage statement in (b) follows from Theorem \ref{thm:gmm-adjustment} and Slutsky's theorem.
\end{proof}

\begin{lem} \label{lemma:inference-variance-convergence}
Impose Assumptions \ref{assumption:linear-rerandomization}, \ref{assumption-gmm}, \ref{assumption-inference}.
Then under $P$ in Definition \ref{defn:pure-stratification}, $\en[\frac{\Di}{\propfn} \assignifarmhesti(\assignifarmhesti)'] = E[\assignifarmh{1}(\assignifarmh{1})'] + \op(1)$ and $\en[\frac{1-\Di}{1-\propfn} \assignifarmhesti(\assignifarmhesti)'] = E[\assignifarmh{0}(\assignifarmh{0})'] + \op(1)$.
Also, we have $\var_n(\scoreadjesti) = \var(\scoreadji) + \opone$.
\end{lem} 

\begin{proof}
Consider the first statement.
We have $\Di\assignifarmhesti=\vard\matgmmest\Di\momesti-\Di\gammaestarm{1}'h_i$ and $\Di\assignifarmhoraclei=\Di\assignifarmh{1}(W_i)$.
We can expand $\en[(\Di / p) \assignifarmhesti(\assignifarmhesti)']$ as 
\[
\en[(\Di / p)\assignifarmhesti(\assignifarmhesti - \assignifarmhoraclei)'] + \en[(\Di / p)(\assignifarmhesti - \assignifarmhoraclei)(\assignifarmhoraclei)'] + \en[(\Di / p) \assignifarmhoraclei (\assignifarmhoraclei)'].
\]
For the first term, we have $\en[(\Di / p)\assignifarmhesti(\assignifarmhesti - \assignifarmhoraclei)'] = \propfn \inv \en[\Di \assignifarmhesti(\Di \assignifarmhesti - \Di \assignifarmhoraclei)']$. 
\begin{align*}
|\Di \assignifarmhesti - \Di \assignifarmhoraclei|_2 &= |\Di \vard \matgmmest \momesti - \Di \vard \matgmm \momi - \Di (\gammaestarm{1} - \gammaoptarm{1})' h_i|_2 \\
&\lesssim |\matgmmest - \matgmm|_2 |\momesti|_2 + |\matgmm|_2 |\momesti - \momi|_2 + |\gammaestarm{1} - \gammaoptarm{1}|_2 |h_i|_2.
\end{align*}
Then using $|xy'|_2 \leq |x|_2 |y|_2$ and triangle inequality, the first term above has 
\begin{align*}
|\en[\Di \assignifarmhesti(\Di \assignifarmhesti - \Di \assignifarmhoraclei)']| &\leq  |\matgmmest - \matgmm|_2 \en[|\Di \assignifarmhesti|_2 |\momesti|_2] + |\matgmm|_2 \en[|\Di \assignifarmhesti|_2 |\momesti - \momi|_2] \\
&+ |\gammaestarm{1} - \gammaoptarm{1}|_2 \en[|\Di \assignifarmhesti|_2 |h_i|_2].
\end{align*}
We claim this term is $\op(1)$. 
Note that $|\matgmmest - \matgmm|_2 = \op(1)$ by assumption and $|\gammaestarm{1} - \gammaoptarm{1}|_2 = \op(1)$ by Theorem \ref{thm:adjustment-general}.  
Then applying Cauchy-Schwarz, it suffices to show $\en[|\Di \assignifarmhesti|_2^2 + |\momesti|_2^2 + |h_i|_2^2] = \Op(1)$ and $\en[ |\momesti - \momi|_2^2] = \op(1)$. 
First, note $\en[|h_i|_2^2] = \Op(1)$ since $E[|h|_2^2] < \infty$.
Next, note $\en[|\Di \assignifarmhesti|_2^2] = \en[|\vard \Di \matgmmest \momesti - \Di \gammaestarm{1}' h_i|_2^2] \leq 2 \en[|\matgmmest \momesti|_2^2] + 2 \en[|\gammaestarm{1}' h_i|_2^2] \leq 2 |\matgmmest|_2^2 \en[|\momesti|_2^2] + 2 |\gammaestarm{1}|_2^2 \en[|h_i|_2^2]$, so suffices to show $\en[|\momesti|_2^2] = \Op(1)$. 

We start by showing that $\en[ |\momesti - \momi|_2^2] = \op(1)$.
By the mean value theorem $\momi(\est) - \momi(\thetatrue) = \frac{\partial \momi}{\partial \theta'}(\tilde \theta_i)(\est - \thetatrue)$, where $\tilde \theta_i \in [\thetatrue, \est]$ may change by row. 
Then we have $\en[|\momi(\est) - \momi(\thetatrue)|_2^2] \leq |\est - \thetatrue|_2^2 \en[|\frac{\partial \momi}{\partial \theta'}(\tilde \theta_i)|_2^2]$, so it suffices to show $\en[|\frac{\partial \momi}{\partial \theta'}(\tilde \theta_i)|_2^2] = \Op(1)$. 
Since $\momi(\theta) = \Di \momonei(\theta) + (1-\Di) \momzeroi(\theta)$ for all $\theta$, $|\frac{\partial \momi}{\partial \theta'}(\tilde \theta_i)|_2^2 \leq 2|\frac{\partial \momonei}{\partial \theta'}(\tilde \theta_i)|_2^2 + 2|\frac{\partial \momzeroi}{\partial \theta'}(\tilde \theta_i)|_2^2$.  
Choose an open ball $U_0$ containing $\thetatrue$ with $U_0\sub U$, and define the event $S_n = \{\est \in U_0\}$.
Since $U_0$ is convex, every $\tilde\theta_i$ lies in $U$ on $S_n$.
Then on $S_n$ we have 
\begin{align*}
&|\frac{\partial \momonei}{\partial \theta'}(\tilde \theta_i)|_2^2 + |\frac{\partial \momzeroi}{\partial \theta'}(\tilde \theta_i)|_2^2 \leq |\frac{\partial \momonei}{\partial \theta'}(\tilde \theta_i)|_F^2 + |\frac{\partial \momzeroi}{\partial \theta'}(\tilde \theta_i)|_F^2 = \sum_{d=0,1} \sum_{k=1}^{\dimmom} |\nabla \mom_{di}^k(\tilde \theta_{ik})|^2_2 \\
&\leq \sum_{d=0,1} \sum_{k=1}^{\dimmom} \sup_{\theta \in U} |\nabla \mom_{di}^k(\theta)|^2_2 \equiv \bar U_i.
\end{align*}
Then $\en[|\frac{\partial \momi}{\partial \theta'}(\tilde \theta_i)|_2^2] \one(S_n) \leq \en[\bar U_i] \one(S_n) = \Op(1)$ since $E[\sup_{\theta \in U} |\nabla \mom_{di}^k(\theta)|^2_2] < \infty$ by assumption. 
Then $\en[|\frac{\partial \momi}{\partial \theta'}(\tilde \theta_i)|_2^2] = \Op(1)$ since $P(S_n^c) \to 0$. 
This finishes the proof of $\en[ |\momesti - \momi|_2^2] = \op(1)$. 
Finally, the claim $\en[|\momesti|_2^2] = \Op(1)$ is clear since $\en[|\momesti|_2^2] \leq 2 \en[ |\momesti - \momi|_2^2] + 2\en[|\momi|_2^2] = \op(1) + \Op(1)$ by the preceding claim. 

Then we have shown $|\en[(\Di / p)\assignifarmhesti(\assignifarmhesti - \assignifarmhoraclei)']| = \op(1)$ and $\en[(\Di / p)(\assignifarmhesti - \assignifarmhoraclei)(\assignifarmhoraclei)'] = \op(1)$ by an identical argument.
This shows that $\en[(\Di / p) \assignifarmhesti(\assignifarmhesti)'] = \en[(\Di / p) \assignifarmhoraclei(\assignifarmhoraclei)'] + \op(1)$.
Next, we claim
\begin{align*}
\en[(\Di / p) \assignifarmhoraclei(\assignifarmhoraclei)'] &= \en[(\Di / p) \assignifarmh{1}(W_i)(\assignifarmh{1}(W_i))'] \\
&= \en[\assignifarmh{1}(W_i)(\assignifarmh{1}(W_i))'] + \op(1) = E[\assignifarmh{1}(\assignifarmh{1})'] + \op(1).
\end{align*}
The first equality is by definition of $\assignifarmhoraclei$ and $\assignifarmh{1}$. 
The second equality is by Lemma \lemmastochasticbalance{} of \cite{cytrynbaum2024} and the third is by the WLLN, both using $E[|\assignifarmhoraclei|_2^2] < \infty$.
This proves the first statement, and the second follows by symmetry. 

Next consider the final statement. 
Note that $\scoreadjesti = \matgmmest \momesti - H_i \gammaest'h_i$ and $\scoreadji =\matgmm \momi(\thetatrue) - H_i \gammaoptmat'h_i$.
Define $\scoreadjonei = \matgmm \momonei - (1/\propfn)\gammaoptmat'h_i$ and $\scoreadjzeroi = \matgmm \momzeroi + (1/(1-\propfn))\gammaoptmat'h_i$, so $\Di\scoreadji=\Di\scoreadjonei$ and $(1-\Di)\scoreadji=(1-\Di)\scoreadjzeroi$.
Then $\Di \scoreadjesti = \Di \matgmmest \momesti - \Di (1/\propfn) \gammaest'h_i$, which is of the form studied above.
Then $\en[\frac{\Di}{\propfn} \scoreadjesti \scoreadjesti'] = E[\scoreadjonei \scoreadjonei'] + \opone$.
Arguing similarly for $\Di = 0$, we have $\en[\scoreadjesti \scoreadjesti'] = \propfn \en[\frac{\Di}{\propfn} \scoreadjesti \scoreadjesti'] + (1-\propfn) \en[\frac{1-\Di}{1-\propfn} \scoreadjesti \scoreadjesti'] = \propfn E[\scoreadjonei \scoreadjonei'] + (1-\propfn) E[\scoreadjzeroi \scoreadjzeroi'] + \opone = E[\Di \scoreadjonei \scoreadjonei'] + E[(1-\Di)\scoreadjzeroi \scoreadjzeroi'] + \opone = E[\scoreadji \scoreadji'] + \opone$.
Moreover, $\en[\scoreadjesti] = \en[\matgmmest \momesti - \hti \gammaest'h_i] = \matgmmest \en[\momesti] + \op(1)$.
Note that $\en[\momesti] = \momest(\est)$ and $\momest(\est) - \momest(\thetatrue) = \momtrue(\est) - \momtrue(\thetatrue) + \op(1) = \op(1)$.
The first equality since $|\momest - \momtrue|_{\thetaspace, \infty} = \op(1)$ and the second by continuous mapping, using Lemma \ref{lemma:ulln}.
Since $E[\scoreadji]=0$, it follows that $\var_n(\scoreadjesti) = E[\scoreadji \scoreadji'] + \opone=\var(\scoreadji)+\opone$.
\end{proof}

\begin{lem} \label{lemma:inference-matching}
Require Assumptions \ref{assumption:linear-rerandomization}, \ref{assumption-gmm}, \ref{assumption-inference}.
Then under $P$ in Definition \ref{defn:pure-stratification}, the estimators defined in Section \ref{section:inference-finite} have $\varestcross \convp E[E[\assignifarmh{1} | \psi] E[\assignifarmh{0} | \psi]']$, and $\varestone \convp E[E[\assignifarmh{1} | \psi] E[\assignifarmh{1} | \psi]']$, and $\varestzero \convp E[E[\assignifarmh{0} | \psi] E[\assignifarmh{0} | \psi]']$.
\end{lem}

\begin{proof}
Let $\varestone^o$ denote the oracle version of $\varestone$ with $\assignifarmhoraclei$ substituted for $\assignifarmhesti$, and similarly define oracle versions $\varestzero^o$, $\varestcross^o$ of $\varestzero, \varestcross$.
Note $\Di \assignifarmhoraclei = \Di \assignifarmh{1}(W_i)$.
In Lemma \lemmainference{} of \cite{cytrynbaum2024}, set $A_i=\assignifarmh{1}(W_i)$ and $B_i=\assignifarmh{1}(W_i)$.
Applying the lemma componentwise gives $\varestone^o \convp E[E[\assignifarmh{1} | \psi] E[\assignifarmh{1} | \psi]']$.
Similarly, we have $\varestzero^o \convp E[E[\assignifarmh{0} | \psi] E[\assignifarmh{0} | \psi]']$, and $\varestcross^o \convp E[E[\assignifarmh{1} | \psi] E[\assignifarmh{0} | \psi]']$. 
Then it suffices to show $\varestone - \varestone^o = \op(1)$, $\varestzero - \varestzero^o = \op(1)$, and $\varestcross - \varestcross^o = \op(1)$.
For the first statement, expand
\begin{align*}
\varestone-\varestone^o=(n\propfn)\inv\sum_{\group\in\widetilde{\groupset}_n}\frac{1}{a_{\group}-1}\sum_{i\ne j\in\group}\Di\Dj(\assignifarmhesti(\assignifarmhestj)'-\assignifarmhoraclei(\assignifarmhoraclej)').
\end{align*}
Expand $\assignifarmhesti(\assignifarmhestj)' - \assignifarmhoraclei(\assignifarmhoraclej)' = \assignifarmhesti((\assignifarmhestj)' - (\assignifarmhoraclej)') + (\assignifarmhesti - \assignifarmhoraclei)(\assignifarmhoraclej)' \equiv A_{ij} + B_{ij}$.
Using triangle inequality, $a_{\group}-1\geq1$, and $\propfn>0$, we calculate $|\varestone-\varestone^o|_2\lesssim n\inv\sum_{\group\in\widetilde{\groupset}_n}\sum_{i,j\in\group}(|A_{ij}|_2+|B_{ij}|_2)\equiv A_n+B_n$.
First consider $B_n$.
Using that $|xy'|_2 \leq |x|_2 |y|_2$, we have 
\begin{align*}
|B_{ij}|_2 &\leq |\assignifarmhesti - \assignifarmhoraclei|_2 |\assignifarmhoraclej|_2 \\
&= |\vard \matgmmest \momesti - \vard \matgmm \momi - \Di (\gammaestarm{1} - \gammaoptarm{1})' h_i - (1-\Di) (\gammaestarm{0} - \gammaoptarm{0})' h_i|_2 |\assignifarmhoraclej|_2 \\
&\leq |\matgmmest - \matgmm|_2 |\momesti|_2 |\assignifarmhoraclej|_2+ |\matgmm|_2 |\momesti - \momi|_2 |\assignifarmhoraclej|_2+ 2 \max_{d=0,1} |\gammaestarm{d}-\gammaoptarm{d}|_2 |h_i|_2 |\assignifarmhoraclej|_2.
\end{align*}
Then $B_n=n\inv\sum_{\group\in\widetilde{\groupset}_n}\sum_{i,j\in\group}(|\matgmmest-\matgmm|_2|\momesti|_2|\assignifarmhoraclej|_2+|\matgmm|_2|\momesti-\momi|_2|\assignifarmhoraclej|_2+2\max_{d=0,1}|\gammaestarm{d}-\gammaoptarm{d}|_2|h_i|_2|\assignifarmhoraclej|_2)\equiv B_{n1}+B_{n2}+B_{n3}$.
Consider $B_{n1}$.
This is 
\begin{align*}
B_{n1}&=|\matgmmest-\matgmm|_2 n\inv\sum_{\group\in\widetilde{\groupset}_n}\sum_{i,j\in\group}|\momesti|_2|\assignifarmhoraclej|_2\leq|\matgmmest-\matgmm|_2(2n)\inv\sum_{\group\in\widetilde{\groupset}_n}\sum_{i,j\in\group}(|\momesti|_2^2+|\assignifarmhoraclej|_2^2) \\
&\leq|\matgmmest-\matgmm|_2(2n)\inv\sum_{\group\in\widetilde{\groupset}_n}|\group|\sum_{i\in\group}(|\momesti|_2^2+|\assignifarmhoraclei|_2^2)\lesssim|\matgmmest-\matgmm|_2\en[|\momesti|_2^2+|\assignifarmhoraclei|_2^2].
\end{align*}
By an identical argument $B_{n3} \lesssim \max_{d=0,1}|\gammaestarm{d}-\gammaoptarm{d}|_2\en[|h_i|_2^2+|\assignifarmhoraclei|_2^2]$.
Then to show $B_{n1} + B_{n3} = \op(1)$, suffices to show $\en[|h_i|_2^2 + |\assignifarmhoraclei|_2^2 + |\momesti|_2^2] = \Op(1)$.
That $\en[|h_i|_2^2 + |\momesti|_2^2] = \Op(1)$ was shown in the proof of Lemma \ref{lemma:inference-variance-convergence}. 
Note $\en[|\assignifarmhoraclei|_2^2] = \en[|\vard \matgmm \momi(\thetatrue) - \Di \gammaoptarm{1}'h_i - (1-\Di) \gammaoptarm{0}'h_i|_2^2] \leq 2 \en[|\matgmm \momi|_2^2] + 2 \en[|\Di \gammaoptarm{1}'h_i + (1-\Di) \gammaoptarm{0}'h_i|_2^2] \leq 2 |\matgmm|_2^2 \en[|\momi|_2^2] + 2 \max_{d\in\{0,1\}} |\gammaoptarm{d}|_2^2 \en[|h_i|_2^2] = \Op(1)$ since $E[|\momi|_2^2] < \infty$ by assumption.
Then $B_{n1} + B_{n3} = \op(1)$.
Finally, consider $B_{n2}$.
By the mean value theorem $\momi(\est) - \momi(\thetatrue) = \frac{\partial \momi}{\partial \theta'}(\tilde \theta_i)(\est - \thetatrue)$, where $\tilde \theta_i \in [\thetatrue, \est]$ may change by row. 
Then we have 
\begin{align*}
B_{n2} &= n \inv \sum_{\group \in \widetilde{\groupset}_n} \sum_{i, j \in \group} |\matgmm|_2 |\momesti - \momi|_2 |\assignifarmhoraclej|_2 \leq |\est - \thetatrue|_2 |\matgmm|_2 \cdot n \inv \sum_{\group \in \widetilde{\groupset}_n} \sum_{i, j \in \group}  |\frac{\partial \momi}{\partial \theta'}(\tilde \theta_i)|_2 |\assignifarmhoraclej|_2 \\
& \lesssim |\est - \thetatrue|_2 |\matgmm|_2 \en[|\frac{\partial \momi}{\partial \theta'}(\tilde \theta_i)|_2^2 + |\assignifarmhoraclei|_2^2] = \op(1).
\end{align*}
The final equality follows since $\en[|\frac{\partial \momi}{\partial \theta'}(\tilde \theta_i)|_2^2]=\Op(1)$, as shown in the proof of Lemma \ref{lemma:inference-variance-convergence}.
Then we have shown $B_n = \op(1)$, and $A_n = \op(1)$ is identical. 
This completes the proof that $\varestone - \varestone^o = \op(1)$, and the proof of $\varestzero - \varestzero^o = \op(1)$, and $\varestcross - \varestcross^o = \op(1)$ are identical. 
\end{proof}

\subsection{Lemmas} \label{proofs:lemmas}

\begin{prop}[L\'evy] \label{prop:levy}
Consider probability spaces $(\Omega_n, \filtrationg_n, P_n)$ and $\sigma$-algebras $\filtrationcand \sub \filtrationg_n$.
We say $A_n \in \mr^d$ has $A_n | \filtrationcand \convwprocess A$ if $\phi_n(t) \equiv E[e^{it'A_n}|\filtrationcand] = E[e^{it'A}|\filtrationcand] + \op(1)$ for each $t \in \mr^d$.
If $g: \mr^d \to \mathbb{C}$ is bounded, measurable, and $P(A \in \{a: g(\cdot) \; \text{discontinuous at} \; a\}) = 0$ then we have
\begin{equation} \label{equation:levy}
E[g(A_n) | \filtrationcand] = E[g(A)] + \op(1).
\end{equation}
\end{prop}

\begin{lem} \label{lemma:modes-of-convergence}
Consider probability spaces $(\Omega_n, \filtrationg_n, P_n)$ and $\sigma$-algebras $\filtrationcand \sub \filtrationg_n$.
Suppose $0 \leq A_n \leq B < \infty$ and $A_n = \op(1)$. 
Then $E[A_n | \filtrationcand] = \op(1)$.
\end{lem}

See \cite{cytrynbaum2021} for the proofs.

\begin{lem} \label{lemma:orthogonalization}
The following statements hold
\begin{enumerate}[label={(\alph*)}, itemindent=.5pt, itemsep=.4pt] 
\item There exists $\gammaoptmat \in \mr^{\dimh \times \dimdiff}$ solving $E[\var(h | \psi)] \gammaoptmat = E[\cov(h, \diff | \psi)]$.
For any solution, we have $E[\var(\diff - \gammaoptmat'h | \psi)] \preceq E[\var(\diff - \gammacoeff'h | \psi)]$ for all $\gammacoeff \in \mr^{\dimh \times \dimdiff}$.  
\item Let $Z = (\zdiff, \zh)$ a random variable with $\var(Z) =  E[\var((\diff, h) | \psi)] \equiv \covjoint$ and define $\zindep = \zdiff - \gammaoptmat'\zh$. 
Then $\cov(\zindep, \zh) = 0$.
In particular, if $(\zdiff, \zh)$ are jointly Gaussian, then $\zindep$ is Gaussian with $\zindep \indep \zh$. 
\end{enumerate}
\end{lem}

\begin{proof}
In the notation of (b), it suffices to show $\covjointhh \gammaoptmat = \covjointhdiff$.
If $\rank(\covjointhh) = 0$ then $\zh = c_h$ a.s.\ for constant $c_h$ and $\covjointhdiff = \cov(\zh, \zdiff) = 0$.
Then any $\gammacoeff \in \mr^{\dimh \times \dimdiff}$ is a solution.
Then suppose $\rank(\covjointhh) = r \geq 1$.
Let $\covjointhh = U \Lambda U'$ be the compact SVD with $U \in \mr^{\dimh \times r}$ and $\rank(\Lambda) = r$, and $U'U = I_r$.
We claim $\zh = UU' \zh$ a.s.
Calculate $\var((UU'-I)\zh) = (UU'-I)U\Lambda U'(UU'-I) = 0$.
Note that $\covjointhh \gammacoeff = \covjointhdiff$ $\iff$ $\var(\zh) \gammacoeff = \cov(\zh, \zdiff)$ $\iff$ $\var(UU' \zh) \gammacoeff = \cov(UU' \zh, \zdiff)$ $\iff$ $U[\var(U' \zh) U' \gammacoeff - \cov(U' \zh, \zdiff)] = 0$.  
Define $\zhbar = U'\zh$ and note $\var(\zhbar) = U'U\Lambda U'U = \Lambda \succ 0$.
Then let $\bar \gammacoeff = \var(\zhbar) \inv \cov(\zhbar, \diff)$ so that $\var(\zhbar) \bar \gammacoeff - \cov(\zhbar, \zdiff) = 0$.
Then it suffices to find $\gammacoeff$ such that $U'\gammacoeff = \bar \gammacoeff$.
Since $U': \mr^{\dimh} \to \mr^{r}$ is onto, there exists $\gammacoeff^k$ with $U'\gammacoeff^k = \bar \gammacoeff^k$. 
Then let $\gammaoptmat^k \in [\gammacoeff^k + \ker(U')]$ and set $\gammaoptmat = (\gammaoptmat^k: k=1, \dots, \dimdiff)$, so that $U'\gammaoptmat = \bar \gammacoeff$.
Then $\covjointhh \gammaoptmat = \covjointhdiff$ by work above.
For the optimality statement, calculate
\begin{align*}
&E[\var(\diff - \gammacoeff'h | \psi)] = \covjointdiff - \covjointdiffh \gammacoeff -\gammacoeff'\covjointhdiff + \gammacoeff'\covjointhh \gammacoeff = \covjointdiff - \covjointdiffh (\gammacoeff - \gammaoptmat + \gammaoptmat) \\
&-(\gammacoeff - \gammaoptmat + \gammaoptmat)'\covjointhdiff + \gammacoeff'\covjointhh \gammacoeff  = \covjointdiff -2 \gammaoptmat' \covjointhh \gammaoptmat - (\gammacoeff - \gammaoptmat)' \covjointhdiff -\covjointdiffh (\gammacoeff - \gammaoptmat) \\
&+ \gammacoeff'\covjointhh \gammacoeff \propto - (\gammacoeff - \gammaoptmat)' \covjointhh \gammaoptmat - \gammaoptmat' \covjointhh (\gammacoeff - \gammaoptmat) + \gammacoeff'\covjointhh \gammacoeff =  - (\gammacoeff - \gammaoptmat)' \covjointhh \gammaoptmat \\
&- \gammaoptmat' \covjointhh (\gammacoeff - \gammaoptmat) + \gammacoeff'\covjointhh \gammacoeff + (\gammacoeff - \gammaoptmat + \gammaoptmat)'\covjointhh (\gammacoeff - \gammaoptmat + \gammaoptmat) \\
&=  \gammaoptmat'\covjointhh \gammaoptmat + (\gammacoeff - \gammaoptmat)'\covjointhh (\gammacoeff - \gammaoptmat).
\end{align*}
Then $E[\var(\diff - \gammacoeff'h | \psi)] - E[\var(\diff - \gammaoptmat'h | \psi)]) = (\gammacoeff - \gammaoptmat)'\covjointhh (\gammacoeff - \gammaoptmat)$ and for any $a \in \mr^{\dimdiff}$ we have $a'(\gammacoeff - \gammaoptmat)'\covjointhh (\gammacoeff - \gammaoptmat) a \geq 0$ since $\covjointhh \succeq 0$.
This proves the claim.
Finally, we have $\cov(\zindep, \zh) = \cov(\zdiff - \gammaoptmat'\zh, \zh) = \covjointdiffh - \gammaoptmat' \covjointhh = 0$.    
The final statement follows from well-known facts about the normal distribution. 
\end{proof}

\begin{lem} \label{lemma:big-op}
$A_n = \Op(1)$ $\iff$ $A_n = \op(c_n)$ for every sequence $c_n \to \infty$.
\end{lem}

\begin{proof}
It suffices to consider $A_n \geq 0$.
The forward direction is clear.
For the backward direction, suppose for contradiction that there exists $\epsilon > 0$ such that $\sup_{n \geq 1} P(A_n > M) > \epsilon$ for all $M$.
Then find $n_k$ such that $P(A_{n_k} > k) > \epsilon$ for each $k \geq 1$.
We claim $n_k \to \infty$.
Suppose not and $\liminf_k n_k \leq N < \infty$. 
Then let $k(j) \to \infty$ such that $n_{k(j)} \leq N$ for all $j$.
Choose $M' < \infty$ such that $P(A_n > M') < \epsilon$ for all $n =1, \dots N$.  
Then for $k(j) > M'$ we have $P(A_{n_{k(j)}} > k(j)) \leq P(A_{n_{k(j)}} > M') < \epsilon$, which is a contradiction.
Then apparently $\lim_k n_k = +\infty$. 
Define $Z_j = \{i: i \geq j\}$. 
Regard the sequence $n_k$ as map $n: \mathbb N \to \mathbb N$.
For $m \in \image(n)$, define $n \mooreinv(m) = \min n \inv(m)$. 
It's easy to see that $n \mooreinv(m_k) \to \infty$ for $\{m_k\}_k \sub \image(n)$ with $m_k \to \infty$. 
Then write  
\begin{align*}
\sup_{k \geq j} P(A_{n_k} > k) &= \sup_{m \in n(Z_j)} \sup_{a \in n\inv(m)} P(A_m > a) \leq \sup_{m \in n(Z_j)} P(A_m > n\mooreinv(m)) 
\end{align*}
Note $A_{m_k} / n\mooreinv(m_k) = \op(1)$ by assumption for any $\{m_k\}_k \sub \image(n)$ with $m_k \to \infty$.  
Then we have
\begin{align*}
\limsup_k P(A_{n_k} > k) &= \lim_j \sup_{k \geq j} P(A_{n_k} > k) = \lim_j \sup_{m \in n(Z_j)} P(A_m > n\mooreinv(m)) = o(1). 
\end{align*}
This is a contradiction, which completes the proof.
\end{proof}

\end{document}